\documentclass[journal,final,twoside]{IEEEtran}
\usepackage[utf8]{inputenc}
\usepackage[T1]{fontenc}
\usepackage{relsize}
\usepackage{flushend}

\usepackage[normalem]{ulem}
\usepackage{multirow}
\usepackage{xfrac}
\usepackage{microtype}

\AtBeginDocument{\hypersetup{pdftitle={Generalized Fast-Convolution-based Filtered-OFDM: Techniques and Application to 5G New Radio}}}
\AtBeginDocument{\hypersetup{pdftitle={Generalized Fast-Convolution-based Filtered-OFDM: Techniques and Application to 5G New Radio}}}

\usepackage{graphicx} 
\DeclareRobustCommand*{\IEEEauthorrefmark}[1]{\raisebox{0pt}[0pt][0pt]
{\textsuperscript{\footnotesize\ensuremath{\ifcase#1\or 1\or 2\or 3\or%
    4\or 5\or 6\or 7\or 8
\else\textsuperscript{\expandafter\romannumeral#1}\fi}}}}
\usepackage[table,svgnames]{xcolor} % JYK 28.06.2013, 11:17
\usepackage[most]{tcolorbox}
\usepackage{tikz}
\usetikzlibrary{svg.path}
\usepackage{scalerel}

\definecolor{orcidlogocol}{HTML}{A6CE39}
\tikzset{
  orcidlogo/.pic={
    \fill[orcidlogocol] svg{M256,128c0,70.7-57.3,128-128,128C57.3,256,0,198.7,0,128C0,57.3,57.3,0,128,0C198.7,0,256,57.3,256,128z};
    \fill[white] svg{M86.3,186.2H70.9V79.1h15.4v48.4V186.2z}
                 svg{M108.9,79.1h41.6c39.6,0,57,28.3,57,53.6c0,27.5-21.5,53.6-56.8,53.6h-41.8V79.1z M124.3,172.4h24.5c34.9,0,42.9-26.5,42.9-39.7c0-21.5-13.7-39.7-43.7-39.7h-23.7V172.4z}
                 svg{M88.7,56.8c0,5.5-4.5,10.1-10.1,10.1c-5.6,0-10.1-4.6-10.1-10.1c0-5.6,4.5-10.1,10.1-10.1C84.2,46.7,88.7,51.3,88.7,56.8z};
  }
}

\newcommand\orcidicon[1]{\,\,\href{https://orcid.org/#1}{\mbox{\scalerel*{
\begin{tikzpicture}[yscale=-1,transform shape]
\pic{orcidlogo};
\end{tikzpicture}
}{|}}}}

\colorlet{spotColor}{gray!100!black}

\usepackage{amsmath}
\usepackage{amssymb}
\usepackage{newtxmath} % Times New Roman for maths
\usepackage[cal=boondoxo,bb=boondox]{mathalfa}

\usepackage[  
colorlinks=true,
breaklinks=true, 
allcolors=DarkBlue,
urlbordercolor={1 0 0}, % hyperlink borders will be red
pdfborderstyle={/S/U/W 0.5},% border style will be underline of width 0.5pt
]{hyperref}
\hypersetup{pdflinkmargin=-0.5pt} % Ignore hyperref for now

\hypersetup{%
  pdfsubject={This paper investigates the application of fast convolution (FC) filtering schemes for flexible and effective waveform generation and processing in 5th generation (5G) systems.},
  pdfkeywords={5G, physical layer, 5G new radio, 5G-NR, multicarrier, waveforms, filtered-OFDM, fast-convolution},
  pdfauthor={Juha Yli-Kaakinen, Toni Levanen, Arto Palin, Markku Renfors, and Mikko Valkama}%
}
\hypersetup{pdflinkmargin=0.5pt}

\usepackage{siunitx}
\usepackage{booktabs} % Typesetting for tables
\usepackage{cite}
 
\providecommand{\iu}{\ensuremath{{\mathop{\mspace{1mu}\mathrm{j}\mspace{0.5mu}}\nolimits}}}
\usepackage{mathtools}
\DeclarePairedDelimiter{\diagpars}{(}{)}

\DeclarePairedDelimiter{\olapars}{(}{)}
\DeclarePairedDelimiter{\conjpars}{(}{)}
\DeclarePairedDelimiter{\vecpars}{(}{)}
\DeclarePairedDelimiter{\norm}{\lVert}{\rVert}
\DeclarePairedDelimiter{\abs}{\lvert}{\rvert}
\newcommand{\diag}{\operatorname{diag}\diagpars}
\newcommand{\bdiag}{\operatorname{bdiag}\diagpars}
\newcommand{\buff}{\operatorname{buff}\diagpars}
\newcommand{\ola}{\operatorname{ola}\olapars}
\newcommand{\conj}{\operatorname{conj}\conjpars}
\renewcommand{\vec}{\operatorname{vec}\vecpars}
\newcommand{\ivec}{\operatorname{vec}^{-1}\vecpars}

\usepackage[framemethod=TikZ]{mdframed}

\makeatletter
\newcommand*{\transpose}{%
  {\mathpalette\@transpose{}}%
}
\newcommand*{\@transpose}[2]{%
  \raisebox{\depth}{$\m@th#1\intercal$}%
}

\def\ps@IEEEtitlepagestyle{
  \def\@oddfoot{\mycopyrightnotice}
  \def\@evenfoot{}
}
\def\mycopyrightnotice{
  {\footnotesize
  \begin{minipage}{\textwidth}
  \centering
  \copyright 2020 IEEE. Personal use of this material is permitted. Permission from IEEE must be obtained for all other users, including reprinting/ republishing this material for advertising or promotional purposes, creating new collective works for resale or redistribution to servers or lists, or reuse of any copyrighted components of this work in other works.
  \end{minipage}
  }
}
\makeatother
\usepackage[nolist,nohyperlinks]{acronym} 

\makeatletter 
\DeclareMathSizes{\@xpt}{\@xpt}{7}{5}
\makeatother

\newlength{\figWidth}
\begin{document} 
% \relscale{0.97}

\title{Generalized Fast-Convolution-based Filtered-OFDM: Techniques and Application to 5G New Radio}
\title{\mbox{Generalized Fast-Convolution-based Filtered-OFDM:} Techniques and Application to 5G New Radio}
\author{Juha Yli-Kaakinen\orcidicon{0000-0002-4665-9332}, Toni Levanen\orcidicon{0000-0002-9248-0835}, 
Arto Palin\orcidicon{0000-0001-8567-5549}, \\ 
Markku~Renfors\orcidicon{0000-0003-1548-6851},~\IEEEmembership{Life Fellow,~IEEE}, and Mikko~Valkama\orcidicon{0000-0003-0361-0800},~\IEEEmembership{Senior~Member,~IEEE}
\thanks{This work was supported in part by Nokia Bell Labs, in part by Nokia Networks, and in part by the Academy of Finland under Project 288670. The work was also supported by the Academy of Finland under the grant 319994.}
\thanks{J. Yli-Kaakinen, T. Levanen, M. Renfors, and M. Valkama are with
Tampere University, FI-33720, Tampere, Finland (e-mail: $\lbrace$juha.yli-kaakinen; toni.levanen; markku.renfors; mikko.valkama$\rbrace$@tuni.fi)}
\thanks{A. Palin is with Nokia Networks, Finland (e-mail: arto.palin@nokia.com)}
\thanks{This article contains multimedia material, available at \url{http://yli-kaakinen.fi/GeneralizedFastConvolution/}}
\thanks{Digital Object Identifier \href{http://dx.doi.org/10.1109/TSP.2020.2971949}{10.1109/TSP.2020.2971949}}
}  

% \IEEEpubid{%
%   {\footnotesize
%     \begin{minipage}{\textwidth}\ \\[12pt]
%       \centering
%       \copyright 2017 IEEE. Personal use of this material is permitted. Permission from IEEE must be obtained for all other users, including reprinting/republishing this material for advertising or promotional purposes, creating new collective works for resale or redistribution to servers or lists, or reuse of any copyrighted components of this work in other works.
%     \end{minipage}
%   }
% }

% \markboth{Submitted to IEEE TRANSACTIONS ON SIGNAL PROCESSING}%
\markboth{IEEE TRANSACTIONS ON SIGNAL PROCESSING}%
{Yli-Kaakinen \MakeLowercase{\text\it{et al.}}: Generalized Fast-Convolution-Based Filtered-OFDM: Techniques and Application to 5G New Radio}

\maketitle  
\begin{abstract}
  This paper proposes a generalized model and methods for \ac{fc}-based waveform generation and processing with specific applications to \ac{5g-nr}.  Following the progress of \ac{5g-nr} standardization in \ac{3gpp}, the main focus is on subband-filtered \ac{cp} \ac{ofdm} processing with specific emphasis on spectrally well localized transmitter processing. Subband filtering is able to suppress the interference leakage between adjacent subbands, thus supporting different numerologies for so-called bandwidth parts as well as asynchronous multiple access. The proposed generalized \ac{fc} scheme effectively combines overlapped block processing with time- and frequency-domain windowing to provide highly selective subband filtering with very low intrinsic interference level.  Jointly optimized multi-window designs with different allocation sizes and design parameters are compared in terms of interference levels and implementation complexity. The proposed methods are shown to clearly outperform the existing state-of-the-art windowing and filtering-based methods.
\end{abstract}    
%\vspace{-3mm}

\section{Introduction}   %----------------------------------------------------------------------
\label{sec:introduction}\bstctlcite{IEEEexample:BSTcontrol}
\IEEEPARstart{O}{rthogonal} frequency-division multiplexing (OFDM)\acused{ofdm} is the dominating multicarrier modulation scheme, being extensively deployed in modern radio access systems. \ac{ofdm} offers high flexibility and efficiency in allocating spectral resources to different users, simple and robust way of channel equalization due to the inclusion of \acf{cp}, as well as simplicity of combining multi-antenna schemes with the core physical layer processing \cite{B:Dahlman2018}.  The main drawback is the limited spectrum localization, especially, in challenging new spectrum use scenarios, like asynchronous multiple access, as well as mixed-numerology cases aiming to use adjustable symbol  and \ac{cp} lengths, \acp{scs}, and frame structures depending on the service requirements \cite{J:20145GNOW,J:2014BanelliModFormatsAndWaveformsFor5G}. 
  
\IEEEpubidadjcol\acused{cp-ofdm,5g,nr}    
\begin{figure}[t!]      
  \centering      
  \includegraphics[trim=0 3 0 0,clip, % trim l b r t
  width=0.8\figWidth]{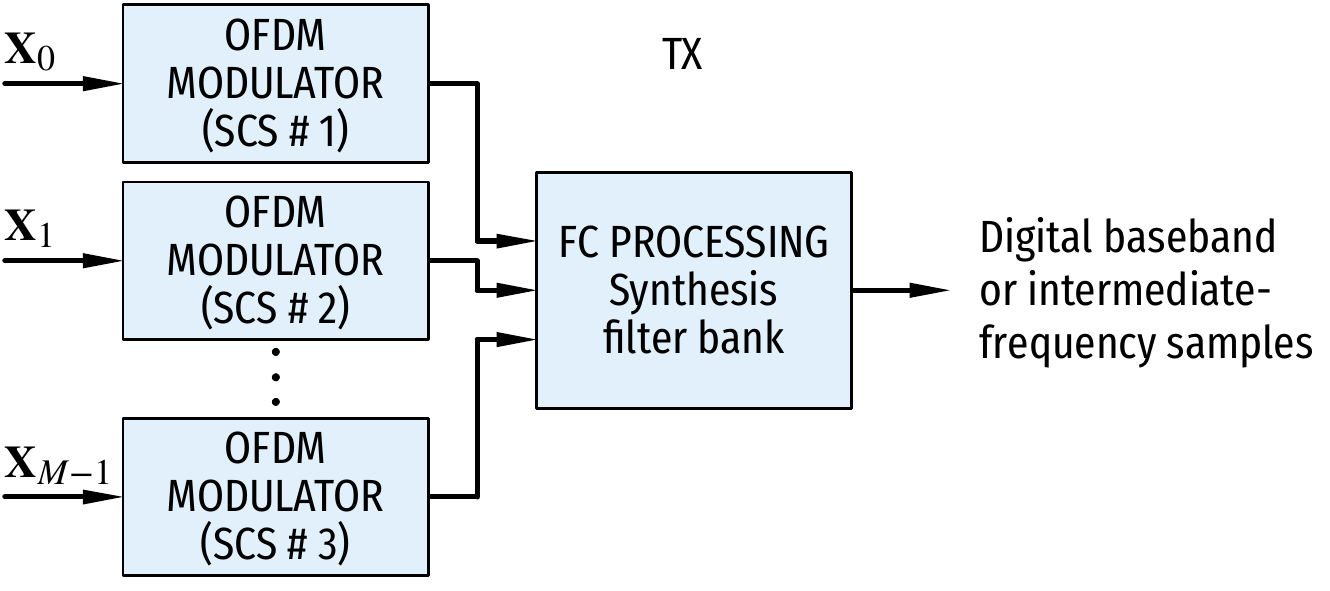}
  \caption{In \acf{fc}-based filtered-OFDM, filtering is applied at subband level, which means one or multiple contiguous \acfp{prb} with same \acf{scs}, while utilizing normal CP-OFDM waveform for the \acsp{prb}. Transmitter processing uses \acs{fc} synthesis filter bank for combining $M$ subbands.}
  \label{fig:Channelization} 
\end{figure}
% \addtolength{\textheight}{-1cm}

In the emerging \acf{5g-nr} mobile networks, building on \ac{ofdma} based radio interface, the ability to control the spectral characteristics of the transmitted waveform and receiver filtering, either over the whole carrier or over the so-called bandwidth parts \cite{S:3GPP:TS38.300} or subbands, is of particular importance. This is primarily due to the requirements to increase the bandwidth utilization, in the form of wider passband for given channel bandwidth, and to support efficient multiplexing of different services with different radio interface numerologies within a carrier. An important concrete example is the frequency multiplexing of \ac{embb} and \ac{urllc} services within the same \acs{nr}\acused{nr} carrier, with different \acp{scs} (e.g., \SI{15}{kHz}, \SI{30}{kHz}, and \SI{60}{kHz}) as illustrated in Fig.~\ref{fig:Channelization}. Such multiplexed signals with different \acp{scs} are not orthogonal, calling for enhanced spectral localization of the signals to minimize interference while allowing high spectral efficiency and flexibility. 

Additionally, it is important to acknowledge that the \ac{nr} specifications \cite[and references therein]{S:3GPP:TS38.300} do not explicitly state how such spectral enhancements are implemented, and that the spectral enhancement methods used in \acp{tx} must be transparent to \acp{rx} (and vice versa) \cite{S:3GPP:TR38.802}. The fundamental definition of transparent waveform processing means that the more advanced \ac{tx} and \ac{rx} solutions have to work with basic \acs{cp-ofdm}\acused{cp-ofdm} \ac{rx} or \ac{tx}, respectively \cite{J:Levanen18:TransparentTxAndRx}. For this reason, in this article, the reference \ac{rx} assumed in the optimization of the transmitter side signal processing methods is a basic \ac{cp-ofdm} \ac{rx}, while the use of more advanced RX schemes is then considered in the numerical radio link performance evaluations. It is also noted that transparent operation does not require completely error or distortion free signal quality with different \ac{tx}-\ac{rx} combinations but that the quality fulfills the requirements set by  the considered technology and standard \cite{S:3GPP:TR38.802}.

In general, there are multiple alternative solutions in the existing literature to improve the spectral characteristics of \acs{cp-ofdm}\acused{cp-ofdm} systems \cite{C:Wang06:WOLA, C:2015_Zhang_f-OFDM_for_5G, C:2016_Coexistence_UFOFDM_CPOFDM, TR:mmMACIG-D4.1, C:Zayani16, J:Zhang18:F-OFDM}. Time-domain windowing combined with overlap-and-add processing, commonly referred to as \ac{wola}, is a computationally inexpensive method to improve the spectral containment to a certain extent \cite{C:Wang06:WOLA,C:Zayani16}. The non-rectangular window with smooth transitions require either the extension of the symbol duration, reducing the spectral efficiency,  or using part of the \ac{cp}, resulting to a reduced tolerance to time dispersion \cite{TR:mmMACIG-D4.1}. Time-domain convolution based \ac{f-ofdm} or \ac{uf-ofdm}, in general, require high-order filters with high complexity to achieve good frequency selectivity with narrow transition bands\cite{C:2015_Zhang_f-OFDM_for_5G,C:2016_Coexistence_UFOFDM_CPOFDM,J:Yli-Kaakinen:JSAC2017}. The complexity can be reduced by using polyphase filter-bank approaches, like \cite{J:Li2014:RB-F-OFDM}, however, in this case the subbands typically have the same bandwidths and thus the configurability is impaired.

In transform-based \acf{fc} filtering solutions, \ac{td} convolution is realized through element-wise multiplication (windowing) of \ac{fd} sequences. This approach provides very high flexibility since the subband frequency response can be directly determined using the \ac{fd} window values, allowing arbitrary bandwidths and center frequencies for the subbands. In our earlier works \cite{C:Renfors2015:fc-f-ofdm,C:Renfors16:adjustableCP,J:Yli-Kaakinen:JSAC2017}, we have developed \ac{fc}-based processing solutions for efficient and flexible implementation of filtered \ac{ofdm} physical layer in \ac{5g-nr}.  In \cite{J:Yli-Kaakinen:JSAC2017}, the complexity and the performance of the \ac{fc} processing are also compared with \ac{wola} and other filtered-\ac{ofdm} schemes.

In this article, we develop and describe generalized \ac{fc}-based processing solutions where, in addition to \ac{fd} windowing, also the \ac{td} windowing is properly combined in the processing, yielding an additional degree of freedom in the overall processing optimization. To the authors' best knowledge, such optimized multi-domain processing solution has not been described in the existing literature. More specifically, the main contributions of this manuscript can be itemized as follows:
\begin{itemize}
\item[$\blacktriangleright$] Generalized \ac{fc} processing based on simultaneously adopting both \ac{td} and \ac{fd} windows is described and proposed. 
\item[$\blacktriangleright$] The \ac{fc}-based filter bank design is formulated as an optimization problem for minimizing the intrinsic passband distortion subject to given subband confinement constraint. Reduced parametrization model for the optimization is also described for simplifying the optimization problem.
\item[$\blacktriangleright$] When both the TD and FD windows are jointly optimized, the proposed method is shown to be an effective way to realize subband-filtered \ac{ofdm} schemes, with considerably improved performance and only slightly increased computational complexity when compared to original \ac{fc}-based approaches with the same baseline processing resolution. 
\item[$\blacktriangleright$] It is shown that the performance improvement is due to the optimization of all the windows simultaneously. Consequently, optimal window functions are not achievable by separate optimization or by analytical means.
\item[$\blacktriangleright$] Generalized \ac{fc}-based approaches with increased \ac{fc} \ac{bs} are shown to have good performance with greatly reduced complexity and processing latency. The possibility of selecting the \ac{ofdm} \ac{scs} and \ac{fc} \ac{bs} independently is also shown to result in increased flexibility in selecting the subband center frequencies. This extension of \ac{fc-f-ofdm} is applicable also in the basic configuration, without time-domain windowing.
\end{itemize}

The remainder of this paper is organized as follows. Section~\ref{sec:mult-fast-conv} reviews the multirate \ac{fc} idea and describes the \ac{cp-ofdm} \ac{tx} processing model for the \ac{fc-f-ofdm}. Then the generalized model for the \ac{fc}-based synthesis filter-bank processing is introduced along with the basic overlap-and-save and overlap-and-add schemes as its special cases. Finally, two possible \ac{cp-ofdm} \ac{rx} configurations are described enabling transparent \ac{rx} processing with basic \ac{cp-ofdm} receiver. Generalized \ac{fc} processing based physical layer waveform design is formulated as an optimization problem in Section \ref{sec:fast-conv-filt-1}. In this problem, the goal is to minimize the subcarrier-level passband \ac{evm} subject to the given subband confinement constraint. Reduced parameterization model is also described improving the convergence of the optimization. In Section \ref{sec:impl-compl}, the implementation complexity of the proposed generalized \ac{fc} processing is analyzed. Section~\ref{sec:fc-based-f} presents numerical results for the optimized scheme with \acf{3gpp} Release 15 \ac{5g-nr} numerology. In addition, the trade-offs between the performance and the complexity for various alternative designs are exemplified. Furthermore, \ac{5g-nr} radio link simulation results are provided verifying the feasibility of the proposed approach. Finally, the conclusions are drawn in Section \ref{sec:conclusions}.

\section*{Notation and Terminology}
In the following, boldface upper and lower-case letters denote matrices and column vectors, respectively. $\mathbf{0}_{q\times p}$ and $\mathbf{1}_{q\times p}$ are the $q\times p$ matrices of all zeros and all ones, respectively. $\mathbf{I}_{q}$ is the identity matrix of size $q$. The entry on the $i$th row and $j$th column of a $q\times p$ matrix $\mathbf{A}$ is denoted by $[\mathbf{A}]_{i,j}$ for $i\in\{1,2,\dots,q\}$ and $j\in\{1,2,\dots,p\}$ and $[\mathbf{A}]_j$ denotes the $j$th column of $\mathbf{A}$. For vectors, $[\mathbf{a}]_j$ denotes the $j$th element of $\mathbf{a}$. The column vector formed by stacking vertically the columns of $\mathbf{A}$ is $\mathbf{a}=\vec{\mathbf{A}}$. $\diag{\mathbf{a}}$ denotes a diagonal matrix with the  elements of $\mathbf{a}$ along the main diagonal. The transpose of matrix $\mathbf{A}$ or vector $\mathbf{a}$ is denoted by $\mathbf{A}^\transpose$ and $\mathbf{a}^\transpose$, respectively. The Euclidean norm is denoted by $\norm{\cdot}$ and $\lvert\cdot\rvert$ is the absolute value for scalars and cardinality for sets. Aperiodic (linear) and cyclic convolution are denoted by $*$ and $\circledast$, respectively. The list of most common symbols is given in Table~\ref{tab:symbols}.

\begin{table}[t]
  \caption{List of Symbols, where subscript $m$ denotes the subband index}
  \label{tab:symbols}
  \centering
  \vspace{-0.8em}
  \footnotesize{
   \begin{tabular}{ccl}
     \toprule
     Notation               & Dimension    & Description \\ 
     \midrule
     $M$                    & $\mathbb{N}$ & Number of subbands \\
     $N$                    & $\mathbb{N}$ & \acs{fc} \acs{ifft} length \\
     $L_m$                  & $\mathbb{N}$ & \acs{fc} \acs{fft} length \\
     $I_m$                  & $\mathbb{N}$ & Interpolation factor in \ac{fc} processing \\
     $L_{\text{S},m}$       & $\mathbb{N}$ & Non-overlapping block length at low rate \\
     $B_{\text{OFDM},m}$    & $\mathbb{N}$ & Number of \acs{ofdm} symbols \\
     $R_m$                  & $\mathbb{N}$ & Number of \ac{fc} processing blocks  \\
     $L_{\text{ACT},m}$     & $\mathbb{N}$ & Number of active subcarriers\\
     $L_{\text{CP},m}$      & $\mathbb{N}$ & Low-rate \acs{cp} length in samples\\
     $L_{\text{OFDM},m}$    & $\mathbb{N}$ & Low-rate \acs{ofdm} \acs{ifft} transform length\\
     $N_{\text{CP},m}$      & $\mathbb{N}$ & High-rate \acs{cp} length in samples\\
     $N_{\text{OFDM},m}$    & $\mathbb{N}$ & High-rate \acs{ofdm} \acs{ifft} transform length \\
     $S_{\text{F},m}$       & $\mathbb{N}$ & Zero padding in block processing\\
     $f_\text{s}$           & $\mathbb{R}$ & Sampling frequency [Hz] \\
     $f_{\text{SCS},m}$     & $\mathbb{R}$ & \ac{ofdm} \acl{scs} [Hz] \\
     $f_{\text{BS},m}$      & $\mathbb{R}$ & \ac{fc} processing bin spacing [Hz] \\
     $\lambda$              & $\mathbb{R}$ & \ac{fc} processing overlap factor \\
     $\mathbf{D}_m$         & $\mathbb{R}^{L_m\times{L_m}}$ & Diagonal \ac{fd} windowing matrix \\
     $\mathbf{A}_m$         & $\mathbb{R}^{L_m\times{L_m}}$ & Diagonal \ac{td} analysis windowing matrix   \\
     $\mathbf{S}$           & $\mathbb{R}^{N\times{N}}$ & Diagonal \ac{td} synthesis windowing matrix   \\
     $\mathbf{W}_p$         & $\mathbb{C}^{p\times{p}}$ & \acs{dft} matrix  \\
     $\mathbf{W}^{-1}_{p}$  & $\mathbb{C}^{p\times{p}}$ & \acs{idft} matrix \\
     \bottomrule
    \end{tabular}}   
\end{table} 

In this article, windowing is an operation of multiplying element-wise the finite-length input sequence by a finite-length window function. Aperiodic convolution refers both to a process of evaluating the convolution sum as
\begin{equation}
  \label{eq:conv}
  y(n) = x(n)\ast h(n)=\sum_{k=-\infty}^{\infty}x(k)h(n-k)
\end{equation}
as well as the result of the above sum. In cyclic or circular convolution, the summation is defined for the input sequences of finite length $N$ as follows:
\begin{equation}
  \label{eq:cconv}
  y(n) = x(n)\circledast h(k)=\sum_{k=0}^{N-1}x(k)h((n-k)\bmod N).
\end{equation}

Finally, following the \ac{3gpp} terminology, we refer to a contiguous set of neighboring subcarriers as \ac{prb}, which in \ac{5g-nr} numerology contains \num{12} subcarriers \cite{B:Dahlman2018}.

% ======================================================================
\section{Multirate Fast-Convolution and Filter Banks}%    
\label{sec:mult-fast-conv} 
% ======================================================================
In general, \acf{fc} refers to the techniques for evaluating the convolution sum faster (with lower complexity) than the direct evaluation of \eqref{eq:conv} or \eqref{eq:cconv}. In the following, the focus is on \ac{fft}-based \ac{fc} algorithms.  These algorithms carry out the circular convolution by element-wise multiplication in frequency domain according to discrete convolution theorem \cite{J:Hunt71}. In this approach, the \ac{td} input sequence and the filter impulse response are transformed to frequency domain using the \ac{fft} and the resulting sequence is converted back to time domain using the \ac{ifft}. Alternatively, according to the theorem, the windowing in time-domain corresponds to the circular convolution in frequency-domain. The concept of filtering in signal processing context is commonly associated with \ac{td} convolution or \ac{fd} windowing due to their ability to remove undesired spectral components from the signal. However, strictly speaking, the time-domain windowing can also be considered as a filtering since it essentially modifies the spectrum through the associated frequency-domain convolution.

In the context of running convolution schemes, \ac{ols} or \ac{ola} processing is typically applied for processing long sequences \cite{B:Rabiner75} as well as for minimizing the \ac{td} aliasing resulting from associated cyclic convolution.  In the conventional \ac{ols} processing, the signal to be processed is first divided into blocks of length $L$ such that each block overlaps with the previous block by $L_{\text{O}}$ samples. Then these blocks are circularly convolved by the filter impulse response, $L_{\text{O}}$ samples are discarded from the resulting output blocks, and the remaining parts are concatenated to form the output signal. Provided that $L_{\text{O}}$ is greater than or equal to the order of the filter, then the above processing evaluates the aperiodic convolution exactly. Alternatively, in traditional \ac{ola} processing, the signal is divided into non-overlapping  blocks of length $L_{\text{S}}=L-L_{\text{O}}$ and then these blocks are zero-padded to length $L$. The zero-padded blocks are circularly convolved by the filter impulse response and the output signal is composed by first overlapping the blocks such that each block overlaps the previous block by $L_{\text{O}}$ samples and then adding the respective samples.

\begin{figure}[t!]      
   \centering  
  \includegraphics[width=1.0\figWidth]{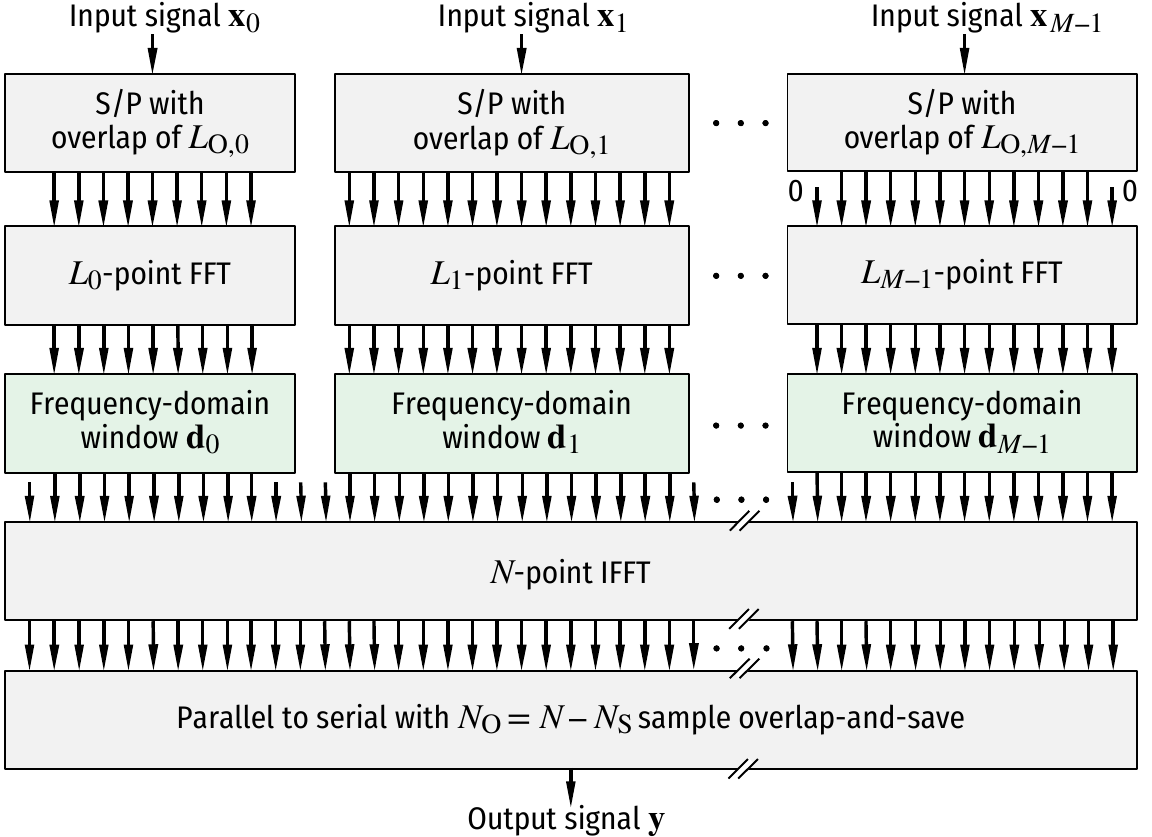}
  \caption{Original \ac{fc}-based \acl{sfb} structure. In this structure, only the \ac{fd} windows are adjustable.}
  \label{fig:Orig_structure}    
\end{figure}   

% ----------------------------------------------------------------------
\subsection{Original Fast-Convolution Filter-Bank Schemes}%
% ----------------------------------------------------------------------
\label{sec:orig-fast-conv}
The application of \ac{fc} to multirate filters has been presented in \cite{PhD:Kwan90}, and \ac{fc} realizations of channelization filters have been considered in \cite{J:Boucheret99, C:Zhang00:Fast-FD-filter, C:Pucker03}. The analysis and optimization methods for \ac{fc}-implementation of nearly perfect-reconstruction filter-bank systems are developed in \cite{J:Renfors14:FC}.  Original \ac{fc}-based synthesis filter-bank structure proposed in \cite{J:Renfors14:FC} is depicted in Fig.~\ref{fig:Orig_structure}, for a case where $M$ incoming low-rate, narrowband signals $\mathbf{x}_m$ for $m=0,1,\dots,M-1$ with adjustable frequency responses and adjustable sampling rates are to be combined into single wideband signal $\mathbf{y}$. This structure efficiently combines the multirate \ac{fc}-based filtering and the straightforward representations of the desired frequency-responses with the \ac{ols} processing to provide, e.g., a generic waveform processing engine for evolving cellular mobile communications systems. The dual structure can be used on the \ac{rx} side as an \ac{afb} for splitting the incoming high-rate, wideband signal into several narrowband signals \cite{J:Yli-Kaakinen16:JSPS}.  Consequently, \ac{fc} approach has been applied for filter-bank multicarrier waveforms in \cite{C:Shao2015ISCAS} and for flexible \ac{sc} waveforms in \cite{C:Renfors2015:ICC}. An optimization-based framework for {\ac{fc-f-ofdm}} waveform processing for \acs{5g-nr} physical layer is developed and evaluated in \cite{J:Yli-Kaakinen:JSAC2017}. 
   
\begin{figure}[t!]       
   \centering 
  \includegraphics[width=1.0\figWidth]{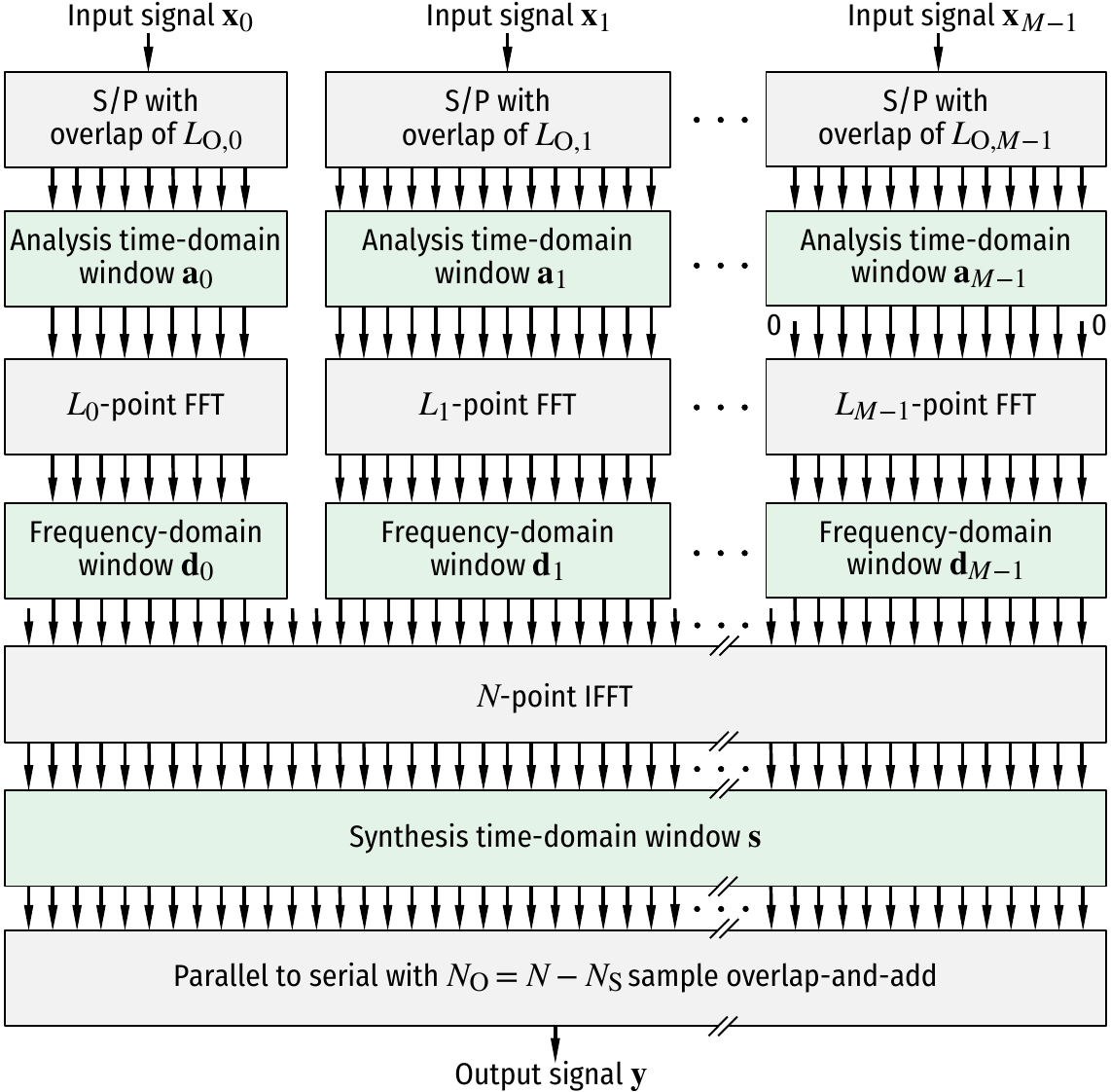}
  \caption{Proposed generalized \ac{fc}-based flexible \acl{sfb} structure. In this structure, the analysis and synthesis \ac{td} windows as well as the \ac{fd} windows are adjustable.}
  \label{fig:Structure}    
\end{figure}   
 
\begin{figure*}[t]        
  \centering 
  \includegraphics[width=0.94\textwidth]{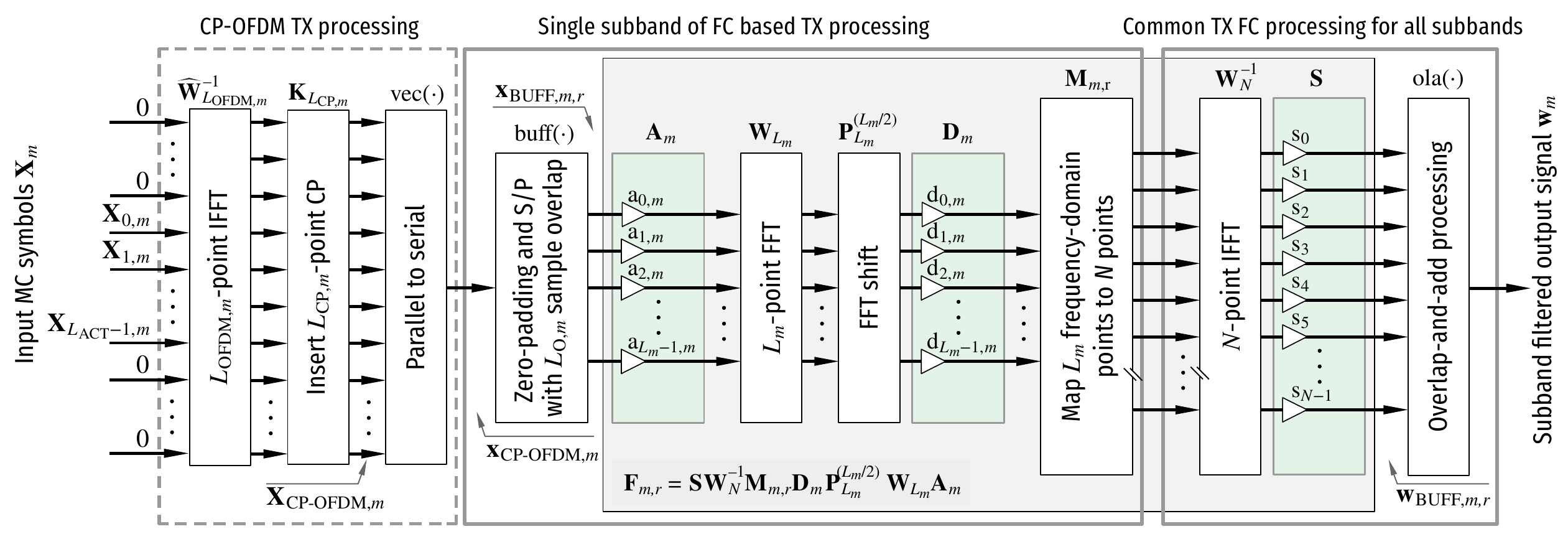} 
  \caption{Proposed generalized \ac{fc-f-ofdm} transmitter. The processing structure for subband $m$ is shown (cf. Fig.~\ref{fig:Structure} for the overall synthesis processing). The long $N$-point \ac{ifft} is common for all the subbands. The \ac{ofdm} \ac{tx} processing shown by the dashed block is the same for both the original and generalized schemes. The \ac{cp-ofdm} \ac{tx} processing is described by \eqref{eq:OFDMmodulation_mtx} and \eqref{eq:OFDMmodulation_vec} while the \ac{tx} \ac{fc} processing is described by \eqref{eq:synth} and \eqref{eq:buff_ola} (or alternatively by \eqref{eq:final_impl}).} 
  \label{fig:FC-F-OFDM_TXblock}    
\end{figure*}  

\subsection{Proposed Generalized Fast-Convolution-based Filtered \ac{ofdm}}%
\label{sec:fast-conv-filt}
% ----------------------------------------------------------------------
In this article, we describe a generalized \ac{fc} processing model providing an improved performance with specific application to subband-filtered \ac{cp-ofdm} \ac{tx} processing. The generalized \ac{fc} processing is optimized and analyzed based on the assumption of a plain \ac{cp-ofdm} \ac{rx}, following the basic principles of transparent signal processing \cite{J:Levanen18:TransparentTxAndRx}. Fig.~\ref{fig:Structure} shows the proposed generalized \ac{fc} synthesis filter-bank structure.  In this structure, each of the $M$ signals to be transmitted is first segmented into overlapping processing blocks of length $L_m$ with overlap of $L_{\text{O},m}$ samples. Let us denote the non-overlapping part by $L_{\text{S},m}=L_m-L_{\text{O},m}$ and the overlap factor by $\lambda=1-L_{\text{S},m}/L_m$. Then, each input block is multiplied element-wise by \ac{td} analysis window $\mathbf{a}_m$ and transformed to frequency domain using $L_m$-point \ac{fft}.  The \ac{fd} bin values of each converted subband signal are multiplied by the \ac{fd} window $\mathbf{d}_m$ corresponding to the \ac{fft} of the finite-length linear filter impulse response.  Finally, the weighted signals are combined and converted back to time domain using $N$-point \ac{ifft} and the resulting \ac{td} output blocks are multiplied by the synthesis window $\mathbf{s}$ and concatenated using the OLA principle such that the overlap between consecutive blocks is $N_{\text{O}}=\lambda N$ samples.
   
In \ac{fc}-based subband-filtered \ac{ofdm} (\ac{fc-f-ofdm}), \ac{fc}-based filtering is applied at subband level, corresponding to one or multiple contiguous \acp{prb} with same \ac{scs}, while utilizing normal \ac{cp}-\ac{ofdm} waveform for the \acp{prb}~\cite{C:Renfors2015:fc-f-ofdm, C:Renfors16:adjustableCP, J:Yli-Kaakinen:JSAC2017}. With \ac{fc-f-ofdm} processing, it is very straightforward to adjust the bandwidths of the subbands individually. This is useful in subband-filtered \ac{ofdm} since there is no need to realize guard bands and filter transition bands between synchronous subbands with the same numerology. In the extreme case, as will be considered in Section \ref{sec:fc-based-f}, the group of filtered \acp{prb} could cover the full carrier bandwidth, and thus the tight channelization filtering for the whole carrier is realized by the \ac{fc} processing. 

Fig.~\ref{fig:FC-F-OFDM_TXblock} illustrates the proposed generalized \ac{fc-f-ofdm} \ac{tx} processing for $m$th subband. The \ac{td} \ac{ofdm} signal $\mathbf{X}_{\text{OFDM},m}$ is obtained from \ac{fd} symbols $\mathbf{X}_{m}$ with $L_{\text{ACT},m}$ active subcarriers by taking the $L_{\text{OFDM},m}$-point \ac{ifft}, that is, with zero-padding of $L_{\text{OFDM},m}-L_{\text{ACT},m}$ subcarriers. Then, \ac{cp} of length $L_{\text{CP},m}$ is inserted and the resulting \ac{cp}-\ac{ofdm} signal $\mathbf{X}_{\text{CP-OFDM},m}$ is converted from parallel to serial for \ac{fc} processing.

Formally, \ac{cp}-\ac{ofdm} \ac{tx} processing of the $m$th subband can be expressed as 
\begin{subequations}
  \label{eq:OFDMmodulation_mtx}
  \begin{equation}
    \label{eq:OFDMmod}
    \mathbf{X}_{\text{CP-OFDM},m}=
    \mathbf{K}_{L_{\text{CP},m}}
    \widehat{\mathbf{W}}_{L_{\text{OFDM},m}}^{-1}\mathbf{X}_m,
  \end{equation}
  where $\mathbf{X}_m\in\mathbb C^{L_{\text{OFDM},m}\times B_{\text{OFDM},m}}$ is the matrix containing the incoming \ac{qpsk} or $\mathcal{M}$-ary \acl{qam} ($\mathcal{M}$-\acs{qam})\acused{qam} symbols (with power normalized to unity),   $\widehat{\mathbf{W}}^{-1}_{L_{\text{OFDM},m}}\in\mathbb C^{L_{\text{OFDM},m}\times L_{\text{OFDM},m}}$ is the \ac{idft} matrix scaled by a factor of $\sqrt{L_{\text{OFDM},m}}$, and $\mathbf{K}_{L_{\text{CP},m}}\in\mathbb N^{(L_{\text{OFDM},m}+L_{\text{CP},m})\times L_{\text{OFDM},m}}$ is the \ac{cp} insertion matrix as given by
  \begin{equation}
    \label{eq:CPins}
    \mathbf{K}_{L_{\text{CP},m}} = 
    \begin{bmatrix}
      \begin{matrix}
        \mathbf{0}_{L_{\text{CP},m}\times(L_{\text{OFDM},m}-L_{\text{CP},m})} & \mathbf{I}_{L_{\text{CP},m}}
      \end{matrix}\\[2pt]
      \mathbf{I}_{L_{\text{OFDM},m}}
    \end{bmatrix}.
  \end{equation}
\end{subequations} 
Here, $\mathbf{I}_{q}$ is $q\times q$ identity matrix and $\mathbf{0}_{q\times p}$ is $q\times p$ zero matrix.

Alternatively, \ac{cp-ofdm} \ac{tx} processing can be represented as
\begin{subequations}
\label{eq:OFDMmodulation_vec}
\begin{equation}
  \mathbf{x}_{\text{CP-OFDM},m} =
  \mathbf{T}_{\text{BD,TX},m}\mathbf{x}_m,
\end{equation}
where $\mathbf{T}_{\text{BD,TX},m}\in\mathbb{C}^{(L_{\text{OFDM},m}+L_{\text{CP},m})B_{\text{OFDM},m} \times L_{\text{OFDM},m}B_{\text{OFDM},m} }$ is block-diagonal \ac{cp-ofdm} modulation matrix given by
\begin{equation}
  \mathbf{T}_{\text{BD,TX},m} =
  \diag[\Big]{
    \underbrace{
      \mathbf{T}_{\text{TX},m},
      \mathbf{T}_{\text{TX},m},
      \dots,
      \mathbf{T}_{\text{TX},m}
    }_{\text{$B_{\text{OFDM},m}$ blocks}}
  }
\end{equation}
with non-overlapping blocks $\mathbf{T}_{\text{TX},m}=\mathbf{K}_{L_{\text{CP},m}}\widehat{\mathbf{W}}_{L_{\text{OFDM},m}}^{-1}$ 
and $\mathbf{x}_m= \vec{\mathbf{X}_m}$ is the column vector formed by stacking vertically the input symbols on subband $m$ as follows:
\begin{equation}
  \label{eq:CP_PtoS}
  \mathbf{x}_m =
  \begin{bmatrix}
    [\mathbf{X}_m]^\transpose_{1} &
    [\mathbf{X}_m]^\transpose_{2} &
    \cdots & 
    [\mathbf{X}_m]^\transpose_{B_{\text{OFDM},m}}
  \end{bmatrix}^\transpose.
\end{equation}
\end{subequations}
Here, $[\mathbf{X}_m]_{k}$ denotes the $k$th column of matrix $\mathbf{X}_m$.   
In the structure of Fig.~\ref{fig:FC-F-OFDM_TXblock}, the \ac{ofdm} \ac{tx} processing module generates samples for the overall symbol duration of $L_{\text{OFDM},m}+L_{\text{CP},m}$ for $B_{\text{OFDM},m}$ symbols. The \ac{fc}-filtering process increases the sampling rate by the factor of 
\begin{equation}
  \label{eqn:Rk} 
	I_m=N/L_{m},  
\end{equation}  
resulting in \ac{ofdm} symbol and \ac{cp} durations of $N_{\text{OFDM},m}=I_m L_{\text{OFDM},m}$ and $N_{\text{CP},m}=I_m L_{\text{CP},m}$, respectively.   
Here $L_{\text{OFDM},m}$ and $L_{\text{CP},m}$ have integer values. It is convenient, but not necessary, that $N_{\text{OFDM},m}$ and $N_{\text{CP},m}$ have integer values as well.

% ----------------------------------------------------------------------
\subsection{Proposed Generalized Model for FC Synthesis Filter Bank}%
\label{sec:generalized-model-fc}
% ----------------------------------------------------------------------
\begin{figure}[t]        
    \centering  
    \includegraphics[width=0.68\figWidth]{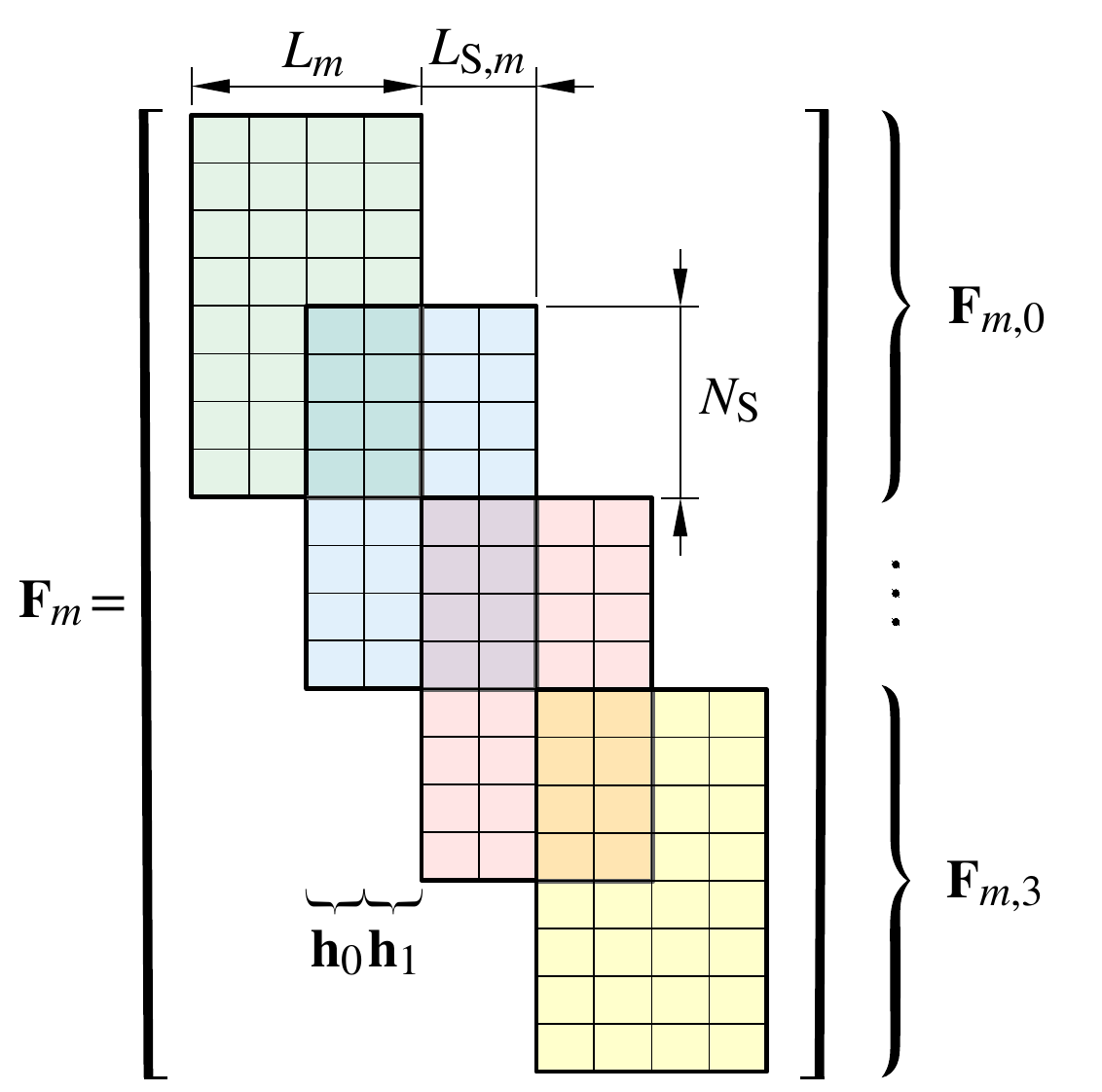}  
    \caption{Structure of block-diagonal synthesis matrix $\mathbf{F}_m$ for four ($R_m= 4$) \ac{fc} processing blocks with $L_m=4$, $N=8$, and $N_\text{S}=2L_{\text{S},m}=4$ ($\lambda=1/2$). The colored $8\times 4$ rectangles illustrate the overlapping processing blocks $\mathbf{F}_{m,r}$ for $r=0,1,2,3$. The impulse responses of the processing are given by the columns $[\mathbf{F}_m]_k$ for $k=3,4,\dots,8$ of the $\mathbf{F}_m$. Here, $h_0$ is given by the columns $[\mathbf{F}_m]_k$ for $k=3,5,7$ and $h_1$ by the columns $[\mathbf{F}_m]_k$ for $k=4,6,8$.}
    \label{fig:synthesisBlock}         
\end{figure}   

In the generalized \ac{fc} \ac{sfb} case, the block processing of $m$th \ac{cp-ofdm} subband signal $\mathbf{x}_{\text{CP-OFDM},m}$ of length
\begin{equation}
  T_m=B_{\text{OFDM},m}(L_{\text{OFDM},m}+L_{\text{CP},m})
\end{equation}
for the generation of high-rate subband waveform $\mathbf{w}_m$ can be represented as 
\begin{subequations}
  \label{eq:BDM}
  \begin{equation}
      \mathbf{w}_m = 
      \mathbf{F}_m \textbf{x}_{\text{ZP},m},
  \end{equation}
  where $\mathbf{F}_m$ is the block diagonal transform matrix of the form
  \begin{align}
    \label{eq:tx_matrix1} 
    \mathbf{F}_m &= \bdiag*{
                   \mathbf{F}_{m,0},
                   \mathbf{F}_{m,1},
                   \dots,
                   \mathbf{F}_{m,R_m-1}
                   }_{L_{\text{O},m},N_\text{O}}
  \end{align}
with overlapping blocks\footnote{Here, $\bdiag{\cdot}_{c,r}$ is an operator for constructing block-diagonal matrix of its arguments. The overlapping between successive blocks is $c$ columns and $r$ rows.} $\mathbf{F}_{m,r}\in\mathbb C^{N\times L_m}$ for $r=0,1,\dots,R_m-1$, as illustrated in Fig.~\ref{fig:synthesisBlock}, and 
\begin{equation}
  \label{eq:zero_padding}
  \mathbf{x}_{\text{ZP},m} = 
  \begin{bmatrix}
    \mathbf{0}_{1\times S_{\text{F},m}} & 
    \mathbf{x}_{\text{CP-OFDM},m}^\transpose & 
    \mathbf{0}_{1\times S_{\text{F},m}}
  \end{bmatrix}^\transpose.
\end{equation}
\end{subequations}
The input signal $\mathbf{x}_{\text{CP-OFDM},m}$ has to be padded in the beginning and end by at least
\begin{equation}
  \label{eq:padding}
  S_{\text{F},m}=L_m-L_{\text{S},m}
\end{equation}
zero-valued samples in order to carry out the processing such that all the incoming samples are facing the same impulse responses (cf. Fig.~\ref{fig:synthesisBlock}). Now, the first sample after the zero-padding is convolved by the impulse response given by $S_{\text{F},m}+1$th column of $\mathbf{F}_{m}$, that is, by $\mathbf{h}_0 = [\mathbf{F}_m]_{S_{\text{F},m}+1}$ whereas the second sample is convolved by $\mathbf{h}_1 = [\mathbf{F}_m]_{S_{\text{F},m}+2}$. Assuming that all the $\mathbf{F}_{m,r}$'s are the same, then the  processing determined by \eqref{eq:BDM} is periodically shift variant in the sense that there are altogether $L_{\text{S},m}$ impulse responses such that every $L_{\text{S},m}$th sample experiences the same impulse response. 

The number of \ac{fc}-processing blocks is given by,
\begin{equation}
  \label{eq:fcblocks}
  R_m = \left\lceil
    \left(
      2S_{\text{F},m} + T_m - L_m
    \right)/L_{\text{S},m}
  \right\rceil+1.
\end{equation} 
The number of blocks can be different for different subbands provided that processing is carried for 
\begin{equation}
  R_{\text{max}}=\max_{m=0,1,\dots,M-1}(R_m)
\end{equation}
blocks. The overall waveform to be transmitted is obtained by summing all the $M$ subband waveforms as 
\begin{equation}
  \label{eq:fcoutput}
  \mathbf{z_{\text{FC-F-OFDM}}}=\sum_{m=0}^{M-1}\mathbf{w}_m.
\end{equation}

The generalized \ac{fc} \ac{sfb} shown in Fig.~\ref{fig:FC-F-OFDM_TXblock} can be represented using block processing by decomposing the $\mathbf{F}_{m,r}$'s as follows:
\begin{subequations}
  \label{eq:synth}
  \begin{equation}
  	\label{eq:ssynth}
    \mathbf{F}_{m,r} = \sqrt{N/L_m}\mathbf{S}\mathbf{U}_{m}\mathbf{C}_{m}\mathbf{A}_{m},
  \end{equation}
  where $\mathbf{A}_{m}\in\mathbb R^{L_m\times L_m}$ and $\mathbf{S}\in\mathbb R^{N\times N}$ are the \ac{td} analysis and synthesis windowing matrices with the analysis and synthesis window weights $\mathbf{a}_m$ and $\mathbf{s}$, respectively, on their diagonals and 
  \begin{equation} 
    \label{eq:circ}
    \mathbf{C}_{m}=\mathbf{W}^{-1}_{L_m} \mathbf{D}_m\mathbf{P}^{(\lceil L_m/2\rceil)}_{L_m} \mathbf{W}_{L_m}
  \end{equation}
  corresponds to the circular-convolution matrix with \ac{dft} shifted processing blocks while 
  \begin{equation}
  	\label{eq:int_mtx}
    \mathbf{U}_{m}=\mathbf{W}^{-1}_{N} \mathbf{M}_{m,r} \mathbf{W}_{L_m}
  \end{equation} 
\end{subequations}
is the interpolation matrix corresponding to zero padding and circular shifting in frequency domain. The first term in \eqref{eq:circ} and the last term in \eqref{eq:int_mtx} cancel out in \eqref{eq:ssynth}, however, those terms are shown here for emphasizing the processing carried out by \eqref{eq:circ} and \eqref{eq:int_mtx}. Here, $\mathbf{W}_{L_m}\in\mathbb C^{L_m\times L_m}$ and $\mathbf{W}_N^{-1}\in\mathbb C^{N\times N}$ are the $L_m\times L_m$ \ac{dft} matrix and $N\times N$ inverse \ac{dft} matrix, respectively.  $\mathbf{P}^{(\lceil L_m/2\rceil)}_m\in\mathbb Z^{L_m\times L_m}$ is the \ac{dft} shift matrix obtained by cyclically shifting the $L_m\times L_m$ identity matrix right by $\lceil L_m/2 \rceil$ positions while the  \ac{fd} mapping matrix $\mathbf{M}_{m,r}\in\mathbb C^{N\times L_m}$ maps $L_m$ \ac{fd} bins of the input signal to \ac{fd} bins $(c_m-\lceil L_m/2\rceil+\ell \bmod N)$ for $\ell=0,1,\dots,L_m-1$ of the output signal. Here $c_m$ is the center bin of the subband $m$. In addition, this matrix rotates the phase of the processing block by
\begin{equation} 
  \label{eq:phase_rot}
  \Theta_m(r) = \exp(\iu 2\pi{r}\theta_{m})\quad\text{with}\quad\theta_m = c_mL_{\text{S},m}/L_m
\end{equation} 
in order to maintain the phase continuity between the consecutive overlapping blocks \cite{J:Renfors14:FC}. The \ac{fd} windowing matrix $\mathbf{D}_m\in\mathbb R^{L_m\times L_m}$ is a diagonal matrix with the \ac{fd} window weights $\mathbf{d}_m$ 
of the subband $m$ on its main diagonal, expressed as 
\begin{equation}
  \label{eq:FDwin} 
  \mathbf{D}_m = \diag{\mathbf{d}_m}.
\end{equation}   
  
The time-domain equivalent of the proposed processing is given by \eqref{eq:ssynth}. As seen from this equation, the analysis windows are applied for each subband separately so this provides a way to adjust the frequency-domain localization of each subband.  The filtering realized through \ac{fc} processing in $\mathbf{C}_m$ provides computationally efficient realization for the convolution with high flexibility in adjusting the bandwidth of the subband.  Considering the interpolation process described by $\mathbf{U}_{m}$, it enables to easily combine and modulate the subband signals to their desired locations with the granularity of \ac{fc} processing bin spacing. However, by inspecting \eqref{eq:int_mtx} it can be noticed that interpolation provided by interpolating \ac{fc} processing corresponds to the frequency-domain zero-padding a.k.a. sinc-interpolation. This equation does not provide any way to control the frequency-response of the interpolator since $\mathbf{M}_{m,r}$ maps $L_m$ frequency-domain bins of the subband signal to $N$ bins of the output signal and these $L_m$ bins are already weighted by the frequency-domain window in $\mathbf{C}_{m}$. On the other hand, the time-domain synthesis window provides additional control for the time-domain localization over the interpolated composite waveform and thus contributes to the spectral localization to certain extent.

The matrix model determined by \eqref{eq:BDM} and \eqref{eq:synth} can be used for analysis and optimization purposes whereas for efficient implementation with real-time hardware or software, it is beneficial to divide the input data into overlapping blocks and carry out the processing for $r=0,1,\dots,R_\text{max}-1$ as
\begin{subequations}
  \label{eq:buff_ola}
  \begin{equation} 
    \label{eq:buff}
    \mathbf{w}_{\text{BUFF},m,r} = \textbf{F}_{m,r} \mathbf{x}_{\text{BUFF},m,r}.
  \end{equation}
  Here, $\mathbf{x}_{\text{BUFF},m,r}$ is the $r$th processing block of $\mathbf{x}_{\text{ZP},m}$ as expressed by
  \begin{equation} 
    [\mathbf{x}_{\text{BUFF},m,r}]_{k} = [\mathbf{x}_{\text{ZP},m}]_{rL_{\text{S},m}+k}
  \end{equation}
  for $k = 1,2,\dots,L_m$. The high-rate subband waveform can be obtained by overlap-and-add processing as given by
  \begin{equation} 
    \mathbf{w}_{m} = \sum_{r=0}^{R_\text{max}-1} \boldsymbol{\Gamma}_{r} \mathbf{w}_{\text{BUFF},m,r},
  \end{equation}
  where
  \begin{equation} 
    \label{eq:ola}
    \boldsymbol{\Gamma}_{r} = 
    \begin{bmatrix}
      \mathbf{0}_{N\times r N_\text{S} } & 
      \mathbf{I}_N & 
      \mathbf{0}_{N \times (R_\text{max}-1)N_\text{S}-r N_\text{S}}
    \end{bmatrix}^\transpose
  \end{equation}
\end{subequations}
overlaps the processed blocks to their desired time-domain locations. The buffering, as expressed by \eqref{eq:buff}, and overlap-and-add processing, as expressed by \eqref{eq:ola}, are denoted in Fig.~\ref{fig:FC-F-OFDM_TXblock} by $\buff{\cdot}$ and $\ola{\cdot}$, respectively. 

By following the above formulation, the overall \acs{fc} \acs{tx} processing can be compactly expressed in three steps: (i) Buffering of the \ac{ofdm} modulated and zero-padded waveforms for $m=0,1,\dots,M-1$ into the overlapping blocks for $r=0,1,\dots,R_\text{max}-1$ and for $k=1,2,\dots,L_m$ as given by 
\begin{subequations}
  \label{eq:final_impl}
  \begin{equation}
    \label{eq:input_buffering}
    [\mathbf{x}_{\text{BUFF},m,r}]_{k} = \begin{bmatrix}
      \mathbf{0}_{S_{\text{F},m}\times 1} \\
      \vec*{\mathbf{K}_{L_{\text{CP},m}}\widehat{\mathbf{W}}_{L_{\text{OFDM},m}}\mathbf{X}_{m}} \\
      \mathbf{0}_{S_{\text{F},m}\times 1} 
    \end{bmatrix}_{rL_{\text{S},m}+k}.
  \end{equation}
  (ii) Processing of the blocks and combining the subbands in frequency domain, expressed as 
  \begin{equation} 
    \label{eq:proc_buff}
    \mathbf{v}_{\text{BUFF},r} = \sum_{m=0}^{M-1}\sqrt{\frac{N}{L_m}}
    \mathbf{M}_{m,r} 
    \mathbf{D}_m\mathbf{P}^{(\lceil L_m/2\rceil)}_{L_m} \mathbf{W}_{L_m}
    \mathbf{A}_{m} 
    \mathbf{x}_{\text{BUFF},m,r}.
  \end{equation}
  (iii) Converting the overall waveform to time-domain and concatenating the processed blocks as given by
  \begin{equation} 
    \mathbf{z}_\text{FC-F-OFDM} = \sum_{r=0}^{R_\text{max}-1} \boldsymbol{\Gamma}_{r} 
    \mathbf{S}\mathbf{W}^{-1}_{N} \mathbf{V}_{\text{BUFF},r},
   \end{equation}
\end{subequations}
where $\boldsymbol{\Gamma}_{r}$ is given by \eqref{eq:ola}.

\begin{figure}[t!]       
  \centering 
  \includegraphics[width=1.0\figWidth]{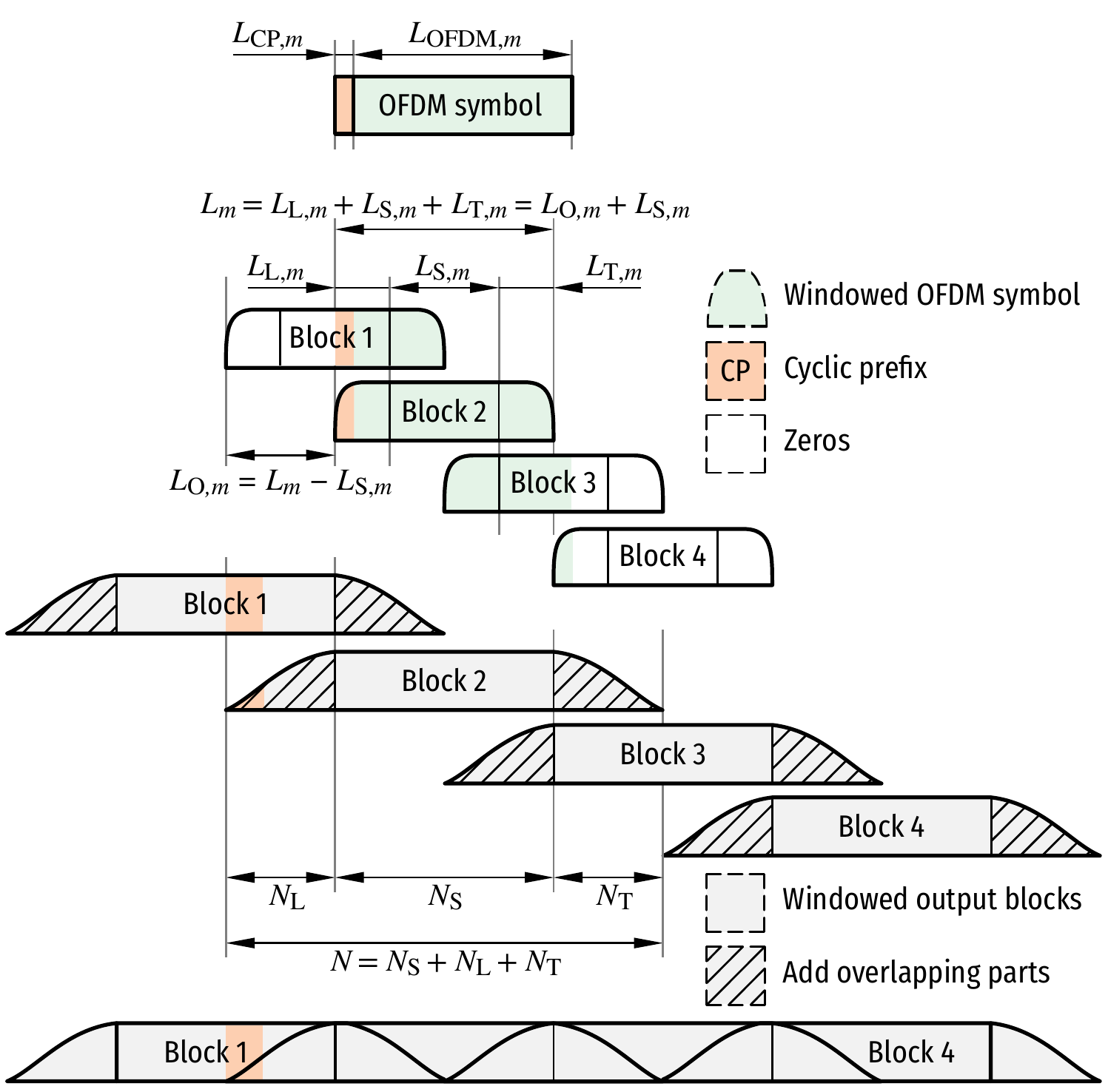} 
  \caption{Example case where four overlapping blocks are required for processing one CP-OFDM symbol with \SI{50}{\%} overlap.} 
  \label{fig:Buffering}     
\end{figure}

Due to the overlapping block processing in \ac{fc}-based filter bank, one \ac{cp}-\ac{ofdm} symbol is divided into multiple \ac{fc} processing blocks and, in general, these processing blocks are not time aligned with the \ac{cp}-\ac{ofdm} symbols in the case of non-zero and possibly varying \ac{cp} lengths. Fig.~\ref{fig:Buffering} depicts an example of block processing for one \ac{cp}-\ac{ofdm} symbol in the case when the ratio of $N$ and $L_m$ is two and the overlap is \SI{50}{\%} ($\lambda=1/2$). For continuous processing of multiple symbols the overhead due to zero-padding and the unmatched \ac{fc} processing block and \ac{cp-ofdm} symbol lengths diminishes as the number of symbols increases and at minimum only one additional \ac{fc}-processing block is needed for overlap of \SI{50}{\%}.

In the multirate version of the conventional \ac{ols} scheme, the incoming signals are first segmented into overlapping data segments of length $L_m$ and then from the processed output segments, $N_\text{O}=N-N_\text{S}$ samples are discarded in order to match the number of input and output samples. Using the above generalized model given by \eqref{eq:synth}, this scheme can be achieved by determining the \ac{td} analysis and synthesis windowing matrices as
\begin{gather}
  \mathbf{A}_{m} = \mathbf{I}_{L_m}\quad\text{and}\quad
  \mathbf{S} = \diag*{
  \begin{bmatrix}
    \mathbf{0}_{1\times N_\text{L}} & 
    \mathbf{1}_{1\times N_\text{S}} & 
    \mathbf{0}_{1\times N_\text{T}} 
  \end{bmatrix}^\transpose},
\end{gather}
respectively. Here, the number of overlapping samples $N_{\text{O}}=N-N_{\text{S}}$ is divided into leading and tailing overlapping parts as follows:
\begin{align}
  N_{\text{L}}=\left\lceil\frac{N-N_{\text{S}}}{2}\right\rceil
  \quad\text{and}\quad 
  N_{\text{T}}=\left\lfloor \frac{N-N_{\text{S}}}{2}\right\rfloor.
\end{align}
The corresponding leading and tailing overlapping parts of the $L_m$ are denoted by $L_{\text{L},m}$ and $L_{\text{T},m}$, respectively.

In the case of multirate version of the conventional \ac{ola} scheme, the non-overlapping input data blocks are zero-padded to length $L_m$ and the output blocks are overlapped and added such that the number of input and output samples match as desired. For the proposed generalized model, this is achieved by selecting the analysis and synthesis windowing matrices as
\begin{gather}
  \mathbf{A}_{m} = \diag*{
  \begin{bmatrix}
    \mathbf{0}_{L_{1\times \text{L},m}} &
    \mathbf{1}_{L_{1\times \text{S},m}} &
    \mathbf{0}_{L_{1\times \text{T},m}} 
  \end{bmatrix}^\transpose}\quad\text{and}\quad
  \mathbf{S} = \mathbf{I}_{N},
\end{gather} 
respectively. 

For the proposed generalized model, in addition to the \ac{fd} window, also the analysis and synthesis \ac{td} windows are adjustable and the \ac{fc}-based filter-bank design is carried out by optimizing all the windows simultaneously. The main target of the \ac{td} analysis window is to improve the spectral localization of the incoming \ac{cp-ofdm} waveform by properly weighting the samples interpolated by the frequency-domain zero-padding in \ac{ofdm} modulation. This windowing may induce some additional replicas in the spectrum which are filtered by the following \ac{fd} windowing. The target of the synthesis window is to smoothen the discontinuities between the overlapping blocks as well as the beginning and end transients since the discontinuities in the output waveform give raise to a high spectral leakage. For latency-critical applications, the zero-padding (cf. \eqref{eq:zero_padding}) in the beginning of the burst can also be reduced and especially in this case, the synthesis windowing of the filtered blocks becomes crucial.

When the \ac{fc}-based filter bank is used for processing the \ac{ofdm} signals, the \ac{ofdm} symbol subcarrier spacing on subband $m$ is determined as 
\begin{equation}
  \label{eq:SCSofdm}
  f_{\text{SCS},m} = \frac{f_{\text{s},m}}{L_{\text{OFDM},m}} = 
  \frac{L_m}{N}\frac{f_\text{s}}{L_{\text{OFDM},m}},
\end{equation}
where, $f_\text{s}$ is the output sampling rate and $f_{\text{s},m}$ is the input sampling rate for subband $m$, which can be selected independently for each subband. The above equation defines $L_m$, given the subcarrier spacing for subband $m$, output sampling rate $f_\text{s}$, and \ac{fc} \ac{ifft} length $N$. It should also be pointed out that the \acf{bs} in \ac{fc} processing as given by
\begin{equation}  
  \label{eq:FCBS}
  f_{\text{BS},m} = \frac{f_\text{s}}{N},
\end{equation}
can be selected to be greater than, smaller than, or equal to subcarrier spacing in \ac{ofdm} processing by properly choosing $N$ and $L_m$. For more details for the parameterization of the original \ac{fc} processing schemes, see \cite{J:Renfors14:FC,C:Renfors2015:fc-f-ofdm,J:Yli-Kaakinen:JSAC2017}. In practice, $f_{\text{BS},m}$ is typically fixed, and the supported \ac{ofdm} subcarrier spacings and sampling rates define the used \ac{fc} \ac{fft} lengths $L_m$.
 
% ----------------------------------------------------------------------
\subsection{OFDM  RX Processing}%
\label{sec:ofdm-rx-processing}
% ----------------------------------------------------------------------
In the case, when \ac{fc} interpolation factor $I_m=N/L_m>1$, the \ac{fc-f-ofdm} waveform can be received transparently with basic \ac{cp-ofdm} receiver by either using the \ac{ofdm} demodulator running at the high input rate or by first decimating the high-rate signal by $N/L_m$ and then using the demodulator on the decimated rate.\footnote{Alternatively, the \ac{fc} processing can be used on the \ac{rx} side for receiving both basic \ac{cp-ofdm} and \ac{fc-f-ofdm} waveforms. However, for presentation clarity and due to limited available space, we do not explicitly address such cases and thus assume basic \ac{cp-ofdm} receiver processing.} In the former case, the basic \ac{cp}-\ac{ofdm} \ac{rx} processing of the $m$th subband on the receiver side can be expressed as 
\begin{subequations} 
  \begin{equation}
    \mathbf{Y}_{m} =
    \widehat{\mathbf{W}}_{N_{\text{OFDM},m}}
    \mathbf{R}_{N_{\text{CP},m}}
    \tilde{\mathbf{Z}}_{\text{FC-F-OFDM}},
  \end{equation}
  where $\widehat{\mathbf{W}}_{N_{\text{OFDM},m}}\in\mathbb C^{N_{\text{OFDM},m}\times N_{\text{OFDM},m}}$ is the \ac{dft} matrix scaled by $1/\sqrt{N_{\text{OFDM},m}}$ and $\mathbf{R}_{N_{\text{CP},m}}\in\mathbb{Z}^{N_{\text{OFDM},m}\times(N_{\text{OFDM},m}+N_{\text{CP},m})}$ is the \ac{cp} removal matrix given by 
  \begin{equation} 
    \mathbf{R}_{N_{\text{CP},m}} = 
    \left[ 
    \begin{matrix}
      \mathbf{0}_{N_{\text{OFDM},m}\times N_{\text{CP},m}} & \mathbf{I}_{N_{\text{OFDM},m}}
    \end{matrix}
    % }
    \right].
  \end{equation} 
  The received signal is modelled by 
    \begin{equation}
    \tilde{\mathbf{Z}}_{\text{FC-F-OFDM}} =
    \ivec{\tilde{\mathbf{z}}_{\text{FC-F-OFDM}}},
  \end{equation}
  which is an $(N_{\text{OFDM},m}+N_{\text{CP},m})\times B_{\text{OFDM},m}$ matrix formed by un-stacking the $(N_{\text{OFDM},m}+N_{\text{CP},m})B_{\text{OFDM},m}$ samples 
  from the received sequence $\tilde{\mathbf{z}}_{\text{FC-F-OFDM}}$. For \ac{tx} performance analysis and 
  optimization purposes, the received sequence is defined as
  \begin{equation}
    \tilde{\mathbf{z}}_{\text{FC-F-OFDM}}= 
    \mathbf{R}_{\text{ZP},m}\mathbf{z}_{\text{FC-F-OFDM}}.
  \end{equation}
  where
  \begin{equation}
    \mathbf{R}_{\text{ZP},m} = 
    \begin{bmatrix}
        \mathbf{0}_{N/L_m S_{\text{F},m} \times N/L_m T_m} \\
        \mathbf{I}_{N/L_m T_m} \\
        \mathbf{0}_{N/L_m S_{\text{F},m} \times N/L_m T_m}
    \end{bmatrix}^\transpose   
  \end{equation}
  discards the zero padding (cf. \eqref{eq:zero_padding}) by removing first and last $N/L_mS_{\text{F},m}$ samples from the high-rate signal before further \ac{rx} processing.
\end{subequations}

In the latter case, when the signal after the discarding of zero padding is decimated by $N/L_m$ before the \ac{ofdm} demodulation, the \ac{ofdm} \ac{rx} processing is expressed as
\begin{subequations}
\begin{equation}
  \mathbf{Y}_{m} =
  \widehat{\mathbf{W}}_{L_{\text{OFDM},m}}
  \mathbf{R}_{L_{\text{CP},m}}
  \bar{\mathbf{Z}}_{\text{FC-F-OFDM}},
\end{equation}
where $\bar{\mathbf{Z}}_{\text{FC-F-OFDM}}$ is an $(L_{\text{OFDM},m}+L_{\text{CP},m})\times B_{\text{OFDM},m}$ matrix containing the received samples, the scaled \ac{dft} matrix is of size $L_{\text{OFDM},m}\times L_{\text{OFDM},m}$, and the \ac{cp} removal matrix is given by
\begin{equation} 
  \mathbf{R}_{L_{\text{CP},m}} = 
  \left[ 
    \begin{matrix}
      \mathbf{0}_{L_{\text{OFDM},m}\times L_{\text{CP},m}} & \mathbf{I}_{L_{\text{OFDM},m}}
    \end{matrix}
    % } 
  \right].
\end{equation} 
\end{subequations}

\begin{figure}[t!]     
  \centering 
  \includegraphics[trim=0 10 0 0,clip, % trim l b r t
  width=\figWidth]{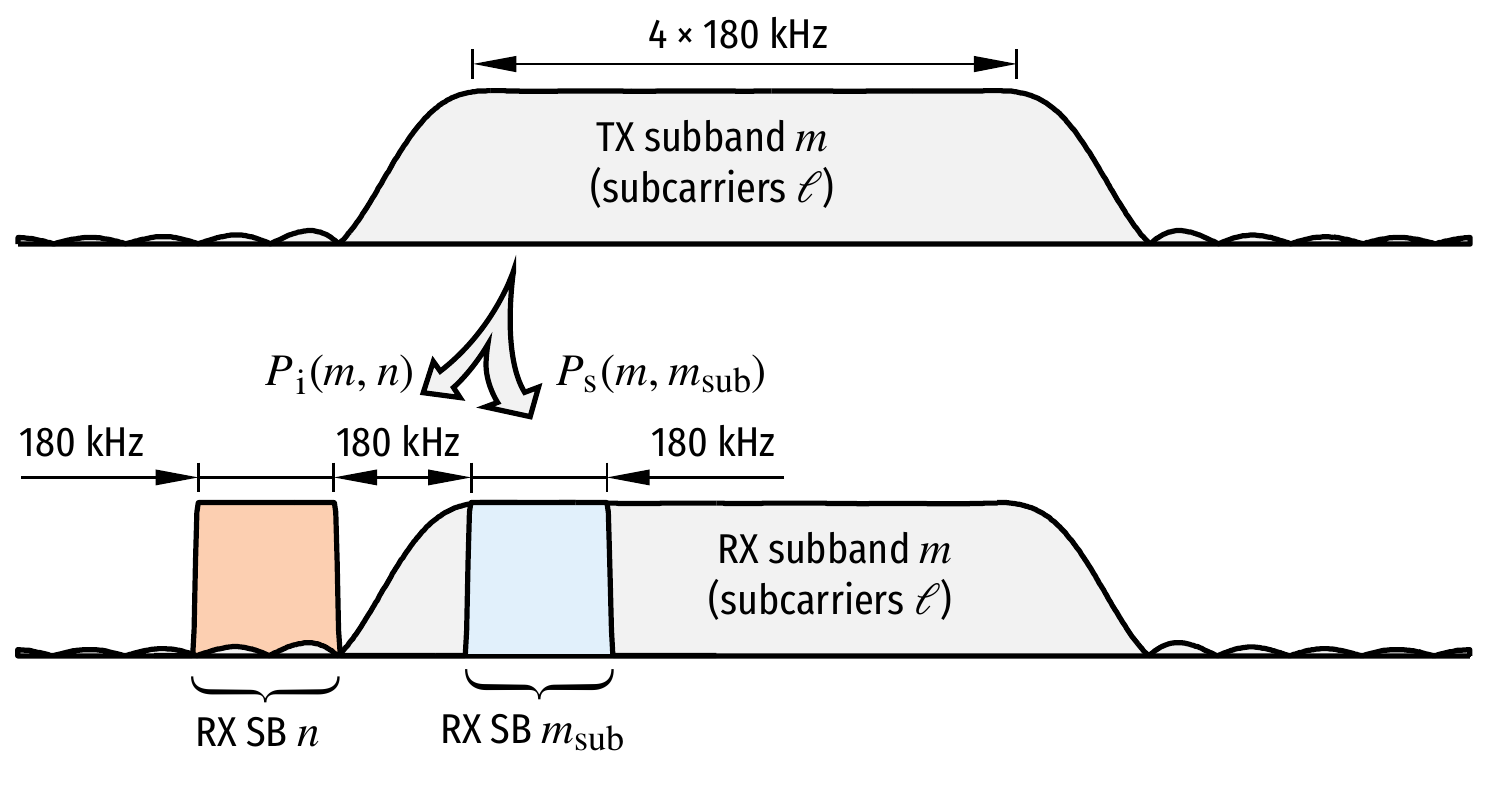} 
  \caption{Illustration of the \acf{scr} evaluation. The level of observable power $P_\text{s}(m,m_\text{sub})$ and leaking power $P_\text{i}(m,n)$ are  measured over the fixed bandwidth of \SI{180}{kHz} which corresponds to a single \ac{prb} when \ac{scs} is \SI{15}{kHz}.}
  \label{fig:SCR}   
\end{figure} 

% ======================================================================
\section{Fast-Convolution Filter-Bank Optimization}%
\label{sec:fast-conv-filt-1}
% ======================================================================
In this article, as formulated in Section II, we are using the generalized \ac{fc} processing for improving the spectral containment of \ac{cp-ofdm} waveform on the \ac{tx} side. The exact \ac{rx} processing is generally not known due to the transparent signal processing assumption \cite{S:3GPP:TR38.802,J:Levanen18:TransparentTxAndRx} and, therefore, when it comes to quantifying the band-limitation of the \ac{tx} signal subbands, we are measuring the \ac{tx} power leakage by using narrow-band \ac{td} measurement filter with bandwidth equal to one \ac{prb}. By using this approach, the \ac{tx} processing frequency selectivity can be accurately measured independent of the exact \ac{rx} processing and subcarrier spacings by locating the unintended \ac{rx} on subband $n$ in the close vicinity of target subband $m$ as illustrated in Fig.~\ref{fig:SCR}. The observable power on target subband $m$ is measured using the same narrow-band filter such that the filter is located at the edge of the subband. The optimization target is then to constrain the in-band unwanted emissions (inside one carrier) and in particular the interference leakage between different subbands, with possibly different numerologies or asynchronous transmissions, without considerably increasing the intrinsic passband distortion induced by the \ac{fc}-based subband filtering process.

% ----------------------------------------------------------------------
\subsection{Performance Metrics}%
\begin{figure}[t!]       
  \centering 
  \includegraphics[trim=0 5 0 0,clip, % trim l b r t
  width=\figWidth]{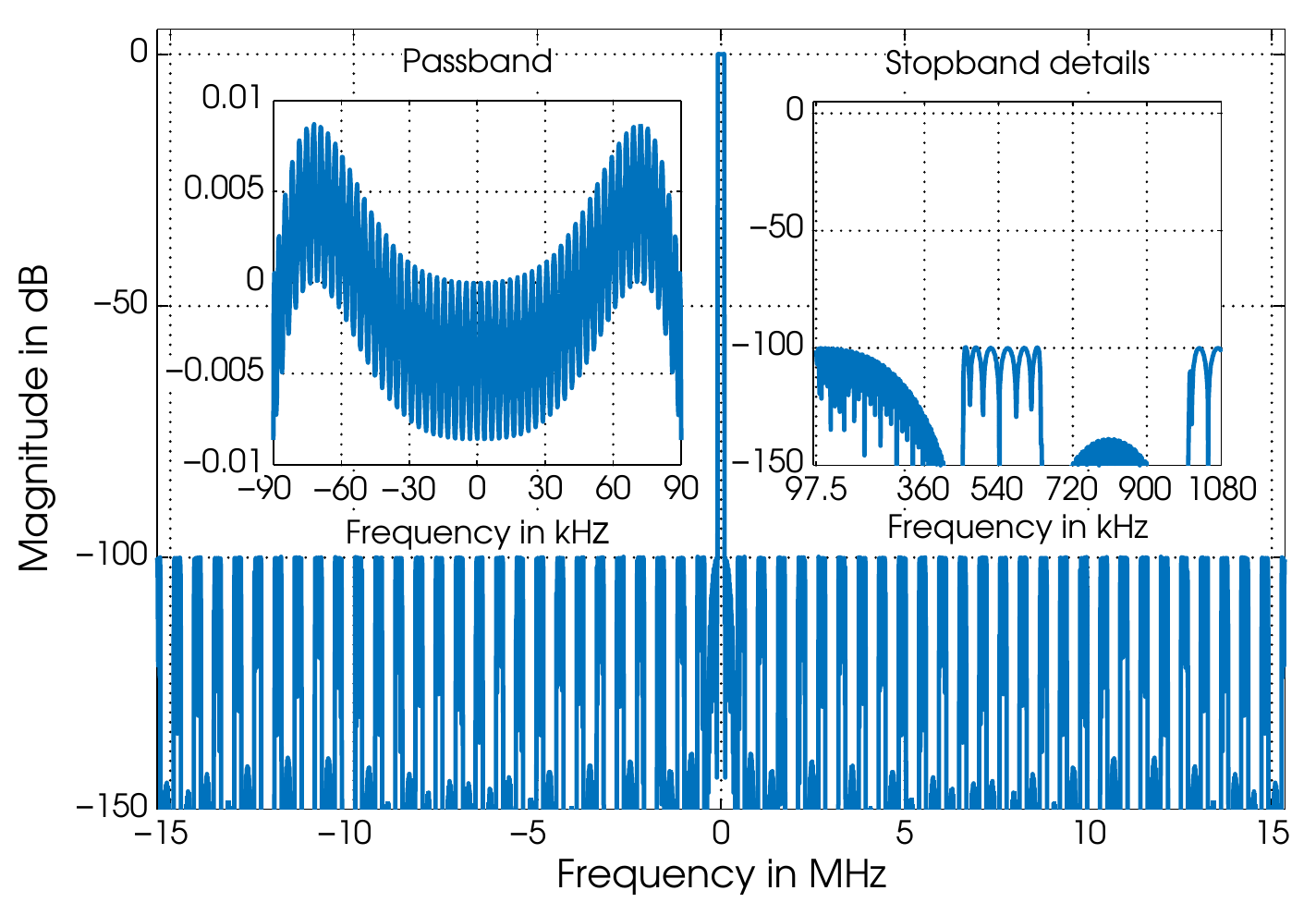} 
  \caption{Magnitude response as well as the passband and stopband details of the measurement \acs{fir} filter for evaluating the \acf{scr}. Here, passband is one \ac{prb} (\SI{180}{kHz}) wide and the transition band is one-half (\SI{7.5}{kHz}) of a baseline subcarrier spacing of \SI{15}{kHz}.}
  \label{fig:analysisFIR}    
\end{figure}

\label{sec:chann-perf-eval}
% ----------------------------------------------------------------------
We define the so-called \acf{scr} as the metric for characterizing unwanted in-band (inside one carrier) or out-of-band emissions.  The power on subband $m$ and $n$ is measured using fixed bandwidth of \SI{180}{kHz} (one \ac{prb} with \SI{15}{kHz} baseline \ac{scs}) with the aid of a narrow-band \ac{fir} measurement filter. The observable power $P_\text{s}{(m,m_{\text{sub}})}$ is obtained by locating the measurement filter over the edge subcarriers while the leaking power $P_\text{i}{(m,n)}$ is measured over the band starting \SI{180}{kHz} from the subband edge. The \SI{180}{kHz} guard band is selected because it corresponds to the resolution by which bandwidth parts can be allocated in the \ac{5g-nr} \cite[Section 7.3]{B:Dahlman2018}, for carrier frequencies below \SI{6}{GHz} \cite[Section 4.4.4.2]{S:3GPP:TS38.211}. Now, \ac{scr} is formally defined as
\begin{equation}  
  \label{eq:scr}   
  \mathrm{SCR}_m= 
  10\log_{10}
  \left(\frac{P_\text{i}{(m,n)}}{P_\text{s}{(m,m_{\text{sub}})}}\right),
\end{equation}
where $P_\text{i}{(m,n)}$ is the power leaking from subband $m$ to subband $n$ and $P_\text{s}{(m,m_{\text{sub}})}$ is the observable power on the subset of subband $m$. In the actual numerical examples and evaluations reported in Section \ref{sec:fc-based-f}, the transition bandwidth of the measurement filter is \SI{7.5}{kHz} (half of the \SI{15}{kHz} baseline \ac{scs}) and the minimum stopband attenuation is $A_\text{s}=\SI{100}{dB}$. The corresponding magnitude response of the measurement \ac{td} filter is show in Fig.~\ref{fig:analysisFIR}.\footnote{This analysis filter is a two-stage design with single-stage equivalent given by $H(z) = H_1(z)H_2(z^{56})$. The orders of the $H_1(z)$ and $H_2(z)$ are \num{362} and \num{351}, respectively.} 

The passband quality on an active subcarrier $\ell$ and on subband $m$ is measured using the \ac{mse} between the (normalized) transmitted and received symbols as follows:
\begin{equation}
  \label{eq:mse} 
  \text{MSE}_{m}(\ell)=\lvert\lvert
  [\mathbf{X}_m]_{\ell,s}-[\mathbf{Y}_m]_{\ell,s}\rvert\rvert^2,
\end{equation} 
where $[\mathbf{X}_m]_{\ell,s}$ and $[\mathbf{Y}_m]_{\ell,s}$, respectively, are the transmitted and received symbols on subcarrier $\ell$ and $s \in \{1,2, \ldots, B_{\text{OFDM},m}\}$. The corresponding \acf{evm} in percents is expressed using \eqref{eq:mse} as 
\begin{equation} 
  \text{EVM}_{m}(\ell) = 100\sqrt{\text{MSE}_{m}(\ell)}.
\end{equation} 
The \ac{mse} and \ac{evm} are measured after executing zero-forcing (ZF) equalization, as defined in \ac{3gpp} \ac{5g-nr} specification in \cite[Annex B]{S:3GPP:TS38.104}. 

The average \ac{mse} is defined as the mean value of the \ac{mse} values on active subcarriers, as given by
\begin{equation}
  \label{eq:MSEavg}
  \text{MSE}_{\text{AVG},m}=
    \frac{1}{{L}_{\text{ACT},m}}
    \sum_{\ell=0}^{{L}_{\text{ACT},m}-1}
    \text{MSE}_{m}(\ell).
\end{equation}  
For the analysis purposes we also quantify the worst-case \ac{mse}. This is determined as a mean value of the \ac{mse} over the edge subcarriers as given by
\begin{equation}
  \label{eq:MSEmax}
  \text{MSE}_{\text{MAX},m}=
    \frac{1}{\lvert \mathcal{E}_\text{L}\cup \mathcal{E}_\text{R}\rvert}
    \sum_{\substack{\ell=\mathcal{E}_\text{L}\cup \mathcal{E}_\text{R}}}
    \text{MSE}_{m}(\ell).
\end{equation}
Here, $\mathcal{E}_\text{L}=\{0,1,\dots,N_\text{edge}-1\}$ and $\mathcal{E}_\text{R}=\{{L}_{\text{ACT},m}-N_\text{edge}, {L}_{\text{ACT},m}-N_\text{edge}+1,\dots, {L}_{\text{ACT},m}-1\}$ denote the left and right edge subcarriers, respectively. The number of edge subcarriers is selected to be $N_\text{edge}=12$. This metric is also in-line with the edge \ac{prb} \ac{evm} measurement defined and presented in \cite{S:3GPP:TR38.803}.

% ----------------------------------------------------------------------
\subsection{Parameterization of Time- and Frequency-domain Windows}%
\label{sec:param-wind}
% ----------------------------------------------------------------------
For simplicity, we consider the optimization of windows with even lengths. For the proposed model, in general, all the weights of the windows are adjustable. However, in the case of \ac{fd} window, it has turned out that it is beneficial to select the window such that the weights on subband $m$ consist of two symmetric transition bands with non-trivial values $\xi_m(p)$ for $p=0,1, \dots,L_{\text{TBW},m}-1$, where $L_{\text{TBW},m}$ also defines the \acl{tbw}. All passband weights are set to one and all stopband weights are set to zero. The number of stopband weights (and the corresponding transform length $L_m$) can be selected to reach a feasible subband oversampling factor. Now the diagonal weight values in \eqref{eq:FDwin} can be expressed as
\begin{align}
  \mathbf{d}_m
  & = \bigl[
  \begin{matrix}
    \mathbf{0}_{1\times(\lceil[L_m-L_{\text{ACT},m}]/2\rceil-L_{\text{TBW},m})} & \xi_{m}(0) & \cdots
  \end{matrix} \nonumber\\
  & \mspace{30mu} \begin{matrix}
    \xi_{m}(L_{\text{TBW},m}-1) & \mathbf{1}_{L_{1\times\text{ACT},m}} &  \xi_{m}(L_{\text{TBW},m}-1) & \cdots
  \end{matrix} \nonumber\\
  &  \mspace{30mu} \begin{matrix}
  \xi_{m}(0) & \mathbf{0}_{1\times(\lfloor[L_m-L_{\text{ACT},m}]/2\rfloor-L_{\text{TBW},m})}
  \end{matrix}
  \bigr]^\transpose
\end{align}
where $L_{\text{ACT},m}$ is the number of active subcarriers on subband~$m$. 
    
Due to the fact that \ac{fc} processing blocks are not time aligned with \ac{cp}-\ac{ofdm} symbols and the fact that the \ac{cp} length can be varying as specified in \cite{S:3GPP:TS38.104}, it is beneficial to synchronize the analysis \ac{td} window with \ac{cp}-\ac{ofdm} symbols in order to process all the blocks in a same manner. Therefore, we define the \ac{td} analysis windowing matrix as
\begin{subequations}
\begin{equation}
  \mathbf{A}_m = \diag{\boldsymbol{\Pi}_{r,m}\hat{\mathbf{a}}_m},
\end{equation} 
where $\hat{\mathbf{a}}_m$ is the analysis window of length $L_{\text{OFDM},m}$ and \begin{equation}
  \boldsymbol{\Pi}_{r,m} = \mathbf{Q}_{r,m} \mathbf{K}_{L_{\text{CP},m}}
\end{equation}
aligns the desired $L_m$ samples of $\hat{\mathbf{a}}_m$ with the $r$th \ac{fc} processing block.
Here, $\mathbf{K}_{L_{\text{CP},m}}$ is given by \eqref{eq:CPins} and $\mathbf{Q}_{r,m}\in\mathbb{N}^{L_m\times (L_{\text{OFDM},m}+L_{\text{CP},m})}$ is expressed as 
\begin{equation}
    [\mathbf{Q}_{r,m}]_{q,p} =
    \begin{cases}
      1, & \text{if $(p-S_{\text{F},m}-rL_{\text{S},m}-1 \bmod L_m)+1 = q$} \\
      0, & \text{otherwise}.
    \end{cases} 
\end{equation}
\end{subequations}

\begin{figure}[t]   
  \centering 
  \includegraphics[trim=0 0 0 0,clip % trim l b r t
  ,width=0.99\figWidth]{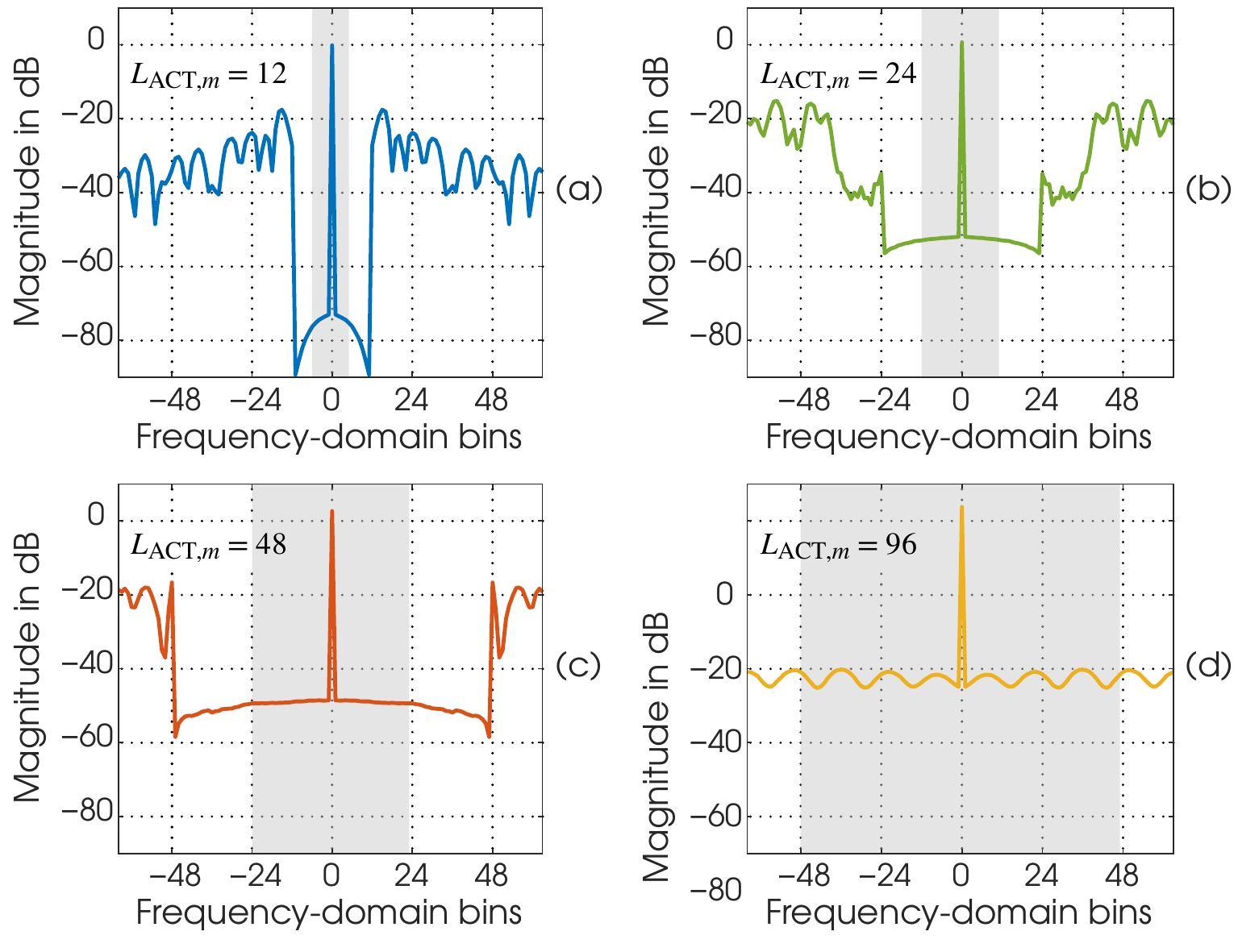}
  \caption{The \ac{fd} representation of \ac{td} analysis window $\hat{\mathbf{a}}_m$ showing the distribution of significant coefficient values. Here, $L_{\text{OFDM},m}=128$ and the number of active subcarriers is (a) $L_{\text{ACT},m}=\num{12}$, (b) $L_{\text{ACT},m}=\num{24}$, (c)  $L_{\text{ACT},m}=\num{48}$, and (d) $L_{\text{ACT},m}=\num{96}$. The grey-colored region in each sub-figure illustrates the bandwidth of the active subcarriers. As seen from this figure, the \ac{fd} representation of the \ac{td} analysis filter is close to zero on $L_{\text{ACT},m}-1$ bins before and after the zero-frequency bin.}
  \label{fig:representations}     
\end{figure}   

In the \ac{fd} representations of the optimized analysis \ac{td} windows, the zero-frequency bin typically has the highest magnitude and other bins have significant values only for bins $[L_{\text{ACT},m}+1, L_{\text{ACT},m}+2, L_{\text{OFDM},m}-L_{\text{ACT},m}-2]$ as illustrated in Fig.~\ref{fig:representations}. This is due to the fact that the \ac{fd} kernel corresponding to analysis time-domain window should keep the signal on active subcarriers essentially intact.  Therefore, these bins with possibility of having significant values are also selected as the optimization parameters. In this approach, the parameter vector $\boldsymbol{\phi}_m$ for the \ac{fd} representation of the \ac{td} analysis window on subband $m$ is composed of zero-frequency bin as well as $L_{\text{OFDM},m}-2L_{\text{ACT},m}$ bins outside the active subcarriers. It is important to note that the proposed generalized \ac{fc} processing with non-trivial \ac{td} windows inherently modifies the interpolation achieved by the \ac{fd} zero-padding in \ac{ofdm} modulation by properly weighting the interpolated samples to provide an interpolant with better spectral localization than using the straightforward rectangular analysis and synthesis \ac{td} windows. 

\begin{figure}
    \centering
    \includegraphics[width=1.0\figWidth]{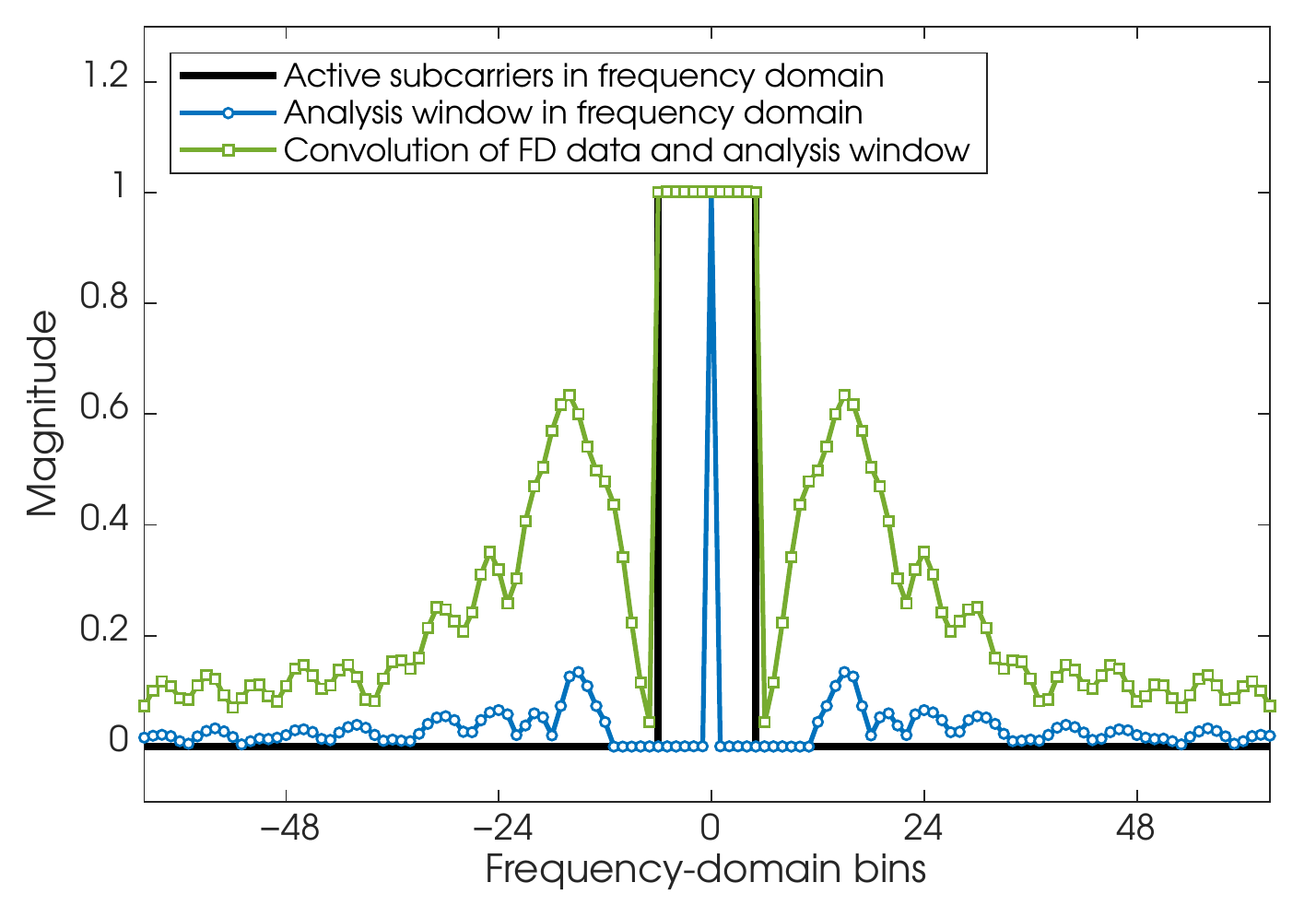}      
    \caption{Illustration of the cyclic convolution of the rectangular window corresponding to waveform with bandwidth equal to $L_{\text{ACT},m}=12$ subcarriers by the frequency-domain representation of the \ac{td} analysis filter optimized for $L_{\text{ACT},m}=12$ active subcarriers.}
    \label{fig:TDconv}
\end{figure}

The requirements for the \ac{fd} representation of the \ac{td} analysis window can also be explained through convolution theorem. The element-wise multiplication of the time-domain \ac{cp-ofdm} waveform by the time-domain analysis window is alternatively expressed by the circular convolution of the corresponding \ac{fd} representations.  Fig.~\ref{fig:TDconv} shows the \ac{fd} representation of the \ac{td} analysis window and a rectangular window corresponding to allocation of $L_{\text{ACT},m}=12$ active subcarriers. In order to carry out the circular convolution such that the result has constant value over the active subcarriers, as shown by the line with square markers in Fig.~\ref{fig:TDconv}, the \ac{fd} representation of the \ac{td} analysis window has to be approximately zero in $L_{\text{ACT},m}-1$ bins on both sides of the zero-frequency bin.

It is assumed that all the windows are real valued and, therefore, when the analysis window length $L_{\text{OFDM},m}$ is even, first and $(L_{\text{OFDM},m}/2+1)$th bins of the corresponding \ac{fd} representation are also real valued. The remaining bins, in general, have complex values such that the \ac{fd} representation is conjugate symmetric with respect to $(L_{\text{OFDM},m}/2+1)$th bin and, therefore, both the real and imaginary parts of the corresponding coefficient values in the lower part of the conjugate-symmetric response need to be included in the optimization. Now, assuming that the $L_{\text{ACT},m}<L_{\text{OFDM},m}/2$, then the lower part of the \ac{fd} representation of the \ac{td} analysis window can be constructed from the parameter vector $\boldsymbol{\phi}_m$ as follows 
\begin{equation}
  \begin{alignedat}{3}
    \alpha_m(0)                      &= \phi_m(0) \\ 
    \alpha_m(L_{\text{ACT},m}+p-1)   &= \phi_m(p) + \\
    &\mspace{20mu}\iu\phi_m(L_{\text{OFDM},m}/2-L_{\text{ACT},m}+1+p) \\ 
    \alpha_m(L_{\text{OFDM},m}/2)   &=\phi_m(L_{\text{OFDM},m}/2-L_{\text{ACT},m}+1) 
  \end{alignedat}
\end{equation}
for $p = 1,2,\dots,L_{\text{OFDM},m}/2-L_{\text{ACT},m}$. The corresponding upper part is obtained by taking the complex conjugate of the resulting lower part, expressed as
\begin{equation}
  \alpha_m(L_m-n) = \conj{\alpha_m(n)}
\end{equation}
for $n = 1,2,\dots,L_m/2-1$. Finally, the analysis window $\hat{\mathbf{a}}_m$ is obtained by taking the \ac{idft} of the \ac{fd} representation as follows:
\begin{equation}
  \hat{\mathbf{a}}_m = \mathbf{W}^{-1}_{L_{\text{OFDM},m}}\boldsymbol{\alpha}_m.
\end{equation}
For $L_{\text{ACT},m}\geq L_{\text{OFDM},m}/2$, the effect of time-domain analysis window becomes negligible due to the fact that the shape of the corresponding \ac{fd} representation is already determined by $2L_{\text{ACT},m}$ zero-valued bins. Overall, this approach considerably reduces the number of parameters to be optimized especially for wider bandwidth allocations when compared with the straightforward parameterization of \ac{td} window values. 

The \ac{td} synthesis window is parametrized in a same manner, however, now the synthesis window is, by definition, common for all the subbands and only the \ac{fd} bins with indices $p=0,1,\dots,\gamma-1$ need to be taken into optimization. Here, $\gamma$ is a small integer, typically \numrange{5}{20} (see, Section~\ref{sec:fc-based-f} for details).\footnote{To our experience, selecting $\gamma=20$ provides sufficient flexibility for the synthesis window in all the considered cases since the target of this window is to smoothen the time-domain transients of the resulting waveform. Increasing $\gamma$ from 20 had no observable effect on performance and in some cases $\gamma=5$ allowed to achieve almost the same performance as $\gamma=20$. Thus, $\gamma$ is not highly sensitive to the selected value, and we have chosen $\gamma=20$ as compromise between obtained performance and optimization complexity.} Therefore, the lower part of the \ac{fd} representation of the \ac{td} synthesis window can be constructed from the parameter vector $\boldsymbol{\psi}$ as follows
\begin{equation}
  \beta(0) = \psi(0) 
  \qquad\text{and}\qquad
  \beta(p) = \psi(p) + \iu\psi(\gamma+p) 
\end{equation} 
for $p = 1,2,\dots,\gamma-1$ and the upper part is obtained by taking the complex conjugate, expressed as
\begin{equation}
  \beta(N-n) = \conj{\beta(n)}
\end{equation}
for $n = 1,2,\dots,N/2-1$. Finally, the \ac{td} synthesis matrix is obtained by taking the \ac{idft} of the \ac{fd} representation, expressed as
\begin{equation}
  \mathbf{S} = \diag{\mathbf{s}}
\quad\text{with}\quad
  \mathbf{s} = \mathbf{W}^{-1}_{N}\boldsymbol{\beta}.
\end{equation}
 
% ----------------------------------------------------------------------
\subsection{Transmultiplexer Optimization for Generalized FC-F-OFDM}%
\label{sec:transm-optim-fc}
% ----------------------------------------------------------------------
The generalized \acl{fc-f-ofdm} system design can now be stated as an optimization problem for finding the optimal values of the aggregate parameter vector containing the adjustable parameters of all the windows as given by
\begin{subequations}
  \label{eq:problem}
\begin{equation}
  \label{eq:parVect}
   \Xi=[\boldsymbol{\xi}_0,\boldsymbol{\xi}_1,\dots,\boldsymbol{\xi}_{M-1},
   \boldsymbol{\phi}_0,\boldsymbol{\phi}_1,\dots,\boldsymbol{\phi}_{M-1},
   \boldsymbol{\psi}] 
\end{equation}
to
\begin{equation}
  \begin{aligned}    
    & \underset{\Xi}{\text{minimize}}
    & & \max_{m=0,1,\dots,M-1}\left(\text{MSE}_{\text{AVG},m}\right) \\ 
    & \text{subject to}
    & & \mathrm{SCR}_m\leq A_\text{des}\quad\text{for \,$m=0,1,\dots,M-1$},
  \end{aligned}
\end{equation}
\end{subequations}
where $A_\text{des}$ is the desired \ac{scr} target. The optimization problem in \eqref{eq:problem} can be straightforwardly solved using non-linear inequality constrained optimization algorithm, e.g., \acl{sqp} \cite{B:Nocedal06:numer_optim}. Due to the non-convex nature of the problem and quite large number of optimization parameters, the convergence to the global optimum can not be unconditionally guaranteed. However, by using different starting points for the optimization, it can be shown that the algorithm  converges reliably to same solution and, therefore, we can assume optimum is also the global one.

It should be noted that the goal here is not to reach the aperiodic convolution exactly through \ac{fc} processing, instead, the optimization target is to keep the time-domain aliasing at a level that does not significantly impact the link error rate performance, such that the non-implementation-related effects are dominating. The main reason for \ac{mse} (or \ac{evm}) degradation is the loss of orthogonality due to the partial suppression of subcarriers, which is unavoidable in all filtered \ac{ofdm} solutions. The proposed approach can be used for finding the desired trade-off between the frequency-selectivity of the processing and the resulting intrinsic interference of the filtering.
  
% ======================================================================
\section{Implementation Complexity}%
\label{sec:impl-compl}
% ======================================================================
The implementation complexity of the proposed generalized \ac{fc} processing consist of forward and inverse \acp{fft} and the \ac{td} and \ac{fd} windowing. The number of real multiplications per \ac{fc} processing block can be expressed as 
\begin{equation} 
  C_{\text{M}}^\text{(GFCB)} = C_{\text{M}}^\text{(TDSW)} + C_{\text{M}}^\text{(IFFT)} + \sum_{m=0}^{M-1} C_{\text{M},m}^\text{(TDAW)} + C_{\text{M},m}^\text{(FDW)} + C_{\text{M},m}^\text{(FFT)},
\end{equation} 
where $C_{\text{M},m}^\text{(TDAW)}$'s, $C_{\text{M}}^\text{(TDSW)}$, and $C_{\text{M},m}^\text{(FDW)}$'s are, respectively, the number of real multiplications required for the \ac{td} analysis and synthesis windows as well as for the \ac{fd} windows while $C_{\text{M},m}^\text{(FFT)}$'s and $C_{\text{M}}^\text{(IFFT)}$, respectively, are the number of real multiplications required for the \ac{fc} short \acp{fft} and long \ac{ifft}. The complexity of the \ac{ofdm} processing consist of only \acp{ifft} and can be expressed as
\begin{equation} 
  C_{\text{M}}^\text{(OFDM)} =\sum_{m=0}^{M-1} C_{\text{M},m}^\text{(IFFT)},
\end{equation}
where $C_{\text{M},m}^\text{(IFFT)}$'s are the complexities of the \ac{ofdm} \acp{ifft} in terms of real multiplications. Now, the total number of real multiplications per \ac{cp-ofdm} symbol can be expressed as
\begin{equation}
  C_{\text{M}} = R_mC_{\text{M}}^\text{(GFCB)}/B_{\text{OFDM},m}  + C_{\text{M}}^\text{(OFDM)},
\end{equation}
where $R_m$, as given by \eqref{eq:fcblocks}, is the number of \ac{fc}-processing blocks. 

The number of real additions can be evaluated by replacing the $C_{\text{M},m}^\text{(FFT)}$'s,  $C_{\text{M},m}^\text{(IFFT)}$'s, and $C_{\text{M}}^\text{(IFFT)}$ with the corresponding $C_{\text{A},m}^\text{(FFT)}$'s,  $C_{\text{A},m}^\text{(IFFT)}$'s, and $C_{\text{A}}^\text{(IFFT)}$ giving the number of real additions required for the transforms and replacing the corresponding windowing complexities by zeros. 

\begin{table}[t!]
  \centering
  \caption{Number of real multiplications (muls) and additions (adds) for the FFT lengths of $N=\widehat N$ and $3\widehat N$ where $\widehat N$ is a power of two value.}
  \label{tab:FFTcomp}
  % Created: 09-May-2016, 12:50
  \vspace{-0.8em}
  {\footnotesize
    \begin{tabular}{lcc}
      \toprule
      \multicolumn{1}{c}{} &
      \multicolumn{1}{c}{$N=\widehat{N}$ } &
      \multicolumn{1}{c}{$N=3\widehat{N}$} \\
      \midrule
      Real muls & $C_\text{M}=\widehat N(\log_2(\widehat N)-3)+4$        & $C_\text{M}=\widehat N(3\log_2(\widehat N)- 7)+12$ \\ 
      Real adds & $C_\text{A}=\widehat N( 3\log_2(\widehat N) -3)  +  4$ & $C_\text{A}=\widehat N( 9\log_2(\widehat N) +3)  + 12$ \\ 
      \bottomrule
    \end{tabular}} 
\end{table}

For a given transform length, the complexities of \ac{fft} and \ac{ifft} are, in general, the same. For power-of-two transform lengths, the split-radix algorithm is considered to be the most efficient one in terms of number of real multiplications \cite{J:Sorensen86:SP-FFT} and the number of real multiplications and additions needed for the transforms are given in Table~\ref{tab:FFTcomp}. Here we consider also transform lengths of the form $3\widehat{N}$ where $\widehat{N}$ is a power-of-two value and the complexities are evaluated for prime-factor \ac{fft} \cite{J:Kolba77}. Generally, the availability of transform lengths other than powers of two increases greatly the flexibility of waveform parametrization. The number of real multiplications and additions for the transform lengths to be considered later on are listed in Table~\ref{tab:evaluatedFFTs}.
 
\begin{table}[t]
  \caption{Number of real multiplications (muls) and additions (adds) per CP-OFDM symbol (or \ac{fc} processing block) for some common transform lengths evaluated according to formulas in Table \ref{tab:FFTcomp}}
  \label{tab:evaluatedFFTs}
  \vspace{-0.8em}
  \centering
  \footnotesize{
    \begin{tabular}{l*6{S[table-format=<4.0]}}
      \toprule
      \multicolumn{1}{c}{$N$} 
                &    16 &    24 &    32 &    48 &    64 &   128 \\
      \midrule
      Real muls &    20 &    28 &    68 &    92 &   196 &   516 \\
      Real adds &   148 &   252 &   388 &   636 &   964 &  2308 \\
      \midrule
      \multicolumn{1}{c}{$N$} 
                &   256 &   384 &   512 &   768 &  1024 &  2048 \\
      \midrule
      Real muls &  1284 &  1804 &  3076 &  4364 &  7172 & 16388 \\
      Real adds &  5380 &  8460 & 12292 & 19212 & 27652 & 61444 \\
      \bottomrule 
    \end{tabular}}  
\end{table}

\begin{table}[t]
  \caption{Example parametrizations for FC-F-OFDM-based \ac{5g-nr} physical layer with \SI{20}{MHz} carrier bandwidth 
  \cite{S:3GPP:TS38.104}}
  \addtolength\tabcolsep{-4.0pt}
  \label{tab:params20MHz}
  \centering
  \vspace{-0.8em}
  \footnotesize{
   \begin{tabular}{cccccccc}
     \toprule
     \multicolumn{1}{c}{Config.}  & 
     \multicolumn{1}{c}{$f_{\text{SCS},m}$}  & 
     \multicolumn{1}{c}{$N_{\text{PRB},m}$}  & 
     \multicolumn{1}{c}{${L}_{\text{OFDM},m}$} & 
     \multicolumn{1}{c}{${L}_{\text{CP},m}$} & 
     \multicolumn{1}{c}{${f}_{\text{s},m}$} & 
     \multicolumn{1}{c}{${L}_m$} & 
     \multicolumn{1}{c}{${f}_{\text{s}}$}  
     \\
     \midrule
     \multirow{3}{*}{\shortstack{Narrow\\alloca-\\tion}}  
     & \SI{15}{kHz} &  $\leq\num{10}$  &  128 &   9 & \SI{1.92}{MHz} & $N/16$ & \SI{30.72}{MHz} \\
     & \SI{30}{kHz} &   $\leq\num{10}$  &  128 &   9 & \SI{3.84}{MHz} & $N/8$  & \SI{30.72}{MHz} \\
     & \SI{60}{kHz} &   $\leq\num{10}$  &  128 &   9 & \SI{7.68}{MHz} & $N/4$  & \SI{30.72}{MHz} \\
     \midrule
     \multicolumn{1}{c}{Config.}  & 
     \multicolumn{1}{c}{$f_{\text{SCS},m}$}  & 
     \multicolumn{1}{c}{$N_{\text{PRB},m}$}  & 
     \multicolumn{1}{c}{${L}_{\text{OFDM},m}$} & 
     \multicolumn{1}{c}{${L}_{\text{CP},m}$} & 
     \multicolumn{1}{c}{${f}_{\text{s},m}$} & 
     \multicolumn{1}{c}{${L}_m$} &
     \multicolumn{1}{c}{${f}_{\text{s}}$}
     \\
     \midrule
     \multirow{3}{*}{\shortstack{Wide\\alloca-\\tion}}  
     & \SI{15}{kHz} & $\leq\num{106}$  & 2048 & 144 & \SI{30.72}{MHz} & $N$ & \SI{30.72}{MHz} \\
     & \SI{30}{kHz} &  $\leq\num{51}$  & 1024 &  72 & \SI{30.72}{MHz} & $N$ & \SI{30.72}{MHz} \\
     & \SI{60}{kHz} &  $\leq\num{24}$   & 512 &  36 & \SI{30.72}{MHz} & $N$ & \SI{30.72}{MHz} \\
     \bottomrule
    \end{tabular}}  
\end{table}

% ======================================================================
\section{Application and Performance Evaluation in 5G-NR Physical Layer}%
\label{sec:fc-based-f}
% ======================================================================
In this section, we provide extensive numerical evaluations in the context of \ac{3gpp} \ac{5g-nr} mobile radio network which utilizes \ac{cp-ofdm} as the baseline waveform. Following the \ac{5g-nr} specification, we assume that the active subcarriers on subband $m$ are always scheduled in \acp{prb} of \num{12} subcarriers \cite{B:Dahlman2018}, and thus $L_{\text{ACT},m}=12N_{\text{PRB},m}$, where $N_{\text{PRB},m}$ is the number of allocated \acp{prb} in subband $m$. In the flexible physical layer numerology available in \ac{5g-nr}, the \ac{ofdm} \ac{scs} is an integer power of two times \SI{15}{kHz}, that is, $f_{\text{SCS},m}=2^{\eta_m}\times\SI{15}{kHz}$ \cite{S:3GPP:TS38.300}. For an example carrier bandwidth of \SI{20}{MHz}, the sampling rate is $f_\text{s}=\SI{30.72}{MHz}$ with $N_{\text{OFDM},m}=2048/2^{\eta_m}$ and $N_{\text{CP},m}=144/2^{\eta_m}$.\footnote{It should be noted that in the exact NR specifications \cite{S:3GPP:TS38.211}, the length of the first \ac{cp} within every \SI{0.5}{ms} is longer.}
  Since $L_{\text{CP},m}$, the \ac{cp} length on the low-rate side, is constrained to be an integer, the minimum \ac{ofdm} \ac{ifft} length becomes $L_{\text{OFDM},m}=\num{128}$. Now, for given maximum number of \acp{prb}, the \ac{ofdm} \ac{ifft} length can be determined as
\begin{subequations} 
  \label{eq:LOFDMest}
  \begin{equation}
    L_{\text{OFDM},m}=\max\left\{2^{\xi_m},128\right\},
  \end{equation}
  where 
  \begin{equation}
    \xi_m=\left\lceil\log_2(12N_{\text{PRB},m})\right\rceil.
  \end{equation}
\end{subequations}
The \ac{fc} processing transform sizes $N$ and $L_m$ are selected to achieve the desired interpolation factor.

Considering the mixed-numerology cases, $L_m$'s should be selected such that the associated $N$ becomes the same for all the numerologies. Table~\ref{tab:params20MHz} shows the example numerologies with \ac{ofdm} \acp{scs} of \SI{15}{kHz}, \SI{30}{kHz}, and \SI{60}{kHz}. By selecting the bin spacing in \ac{fc} processing, as given by (\ref{eq:FCBS}), to be larger than the \ac{scs} in \ac{ofdm} processing, as given by (\ref{eq:SCSofdm}), the complexity of the \ac{fc} processing can be significantly reduced as shown later in this section. In addition, if the \ac{ofdm} \ac{scs} and the \ac{fc} bin spacing do not share a common factor, then increased flexibility is achievable in selecting the subband center frequencies. For example, by shifting the signal three bins to the right with $f_{\text{SCS},m}=\SI{15}{kHz}$ in \ac{ofdm} processing domain and one bin to the left with $f_{\text{BS},m}=\SI{40}{kHz}$ bin spacing in \ac{fc} processing domain, an overall shift of \SI{5}{kHz} can be achieved if desired.

\begin{table}[t]  
  \caption{Passband \acp{mse} in Example 1. \ac{fd} window is adjustable in all cases. Here, $\lvert\Xi\rvert$ denotes the number of adjustable parameters in optimization}
  \label{tab:Example1}
  \centering
  \vspace{-0.8em}
  \footnotesize{ 
    \begin{tabular}{ccccccc}
      \toprule
      \multirow{2}{*}{Case} &
      \multicolumn{1}{c}{\ac{td}} &
      \multicolumn{1}{c}{\ac{td}} &
      \multicolumn{1}{c}{MSE} & 
      \multicolumn{1}{c}{MSE} &      
      \multirow{2}{*}{$\abs{\Xi}$} &
      \multicolumn{1}{c}{\acs{cpu}} \\
      &  
      \multicolumn{1}{c}{synthesis} &
      \multicolumn{1}{c}{analysis} &
      \multicolumn{1}{c}{$(\lambda=\text{\sfrac{1}{2}})$} &
      \multicolumn{1}{c}{$(\lambda=\text{\sfrac{1}{4}})$} &
      &      
      \multicolumn{1}{c}{time} \\
      \midrule
      I   &              &              & \SI{-39.6}{dB} & \SI{-31.5}{dB} &   8 & \SI{120}{s} \\
      II  & $\checkmark$ &              & \SI{-40.6}{dB} & \SI{-31.5}{dB} & 136 & \SI{1.5}{h} \\
      III &              & $\checkmark$ & \SI{-57.0}{dB} & \SI{-40.4}{dB} & 136 & \SI{0.8}{h} \\
      IV  & $\checkmark$ & $\checkmark$ & \SI{-58.4}{dB} & \SI{-42.7}{dB} & 264 & \SI{6.1}{h} \\
      V   & $\checkmark$ & $\checkmark$ & \SI{-56.6}{dB} & \SI{-41.4}{dB} & 131 & \SI{1.5}{h} \\
      \bottomrule 
    \end{tabular}}
    \vspace{-1em}
\end{table}

\begin{figure}
    \centering
    \includegraphics[width=1\figWidth]{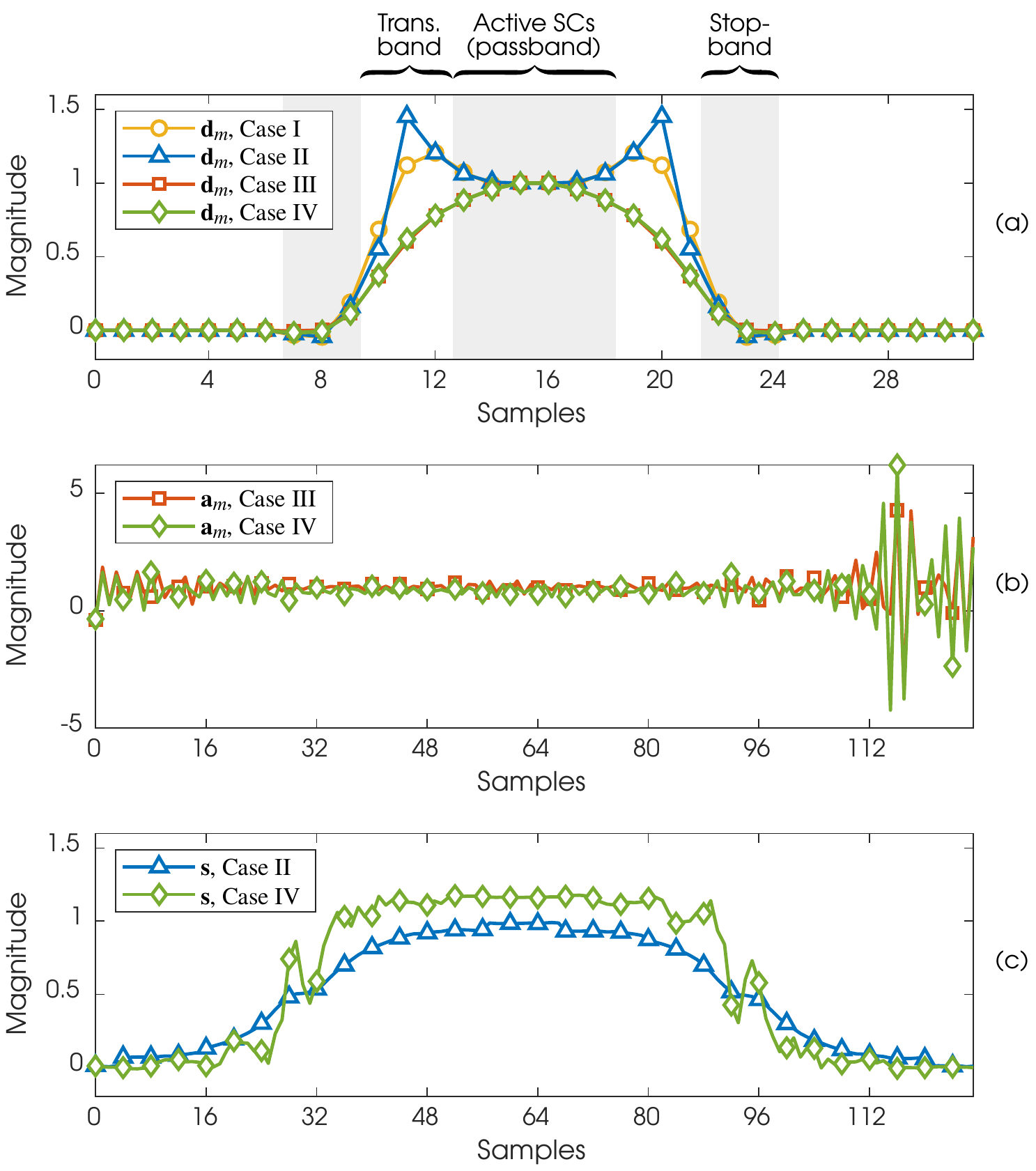}
    \caption{Optimized \ac{fd} and \ac{td} windows in Example 1 with \SI{50}{\%} overlap. (a) \ac{fd} windows. The passband and stopband regions with respect to \ac{fd} windows
are also illustrated in this figure. (b) Analysis \ac{td} windows. (c) Synthesis \ac{td}
windows}
    \label{fig:Ex1}
\end{figure}  

% ----------------------------------------------------------------------
\subsection{Example 1}% 
\label{sec:example-1}
% ----------------------------------------------------------------------
%%%%%%%%%%%%%%%%%%%%%%%%%%%%%%
Here we first compare the performance of the various \ac{fc} processing alternatives by optimizing the windows in five cases shown in Table~\ref{tab:Example1}. In this table, the check marks denote the windows which are adjustable in the optimization. Case I corresponds to the original \ac{fc} processing in \cite{J:Yli-Kaakinen:JSAC2017} with only adjustable \ac{fd} window and Cases IV and V correspond to generalized processing where all the windows are adjustable. In Cases \mbox{I--IV}, all the values of the \ac{td} windows are adjustable as well as that of the \ac{fd} window on passband and transition-band regions whereas in Case V, the parameter reduction techniques of Section \ref{sec:param-wind} with $\gamma=20$ are utilized.  

In this example, the number of subbands is $M=1$ while the number of active \acp{prb} is $N_{\text{PRB},0}=2$ (\num{24} active subcarriers), the \ac{ofdm} \ac{ifft} length is $L_{\text{OFDM},0}=\num{128}$ while the \ac{fc} \ac{fft} and \ac{ifft} lengths are $L_{0}=32$ and $N=4L_{0}=128$, respectively, corresponding to interpolation factor of $I_{0}=4$ and \ac{fc} bin spacing of $f_{\text{BS},0}=4f_{\text{SCS},0}$. The desired \ac{scr} target is $A_\text{des}=\SI{-50}{dB}$ and the overlap in \ac{fc} processing is either \SI{50}{\%} or \SI{25}{\%} ($\lambda=0.5$ or $\lambda=0.25$). 
  
The resulting passband \ac{mse} values are shown in Table~\ref{tab:Example1}. As can be observed from these figures, the effect of \ac{td} synthesis window together with the \ac{fd} window is negligible to \ac{mse} for both overlaps whereas the optimization of all the windows together reduces the \ac{mse} more than \SI{15}{dB} for the overlap of \SI{50}{\%} and more than \SI{10}{dB} for the overlap of \SI{25}{\%}. However, since the synthesis window targets on smoothing the time-domain transients, this window mostly contributes on reducing the spectral leakage. It should be further pointed out that since the \ac{scr} target for the optimization is selected such that it can be met with plain \ac{fc} processing, the full benefits of the synthesis window are not shown in this case. The optimized \ac{fd} and \ac{td} windows are shown in Fig.~\ref{fig:Ex1} and, as seen from these responses, the optimized windows are not typical analytic window designs and, therefore, not feasible to devise without optimization techniques. In addition, these responses are different for all cases essentially meaning that the simultaneous optimization of all the windows is required for the best performance.

The number of variables in the optimization and the corresponding \acs{cpu} times running Matlab R2018a on Intel Xeon E5-2620 are also given in Table~\ref{tab:Example1} for reference. As seen from these optimization times, the parameter reduction techniques of Section \ref{sec:param-wind} considerably reduces the optimization time with only slight degradation to \ac{mse} performance. However, for longer \ac{ofdm} transform sizes, the computational complexity of the fully parameterized optimization increases dramatically. The optimization of \ac{fd} window together with the \ac{td} analysis window allows in this particular case to reach faster essentially the same performance as in Case V with the reduced parameter set.

% ----------------------------------------------------------------------
\subsection{Example 2}%
\label{sec:example-2}
% ----------------------------------------------------------------------
In this example, we evaluate the passband \ac{mse} performance of original and generalized \ac{fc-f-ofdm} for different narrow-band configurations with $N_{\text{PRB},0}=\{1,2,4,8\}$ active \acp{prb} (12, 24, 48, or 96 active subcarriers) and \SI{15}{kHz} \ac{scs} for the \ac{ofdm} processing. The purpose of this example is to more extensively compare the performance and the complexity of the original \ac{fc-f-ofdm} processing with the proposed generalized one. In addition, the goal is to exemplify the effect of \ac{fc} processing bin spacing to filtering performance and implementation complexity.
  
According to \eqref{eq:LOFDMest}, the \ac{ofdm} processing \ac{ifft} length can be selected as $L_{\text{OFDM},0}=128$.  For sampling rate of $f_\text{s}=\SI{30.72}{MHz}$, the interpolation factor of $I_{0}=16$ is needed to achieve the desired overall symbol duration. Assuming that $f_{\text{BS},0}\geq f_{\text{SCS},0}$, then $L_{0}\leq L_{\text{OFDM},0}$ and $N=16L_{0}$. Here, we have chosen the \ac{fc} short transform lengths of $L_{0}=\{16,24,32,48,64,128\}$  corresponding to \ac{fc} processing bin spacings of $f_{\text{BS},0}=\{120,80,60,40,30,15\}$\,kHz, respectively. The \ac{scr} targets in optimization are $A_\text{des}=\{\SI{-40}{dB}, \SI{-45}{dB}, \SI{-50}{dB}, \SI{-55}{dB}, \SI{-60}{dB}, \SI{-65}{dB}\}$ for transform lengths of $L_{0}=\{16,24,32,48,64,128\}$, respectively, and the overlap factor in \ac{fc} processing is selected to be \SI{50}{\%}. The number of transition band weights $L_{\text{TBW},0}$ is selected to cover \SI{360}{kHz} bandwidth corresponding to \SI{180}{kHz} guard band and \SI{180}{kHz} measurement bandwidth, that is, $L_{\text{TBW},0}=\{3,5,6,9,12,24\}$ for $L_{0}=\{16,24,32,48,64,128\}$, respectively.  

\begin{figure} 
    \centering
    \includegraphics[width=1\figWidth]{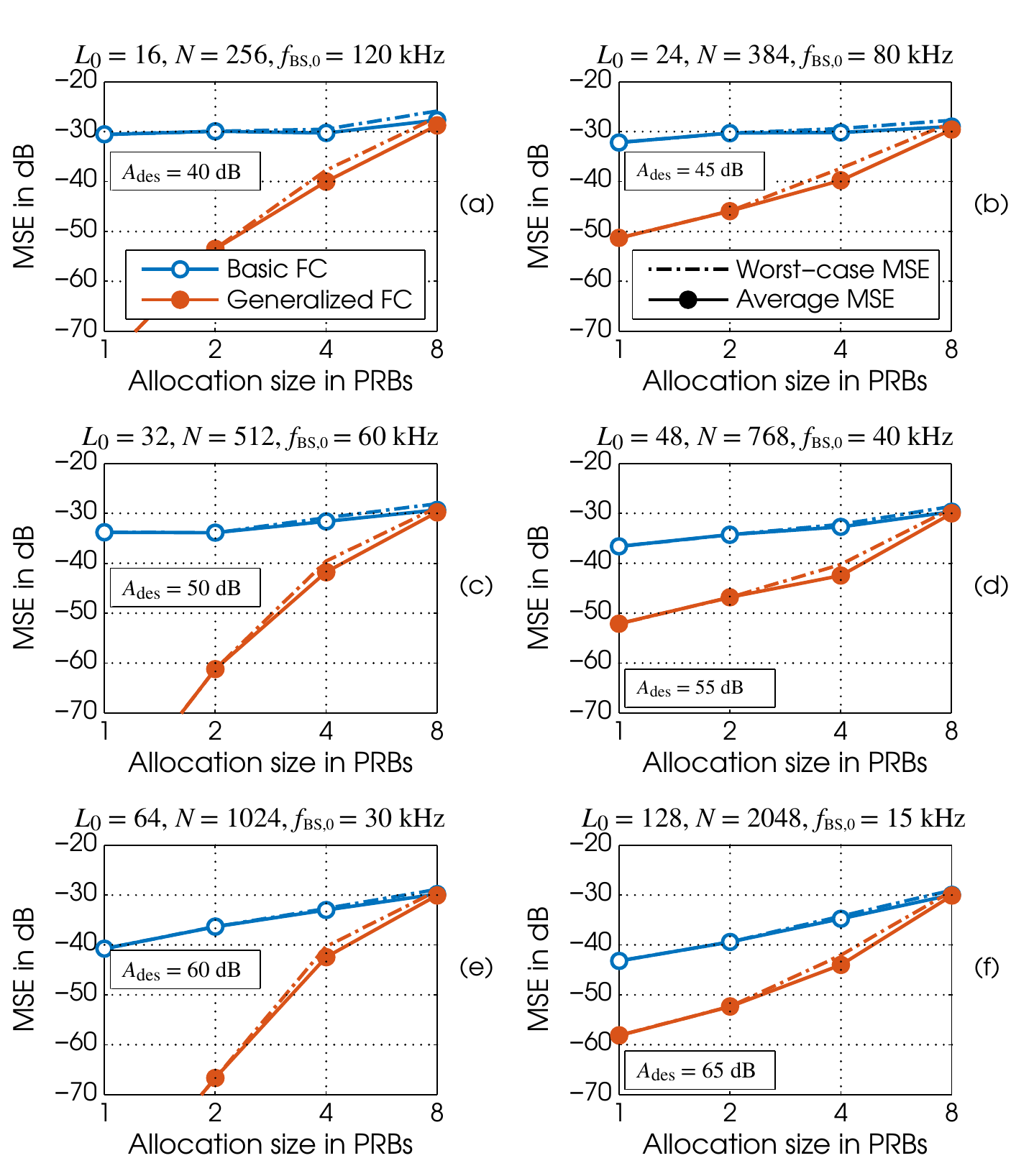}       
    \caption{Average and worst-case passband \acp{mse} for a filtered group of $N_{\text{PRB},0}=\{1,2,4,8\}$ \acp{prb} with \SI{50}{\%} overlap in \ac{fc} processing.}
    \label{fig:IBI_perf}  
\end{figure} 

The results are shown in Figs.~\ref{fig:IBI_perf}(a)--(f) for the given \ac{fc} processing bin spacings. In these evaluations, the average and worst-case \acp{mse}, as given by \eqref{eq:MSEavg} and \eqref{eq:MSEmax}, are used as figure of merit of passband quality. The results are shown for both the proposed generalized processing and for the original \ac{fc} processing in \cite{J:Yli-Kaakinen:JSAC2017} (optimized and evaluated according to same criteria). We can see that both the worst-case and average \acp{mse} are considerably better for the proposed generalized \ac{fc} processing when compared to original scheme with only \ac{fd} windowing.  For both schemes, the worst-case \ac{mse} is slightly higher than the corresponding average. This is obviously due to the fact that on the edge subcarriers, the strict orthogonality is impaired.  This contribution to average \ac{mse} is slightly higher with wider allocations. In addition, it can be observed that \ac{mse} errors increase as the allocation size increases. Furthermore, the difference in performance between these two schemes reduces for increasing allocation size. This is due to the fact that for wider allocations there are less interpolated samples (due to the \ac{fd} zero padding in \ac{ofdm} modulation) which can be used for controlling the spectral localization. Moreover, the processing where $L_{0}=3P$ with $P$ being the power of two has systematically higher \ac{mse} values for narrow allocations when compared with the power-of-two transform lengths. These results can be interpreted in the context of the \ac{evm} requirements of \ac{5g-nr}, stated as \{\SI{17.5}{\%}, \SI{12.5}{\%}, \SI{8}{\%}, \SI{3.5}{\%}\} or \{\SI{-15}{dB}, \SI{-18}{dB}, \SI{-22}{dB}, \SI{-29}{dB}\} for \{QPSK, 16-QAM, 64-QAM, 256-QAM\} modulations, respectively \cite{S:3GPP:TS38.104}. We can conclude that for 256-QAM, \SI{120}{kHz} \ac{scs} can be considered sufficient from the average \ac{evm} point of view.

The implementation complexities of the original and proposed schemes are compared in Table~\ref{tab:complexities}. In this table, the figures annotated with asterisks correspond to original processing where the baseline \ac{fc} processing bin spacing is the same as the \ac{ofdm} processing \ac{scs} \cite{J:Yli-Kaakinen:JSAC2017}, while the complexities for the original processing with increased \ac{fc} processing bin spacing are also given. The numbers typeset in bold face give the corresponding minimum complexities for the proposed scheme. It can be observed from this table that for the proposed scheme the number of real multiplications increases approximately \SI{25}{\%} or \SI{40}{\%} for the \ac{fc} processing bin spacing of \SI{15}{kHz} and \SI{120}{kHz}, respectively. However, overall, the complexity of the proposed processing with increased bin spacing is at least \SI{15}{\%}  lower than that of the original with \SI{15}{kHz} baseline processing \cite{J:Yli-Kaakinen:JSAC2017}. 

\begin{figure}
  \centering
  \includegraphics[width=1\figWidth]{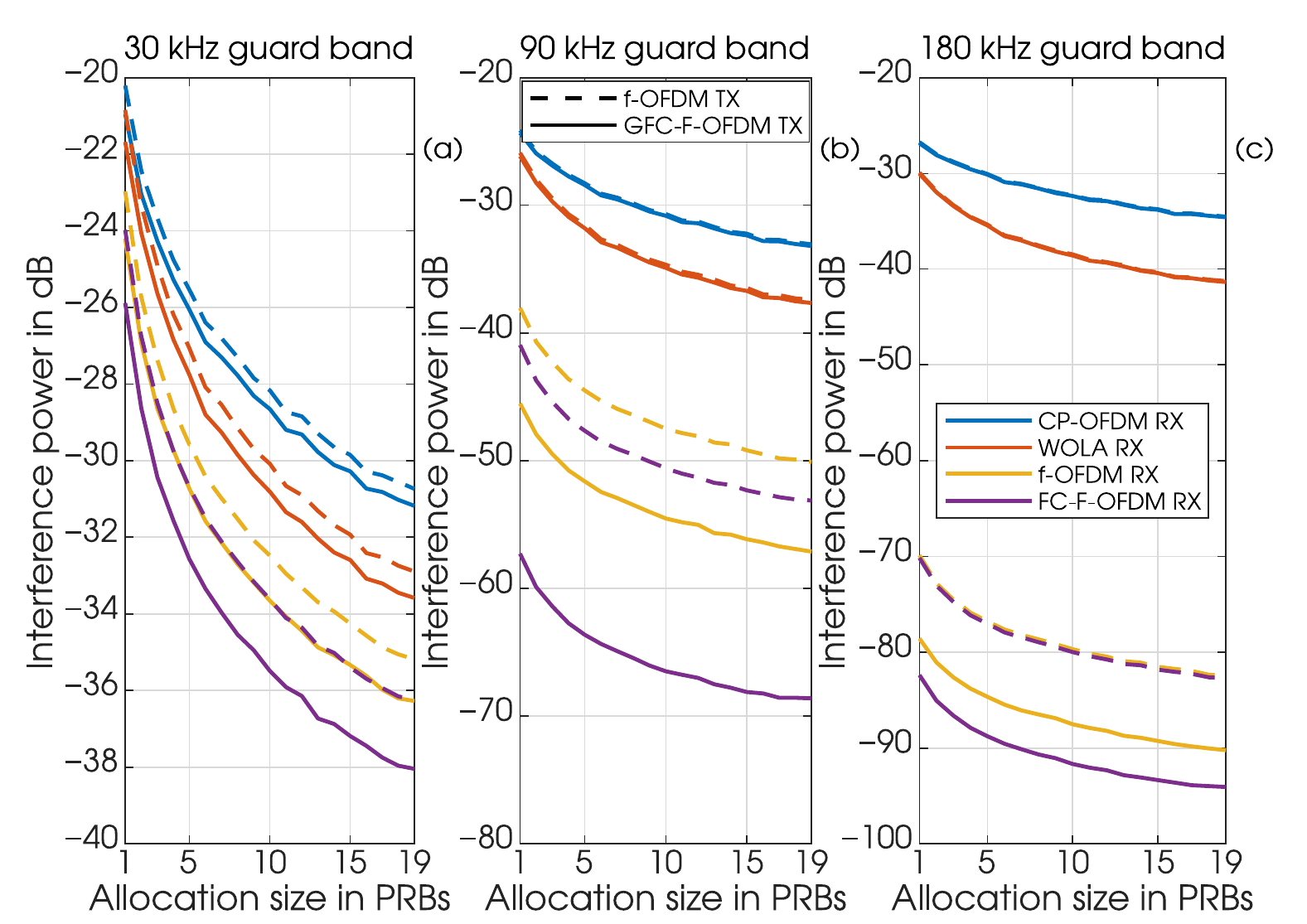}     
  \caption{Inter-numerology interference (INI) seen by the victim \ac{rx} averaged over the allocation size. On the \ac{tx} side, either \ac{td} convolution-based \acs{f-ofdm} or generalized \ac{fc}-based \ac{F-ofdm} processing is used while, on the \ac{rx} side, plain \ac{cp-ofdm}, \ac{wola}, \ac{f-ofdm}, or \ac{fc-f-ofdm} is used. The guard band between the numerologies is (a) \SI{30}{kHz}, (b) \SI{90}{kHz}, or  (c) \SI{180}{kHz}.}
  \label{fig:Leakage}
\end{figure}
  
The complexities of the plain \ac{cp-ofdm}, \ac{wola}, and \ac{td} convolution-based \ac{f-ofdm} are also included in Table~\ref{tab:complexities}. The overlapping extension in \ac{wola} is assumed to be half the \ac{cp} length (\num{72} samples). For \ac{f-ofdm}, the filter length is half the \ac{ofdm} symbol length as defined in \cite{C:2015_Zhang_f-OFDM_for_5G}, that is,  $N_\text{FILT}=N_\text{OFDM}/2$. Therefore, the overall complexity in terms of real multiplications per \ac{ofdm} symbol is $C_\text{M}=N_\text{FILT}(N_{\text{OFDM}}+N_{\text{CP}}) +C_{\text{M}}^\text{(IFFT)}$ when coefficient symmetry is utilized or $C_\text{M}=2N_\text{FILT}(N_{\text{OFDM}}+N_{\text{CP}})/I_0 +C_{\text{M}}^\text{(IFFT)}$ when commutative model for the polyphase interpolator is used. As seen from this table, the complexity of the proposed processing scheme is approximately two times the complexity of \ac{cp-ofdm} or \ac{wola} and \SI{1.42}{\%} or \SI{11.4}{\%} that of the \ac{f-ofdm}.

\begin{table*}[t]
  \addtolength\tabcolsep{-2pt}
  \caption{Computational complexity of the proposed and original schemes for different numbers of processed \acs{cp-ofdm} symbols and \acs{fc} processing \acp{scs} in Example 2. \acs{cp-ofdm}, \acs{wola}, and \ac{td} convolution-based \ac{f-ofdm} (non-interpolating and interpolating processing) are shown here for reference.} 
  \label{tab:complexities}
  \vspace{-0.8em}
  \centering%\newcommand{\wbalsup}[1]
  \newcommand{\mr}[1]{\multirow{2}{*}{#1}}
  \footnotesize{ 
    \begin{tabular}{ccccccc*2{S[table-format=<4.0]}ccccc}
      \toprule
      \multicolumn{1}{c}{\multirow{2}{*}{$B_{\text{OFDM},0}$}} &
      \multicolumn{1}{c}{\multirow{2}{*}{$L_{\text{OFDM},0}$}} &
      \multicolumn{1}{c}{\multirow{2}{*}{$L_{\text{CP},0}$}} &
      \multicolumn{1}{c}{\multirow{2}{*}{$L_{0}$}} &
      \multicolumn{1}{c}{\multirow{2}{*}{$N$}} &
      \multicolumn{1}{c}{\multirow{2}{*}{$f_{\text{SCS},0}$}} &
      \multicolumn{1}{c}{\multirow{2}{*}{$f_{\text{BS},0}$}} & 
      \multicolumn{1}{c}{$C_{\text{M}}$} & 
      \multicolumn{1}{c}{$C_{\text{M}}$} & 
      \multicolumn{1}{c}{\multirow{2}{*}{$N_{\text{OFDM}}$}} &
      \multicolumn{1}{c}{\multirow{2}{*}{$N_{\text{CP}}$}} &
      \multicolumn{1}{c}{$C_{\text{M}}$} &  
      \multicolumn{1}{c}{$C_{\text{M}}$} &  
      \multicolumn{1}{c}{$C_{\text{M}}$} \\
      &&&&&
      & 
      & 
      \multicolumn{1}{c}{(Original)} &
      \multicolumn{1}{c}{(Proposed)} &&&
      \multicolumn{1}{c}{(CP-OFDM)} &
      \multicolumn{1}{c}{(WOLA)} &
      \multicolumn{1}{c}{(f-OFDM)} \\
      \midrule
      \mr{1}  & \mr{128} & \mr{9} & 128 & 2048 & \mr{\SI{15}{kHz}} &  \SI{15}{kHz}  & 68132\textsuperscript{*}                      &         84772 & \mr{2048} & \mr{144} & \mr{\num{16388}}  & \mr{\num{16676}} & {\num{2260996}} \\
      &&                                                & 16 & 256 && \SI{120}{kHz}  & 25292\hphantom{\textsuperscript{*}}           & \textbf{35\,276} & & & & & \num{280576} \\
      \midrule
      \mr{7}  & \mr{128} & \mr{9} & 128 & 2048 & \mr{\SI{15}{kHz}} &  \SI{15}{kHz}  & 39154\textsuperscript{*}                      &         48772 & \mr{2048} & \mr{144} & \mr{\num{16388}}  & \mr{\num{16676}} & \num{2260996} \\
      &&                                                & 16 & 256 && \SI{120}{kHz}  & 23057\hphantom{\textsuperscript{*}}           & \textbf{32\,163}  & & & & & \num{280576} \\
      \midrule 
      \mr{14} & \mr{128} & \mr{9} & 128 & 2048 & \mr{\SI{15}{kHz}} &  \SI{15}{kHz}  & 37946\textsuperscript{*}                      &         47272 & \mr{2048} & \mr{144} & \mr{\num{16388}}  & \mr{\num{16676}} & \num{2260996} \\
      &&                                                & 16 & 256 && \SI{120}{kHz}  & 22963\hphantom{\textsuperscript{*}}           & \textbf{32\,033}  & & & & & \num{280576} \\
      \bottomrule 
    \end{tabular}}  
\end{table*}

The increased bin spacing in \ac{fc} processing also reduces the overhead in short burst transmissions since the relative part of the time-domain zero padding (cf. \eqref{eq:padding}) with respect to the \ac{ofdm} symbol length reduces. For example, with $f_{\text{BS},0}=\SI{120}{kHz}$ \ac{fc} processing bin spacing, the number of output samples to be processed for one \ac{ofdm} symbol is only $R_{0}N_\text{S}=20\times128=2560$ which is half the samples required with $f_{\text{BS},0}=\SI{15}{kHz}$ bin spacing ($R_{0}N_\text{S}=5\times1024=5120$). The latency of the processing is thus also reduced to half since the evaluation of the last samples corresponding to the current \ac{ofdm} symbol is completed two times faster.

In order to further exemplify the performance of the generalized \ac{fc} \ac{F-ofdm}, while also comparing it against \ac{td} filtering-based approach, we consider the following scenario: Four \acp{prb} with \SI{30}{kHz} \ac{scs} are transmitted either using the generalized \ac{fc}-based processing or \ac{td} convolution-based \ac{f-ofdm} processing. The \ac{ofdm} symbol length is $N_{\text{OFDM},0}=1024$ and \ac{cp} length is $L_{\text{CP},0}=72$. The bin-spacing in the \ac{fc} processing is $f_{\text{BS},0}=\SI{15}{kHz}$. Non-active allocation with \SI{15}{kHz} \ac{scs} is located next to active allocation such that the guard band between the allocations is either \SI{30}{kHz}, \SI{90}{kHz}, or \SI{180}{kHz}. The bandwidth of the non-active allocation is adjusted from \num{1} \ac{prb} to \num{19} \acp{prb}. On the \ac{rx} side, either plain \ac{cp-ofdm}, \ac{wola}, \ac{f-ofdm}, or \ac{fc} filtered-\ac{ofdm} is used. In the optimization, the objective function and the constraints are now interchanged such that the \ac{scr} is minimized subject to constraint that in-band \ac{mse} has to be at least \SI{-37.0}{dB}.\footnote{Here, the \ac{mse} target for optimization is chosen based on realized \ac{mse} of \ac{td} convolution-based \ac{f-ofdm} \ac{tx}/plain \ac{cp-ofdm} \ac{rx} pair.} Fig.~\ref{fig:Leakage} shows the interference power evaluated over the non-active subcarriers for each \ac{tx}/\ac{rx} processing alternative. As seen from this figure,  generalized \ac{fc}-based \ac{tx} processing has considerably better performance when compared with \ac{f-ofdm} \ac{tx} while \ac{f-ofdm} \ac{rx} can also be used with generalized \ac{fc}-based \ac{tx} processing. Overall, the results in Fig.~\ref{fig:Leakage} clearly illustrate the excellent bandlimitation properties of the generalized FC based transmitter processing.
 
% ----------------------------------------------------------------------
\subsection{Example 3}% 
\label{sec:example-3}
% ----------------------------------------------------------------------
Here, we consider a wideband example where the \ac{fc} processing carries out the channelization filtering of an overall \SI{5}{MHz} \ac{ofdm} carrier with $n_{\text{ACT}}=300$ active subcarriers ($N_{\text{PRB},0}=25$) and sampling rate of $f_{\text{s}}=\SI{7.68}{MHz}$ and at the same time the signal is interpolated to the output rate of \SI{122.88}{MHz}. Now, the \ac{ofdm} processing \ac{ifft} length has to be at least $L_{\text{OFDM},0}=512$ according to \eqref{eq:LOFDMest}. Assuming that the \ac{fc} \ac{bs} is chosen as $f_{\text{BS},0}=2f_{\text{SCS},0}=\SI{30}{kHz}$, the \ac{fc} transform sizes are $L_{0}=256$ and $N=4096$. The average and worst-case \ac{mse} values for the original and generalized \ac{fc} processing are shown in Table \ref{tab:paramsEx3}. As seen from this table, the generalized model only slightly improves the performance with respect to original processing for the case when the allocation size is larger than the \ac{ofdm} \ac{ifft} size divided by two. 

If the \ac{ofdm} \ac{ifft} size is increased to $L_{\text{OFDM},0}=1024$ and \ac{fc} processing inverse transform size is reduced to $N=2048$ while keeping $L_{\text{OFDM},0}=512$, then the proposed generalized model achieves more than \SI{5}{dB} better average \ac{mse} when compared to the original scheme with same parameterization and \SI{9}{dB} improvement when compared to original \ac{fc-f-ofdm} scheme with \ac{ofdm} \ac{ifft} size of $L_{\text{OFDM},0}=512$. These improved \ac{mse} values are indicated in bold typeface in Tab.~\ref{tab:paramsEx3}.

\begin{table}[t] 
  \caption{Passband \acp{mse} in Example 3. 
  }
  \label{tab:paramsEx3} 
  \vspace{-0.8em}
  \centering
  \footnotesize{
   \begin{tabular}{@{}lcccc@{}}
     \toprule & 
     \multicolumn{2}{p{3.2cm}}{\shortstack{${L}_{\text{OFDM},0}=512$,\\ ${L}_{0}=256$, and $N=4096$}} & 
     \multicolumn{2}{p{3.2cm}}{\shortstack{${L}_{\text{OFDM},0}=1024$,\\ ${L}_{0}=256$, and $N=2048$}} \\ 
     \midrule
     & Original & Proposed  & Original & Proposed  \\
     \midrule 
     $\text{MSE}_{\text{AVG},0}$ & \SI{-33.9}{dB} & \SI{-34.3}{dB} & \SI{-37.3}{dB} & \textbf{\SI{42.9}{dB}} \\
     $\text{MSE}_{\text{MAX},0}$ & \SI{-30.1}{dB} & \SI{-31.7}{dB} & \SI{-31.7}{dB} & \textbf{\SI{34.8}{dB}} \\
    \bottomrule 
    \end{tabular}}   
\end{table}

% ----------------------------------------------------------------------
\subsection{Example 4}% 
\label{sec:example-4}
% ----------------------------------------------------------------------

\begin{figure*}  
    \centering
    \includegraphics[width=0.999\figWidth]{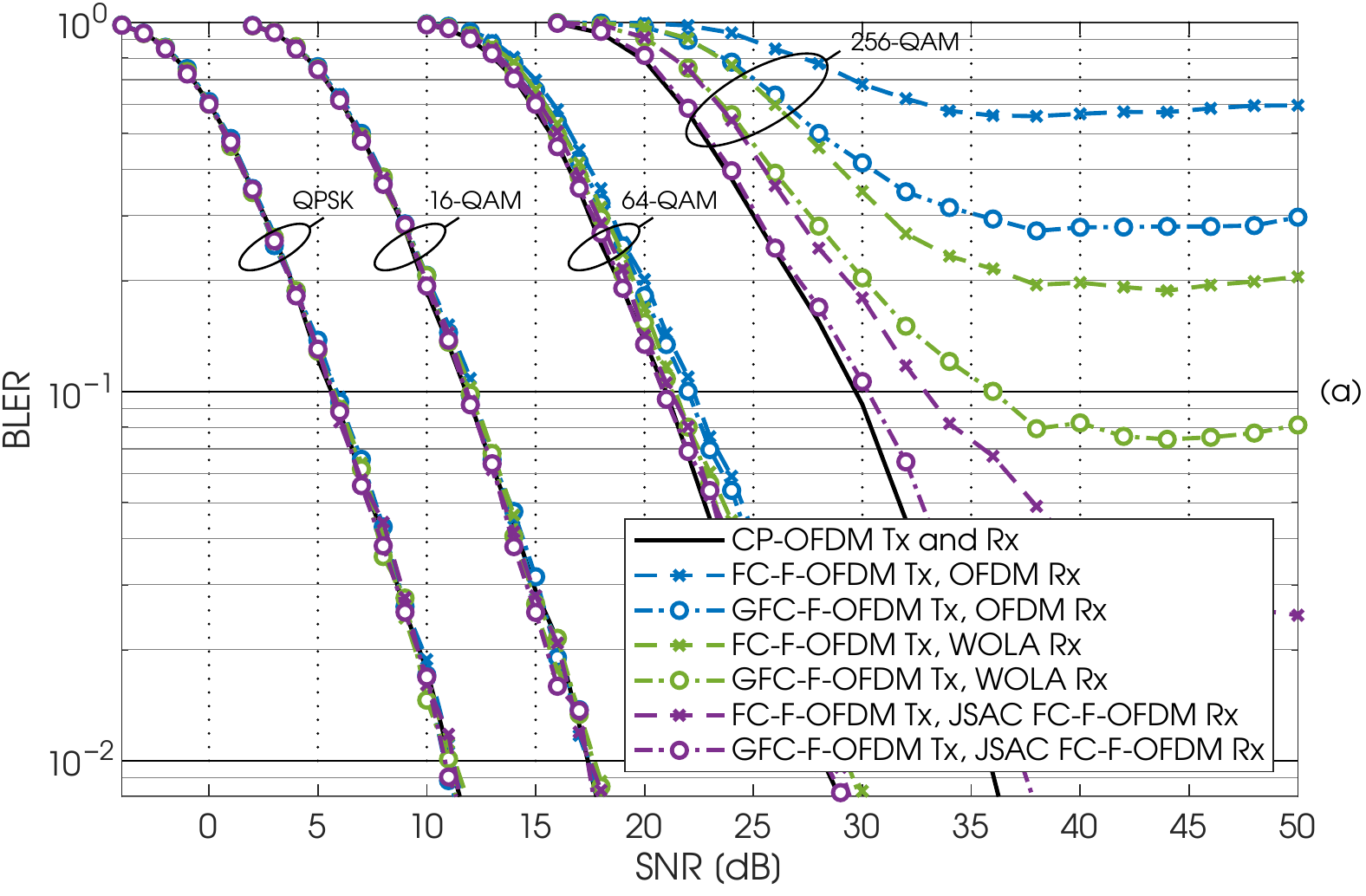}      
    \includegraphics[width=0.999\figWidth]{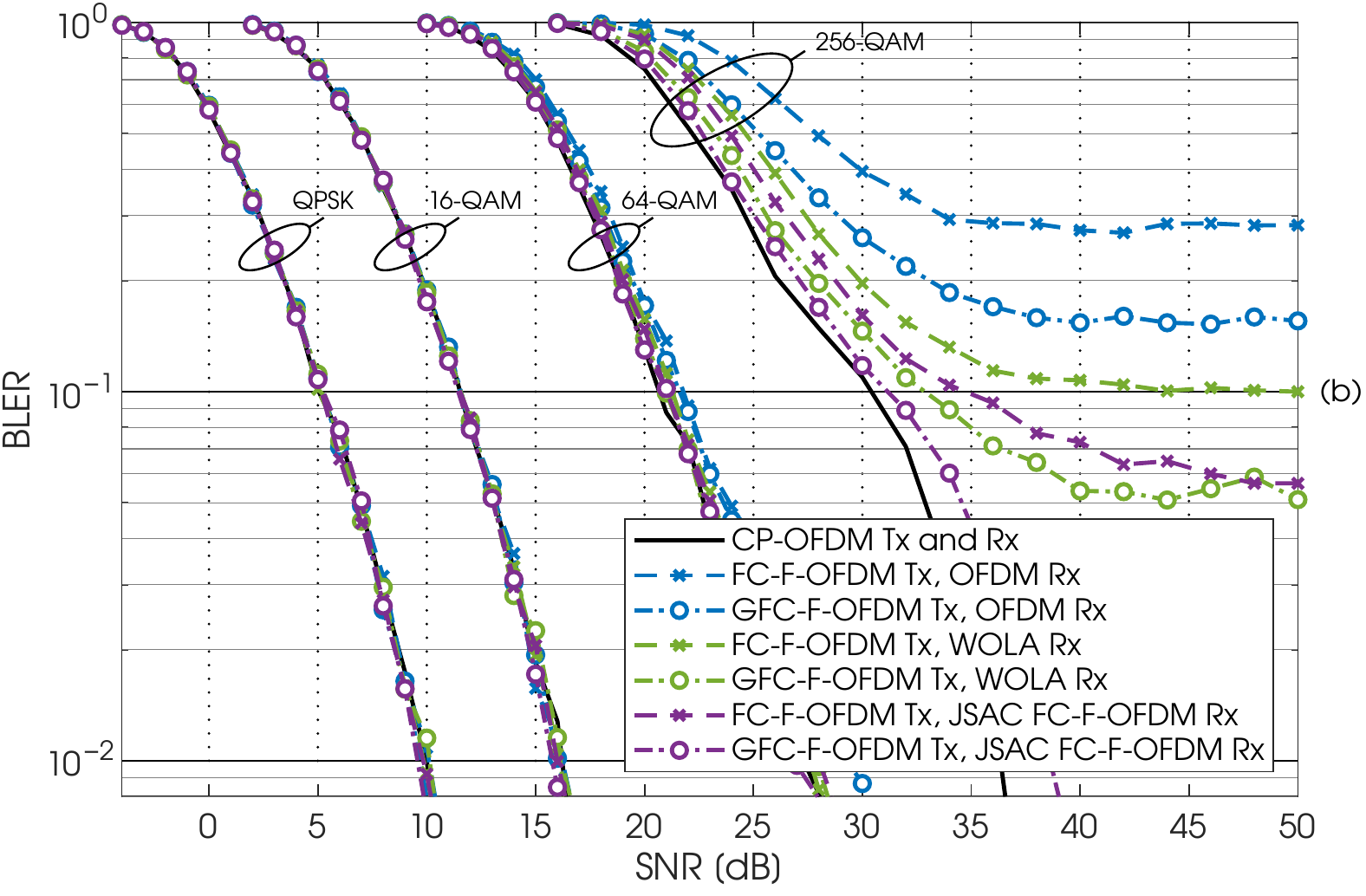}      
    \caption{Block error rate without CFO in downlink mixed numerology interference scenario with (a) 1 PRB and (b) 4 PRB allocation for the desired signal with 1 PRB guard band between the signals. The desired signal is assumed to use \SI{15}{kHz} SCS and the interfering signal is assumed to use \SI{30}{kHz} SCS. CP-OFDM curves refer to ideal reference case with desired signal only.}
    \label{fig:linkPerf_withoutCFO} 
\end{figure*}  

\begin{figure*}  
    \centering
    \includegraphics[width=0.999\figWidth]{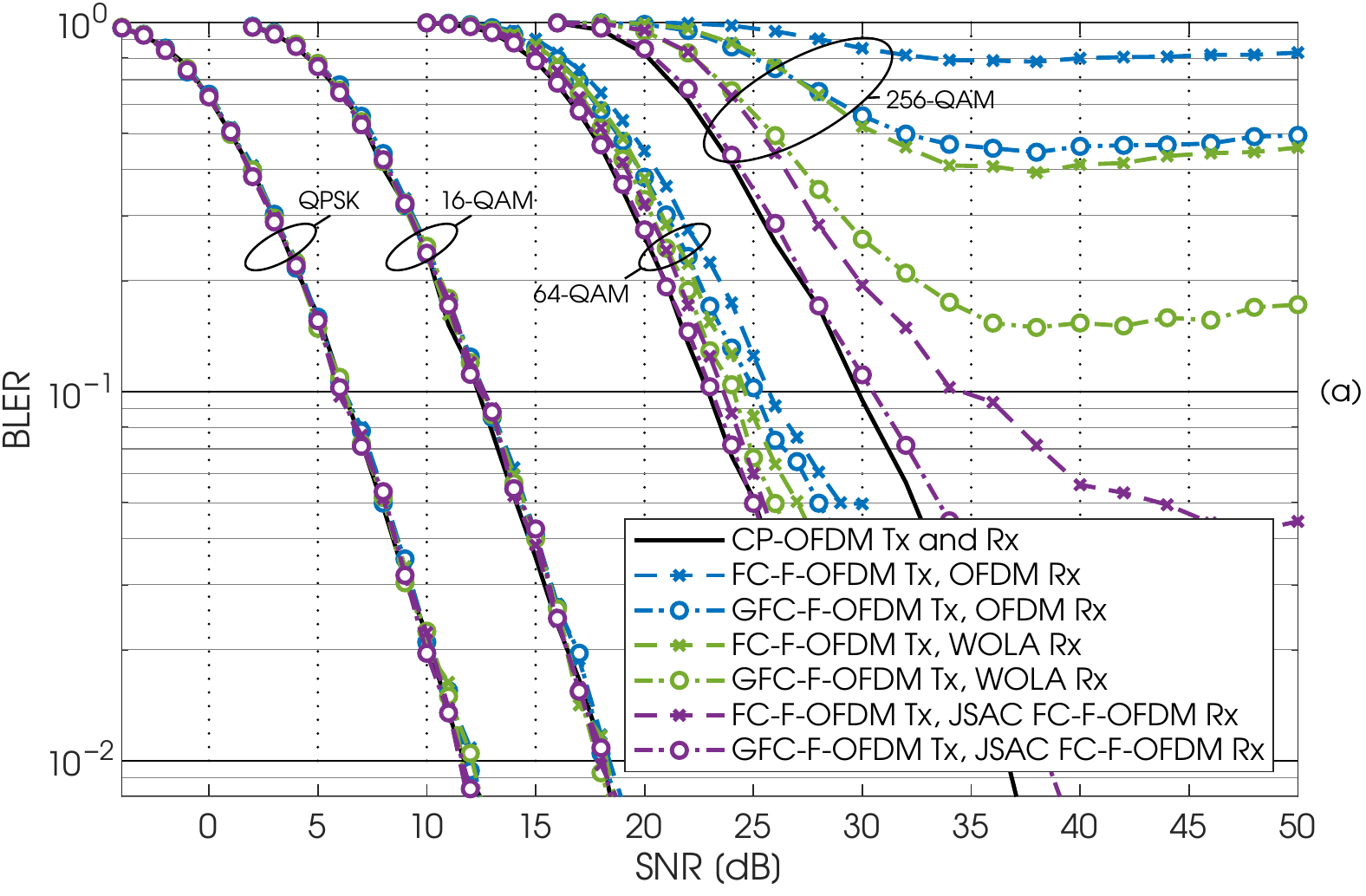}      
    \includegraphics[width=0.999\figWidth]{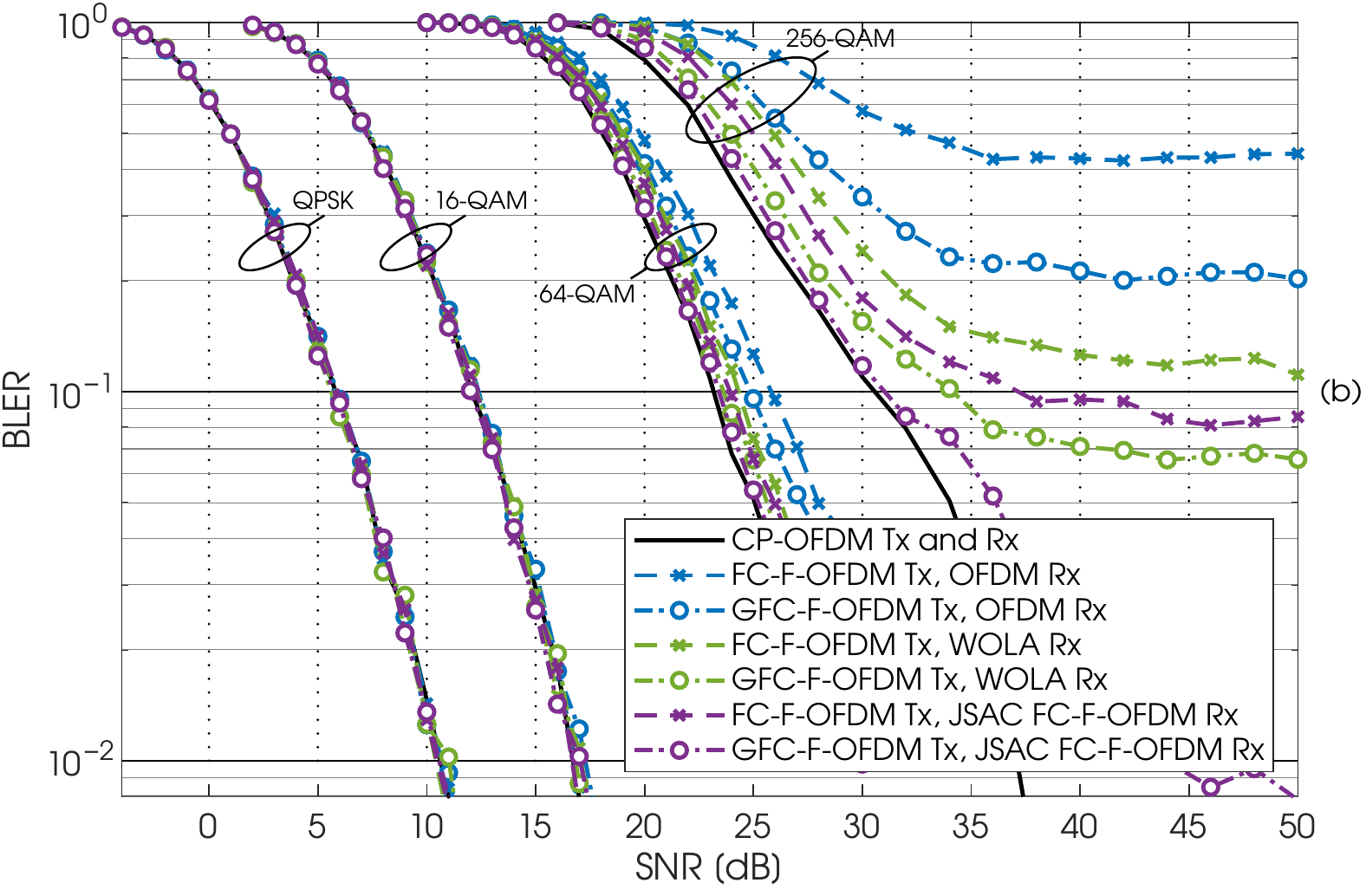}   
    \caption{Block error rate with CFO in downlink mixed numerology interference scenario with (a) 1 PRB and (b) 4 PRB allocation for the desired signal with 1 PRB guard band between the signals. The desired signal is assumed to use \SI{15}{kHz} SCS and the interfering signal is assumed to use \SI{30}{kHz} SCS. CP-OFDM curves refer to ideal reference case with desired signal only.}
    \label{fig:linkPerf_withCFO} 
\end{figure*}  

In this final example, the actual \ac{dl} mixed-numerology radio link performance is evaluated in terms of coded \ac{bler}. We assume that $f_{\text{BS},m}=\SI{120}{kHz}$ is used in both the original and generalized \ac{fc} processing on the \ac{tx} side. This corresponds to the scenario presented in Fig.~\ref{fig:IBI_perf}(a), which allows to minimize the complexity and the latency of the \ac{fc} processing as indicated in Table \ref{tab:complexities}. On the \ac{rx} side either plain \ac{cp-ofdm} receiver, \ac{wola} \cite{C:Wang06:WOLA,C:Zayani16} based \ac{rx} waveform processing, or original \ac{fc-f-ofdm}-based receiver waveform processing \cite{J:Yli-Kaakinen:JSAC2017} is applied. With \ac{wola}, a rising or falling window slope length of $L_{\text{CP},m}/4$ samples with $m=0,1$ is assumed and the slope follows the well known raised-cosine response. With original \ac{fc-f-ofdm} \ac{rx} operating with \SI{15}{kHz} bin spacing, we have used the filter design presented in \cite{J:Yli-Kaakinen:JSAC2017} to optimize \ac{rx} \ac{fd} window with \num{12} transition-band bins and \SI{10}{dB} attenuation target. 

The radio link performance of the desired \ac{cp-ofdm} signal with $f_{\text{SCS},0} = \SI{15}{kHz}$ and with either 1~\ac{prb} or 4~\ac{prb} allocation is measured while being interfered from one side by another \ac{cp-ofdm} signal with $f_{\text{SCS},1}=\SI{30}{kHz}$ and with a fixed 4~\ac{prb} allocation. The guard band between the two different numerologies is assumed to be \SI{180}{kHz}, following the earlier discussion. As we are modeling \ac{dl} mixed numerology interference, it is assumed that the base station transmitter applies the same waveform processing on both transmitted signals. The evaluated modulations correspond to \ac{qpsk}, 16-\ac{qam}, 64-\ac{qam}, and 256-\ac{qam} with coding rates $R=602/1024$, $R=658/1024$, $R=873/1024$, and $R=885/1024$, respectively. These correspond to \ac{mcs} indices 4, 10, 19, and 25 in the \ac{mcs} index Table 2 defined in \cite[Table 5.1.3.1-2]{S:3GPP:TS38.214}. A wide range of evaluated \acp{mcs} allows to understand which modulations are sensitive to the \ac{mse} levels induced by considered \ac{tx} processing solutions and also highlight how different \ac{rx} processing solutions affect the performance. In addition, 256-\ac{qam} modulation based operation point is included to illustrate how generalized \ac{fc} processing allows us to use high \ac{mcs} even with narrow allocations and minimal complexity increase compared to original \ac{fc} processing. The assumed slot length for both subband signals is \num{14} \ac{ofdm} symbols and the performance is averaged over \num{5000} independent channel realizations. The assumed radio propagation model is a \ac{tdl} C channel \cite{S:3GPP:TR38.900} with \SI{300}{ns} root-mean-squared delay spread. The evaluations are based on a \ac{5g-nr} compliant link simulator.

In Fig.~\ref{fig:linkPerf_withoutCFO}, illustrating the link performance without \ac{cfo}, the relatively small \ac{mse} induced by the subband filtering does not have significant effect with modulation orders smaller than 256-\ac{qam}. With 64-\ac{qam}, approximately \SI{2}{dB} difference in the required \ac{snr} for $\text{BLER}=10^{-1}$ is observed when comparing plain \ac{ofdm} \ac{rx} performance to \ac{wola} or \ac{fc-f-ofdm}-based \ac{rx} performance. This difference is mainly from the assumed \ac{rx} processing. In the case of 256-\ac{qam}, the effect of using generalized \ac{fc} or original \ac{fc} processing in the \ac{tx} is more clearly visible, and the error floor behaviour with generalized \ac{fc} \ac{tx} processing is defined by the considered \ac{rx} waveform processing solution. The \ac{cp-ofdm}-based \ac{rx} collects interference from neighboring subband and therefore provides the worst performance in all cases. With \ac{wola}-based \ac{rx}, the performance is clearly improved, but due to the limited selectivity the performance is not as good as with \ac{fc}-based \ac{rx}. As expected, lower order modulations are not as sensitive to \ac{tx} \ac{mse} induced by the subband filtering, whereas with 256-\ac{qam} clear differences can be observed. The \ac{rx} waveform processing has a significant impact on the link performance, although highly selective and low in-band distortion enabling generalized \ac{fc} processing is applied in \ac{tx}. Another way to look at these results, is to note that using generalized \ac{fc}-based \ac{tx} processing allows the best possible performance from the \ac{tx} side, and the experienced link performance depends on the \ac{rx} implementation, which can be improved in the future device generations if highly selective subband filtering is applied in the devices. Based on Fig. \ref{fig:Leakage}, it can also be presumed that the effect of filtering becomes more profound when the guard band between numerologies is reduced from \SI{180}{kHz} and that the effect becomes visible also with lower modulation orders.

To highlight the possible effect of \ac{cfo} on the \ac{dl} link performance, results with \ac{cfo} are shown in Fig. \ref{fig:linkPerf_withCFO}. As we are concentrating on \ac{dl} performance, the \ac{cfo} between the base station and the user equipment affects both subbands similarly. Thus, the receiving device observes the same frequency offset in the desired signal and in the interfering signal. In these evaluations, we have assumed a \SI{350}{Hz} \ac{cfo}, corresponding to the 0.1~part-per-million (\SI{0.1}{ppm}) accuracy required from the base station transmitter, as defined in \cite{S:3GPP:TS38.104}, while noting that the performance evaluations are done at the \SI{3.5}{GHz} carrier frequency. Based on the results shown in Fig.~\ref{fig:linkPerf_withCFO}, we can observe that with 64-\ac{qam}, \ac{cp-ofdm} \ac{rx}, \ac{wola} \ac{rx}, and FC-F-OFDM \ac{rx} require approximately \SI{2.5}{dB}, \SI{1.5}{dB}, and \SI{0.5}{dB} larger \ac{snr} at \SI{10}{\%} \ac{bler} target level, respectively, when using FC-F-OFDM \ac{tx}, and \SI{2}{dB}, \SI{1}{dB}, and \SI{0.1}{dB} larger \ac{snr} at \SI{10}{\%} \ac{bler} target level, respectively, when using GFC-F-OFDM \ac{tx}. It is noted that with 256-\ac{qam}, due to its larger sensitivity to \ac{cfo} \cite{J:Pollet1995:OFDMsensitivityCFOandPN}, we have applied a \ac{dmrs}-based fine-frequency tracking in the receiver. From Fig. \ref{fig:linkPerf_withCFO} (a), we can observe that in the case of 1~\ac{prb} allocation, \ac{fc}-based \ac{rx} processing is required to support 256-\ac{qam} modulation, and with generalized \ac{fc}-based \ac{tx} the performance is very close to the interference free reference. In the case of 4~\acp{prb}, as shown in Fig. \ref{fig:linkPerf_withCFO} (b), it is interesting to note that by applying generalized \ac{fc} processing on the \ac{tx} side, we are able to support 256-\ac{qam} modulation also with \ac{wola}-based \ac{rx}. Comparing the 256-\ac{qam} results shown in Fig. \ref{fig:linkPerf_withCFO} with Fig. \ref{fig:linkPerf_withoutCFO}, we can observe that the clear in-band \ac{mse} improvement provided by the genralized \ac{fc} \ac{tx} processing is maintained also under \ac{cfo}. 
Generalized \ac{fc} processing is thus shown to be the most efficient processing solution to flexibly allocate different numerologies with only \SI{180}{kHz} guard band apart from each other while simultaneously supporting full range of different \ac{mcs} values currently defined in the \ac{5g-nr} specifications and allowing more degrees of freedom in the implementation.

% ======================================================================
\section{Conclusions}%
In this article, focusing on non-orthogonal multiple-access scenarios with CP-OFDM waveform, we presented a novel spectrum enhancement method combining analysis and synthesis time-domain windowing with subband filtering implemented through the fast-convolution process. It was found that joint optimization of the time- and frequency-domain windows can offer greatly enhanced performance over basic subband filtering, and also over straightforward combinations of filtering and time-domain windowing.  This conclusion was verified also by detailed link-level simulations in typical \ac{5g-nr} downlink mixed-numerology scenarios. The performance gain over original \ac{fc-f-ofdm} is pronounced with relatively narrow allocations using modulation and coding schemes aiming at high spectral efficiency. Generally, the demonstrated performance gains help to enhance the spectrum utilization efficiency of \ac{cp-ofdm}-based multiservice wireless networks, like the emerging \ac{5g-nr}. In this article, we also extended the parametrization alternatives of original and generalized \ac{fc-f-ofdm} by considering independent choice of the subcarrier spacing and the \ac{fft} bin spacing in \ac{fc} processing. This was found to support feasible performance with greatly reduced complexity and processing latency, providing additional degrees of freedom in the design and implementation.  

Strong emphasis in this work was on the transparent spectrum enhancement schemes, which allows fast initial deployment and backwards compatible enhancements of \ac{5g-nr} technology. The proposed scheme was applied for the transmitter side of the link, while plain \ac{cp-ofdm} receiver was assumed in the joint optimization of the time- and frequency domain window coefficients. Throughout the different examples, the received signal quality and link performance were shown to improve with different transparent receiver waveform processing solutions when generalized \ac{fc-f-ofdm} was applied in the transmitter. However, the idea of combined joinly-optimized time- and frequency-domain windowing can be applied on the receiver side as well. Therefore, the optimization of receiver-side processing in transparent way, and joint transmitter-receiver optimization using time- and frequency-domain windows will be the main topics for our future studies.    

\label{sec:conclusions}

\bibliographystyle{IEEEtran} 
\bibliography{jour_short,conf_short,References} 

% ======================================================================
\begin{IEEEbiography}[{\includegraphics[width=1in,height=1.25in,clip,keepaspectratio]{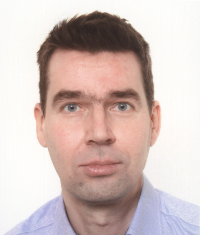}}]{Juha Yli-Kaakinen} received the degree of Diploma Engineer in electrical engineering and the Doctor of Technology (Hons.) degree from the Tampere University of Technology (TUT), Tampere, Finland, in 1998 and 2002, respectively.

Since 1995, he has held various research positions with TUT. His research interests are in digital signal processing, especially in digital filter and filter-bank optimization for communication systems and very large scale integration implementations.
\end{IEEEbiography}

\begin{IEEEbiography}[{\includegraphics[width=1in,height=1.25in,clip,keepaspectratio]{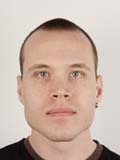}}]{Toni Levanen} received the M.Sc. and D.Sc. degrees from Tampere University of Technology (TUT), Finland, in 2007 and 2014, respectively. He is currently with the Department of Electrical Engineering, Tampere University. 

In addition to his contributions in academic research, he has worked in industry on wide variety of development and research projects. His current research interests include physical layer design for 5G NR, interference modelling in 5G cells, and high-mobility support in millimeter-wave communications.
\end{IEEEbiography}

\begin{IEEEbiography}[{\includegraphics[width=1in,height=1.25in,clip,keepaspectratio]{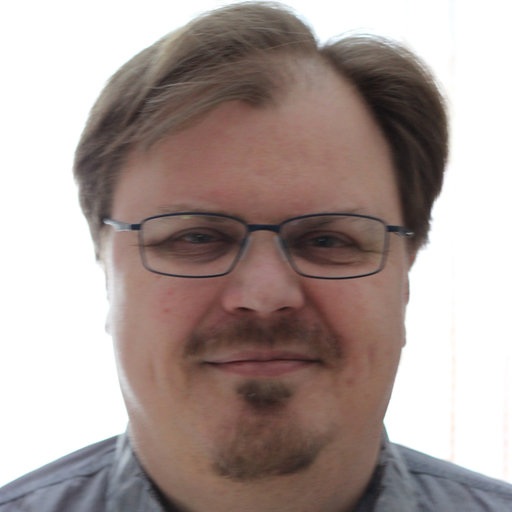}}]{Arto Palin} has long industrial experience in wireless technologies, covering cellular networks, broadcast systems and local area communications. He holds an MSc. (Tech.) degree from earlier Tampere University of Technology, and is currently working as Technical Leader at Nokia Mobile Networks, Finland, in the area of 5G SoC architectures.
\end{IEEEbiography}

% \newpage

\begin{IEEEbiography}[{\includegraphics[width=1in,height=1.25in,clip,keepaspectratio]{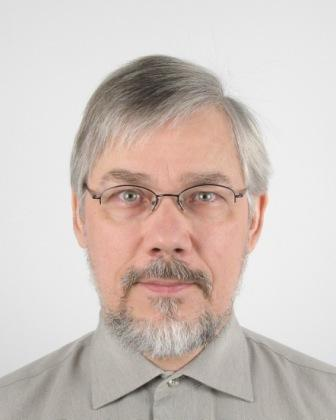}}]{Markku Renfors} (S’77--M’82--SM’90--F’08) received the D.Tech. degree from the Tampere University of Technology (TUT), Tampere, Finland, in 1982.
Since 1992, he has been a Professor with the Department of Electronics and Communications Engineering, TUT, where he was the Head from 1992 to 2010. His research interests include filter-bank based multicarrier systems and signal processing algorithms for flexible communications receivers and transmitters.

Dr.~Renfors was a corecipient of the Guillemin Cauer Award (together with T.~Saramäki) from the IEEE Circuits and Systems Society in 1987.
\end{IEEEbiography}

\begin{IEEEbiography}[{\includegraphics[width=1in,height=1.25in,clip,keepaspectratio]{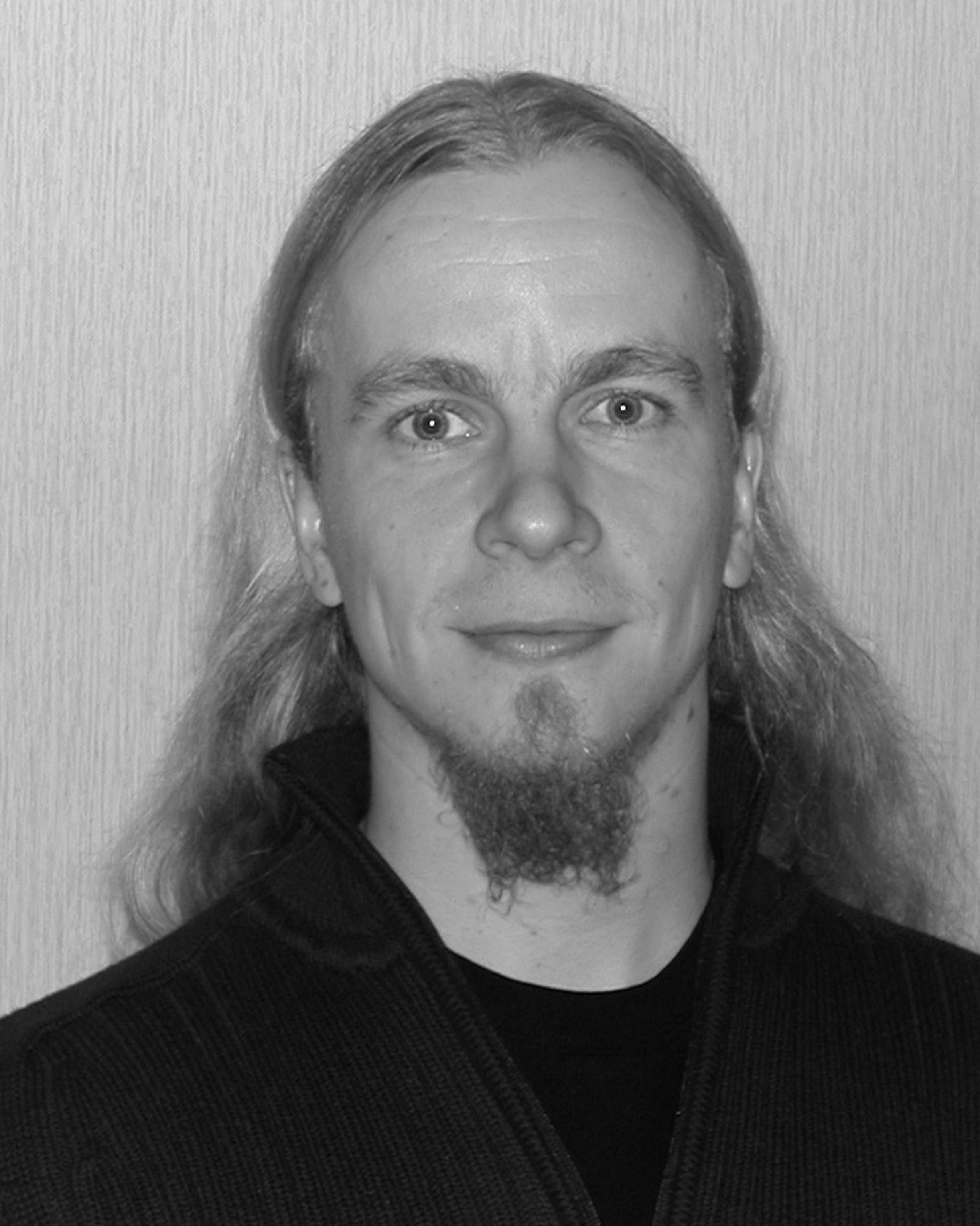}}]{Mikko Valkama} received the D.Sc. (Tech.) degree (with honors) from Tampere University of Technology, Finland, in 2001. 
In 2003, he was with the Communications Systems and Signal Processing Institute at SDSU, San Diego, CA, as a visiting research fellow. Currently, he is a Full Professor and Department Head of Electrical Engineering at newly formed Tampere University (TAU), Finland. His general research interests include radio communications, radio localization, and radio-based sensing, with particular emphasis on 5G and beyond mobile radio networks.
\end{IEEEbiography}

% Acronyms that won't appear in the printed list
%
% Acronyms for the printed list
% Rewritten using lowercase
\begin{acronym}[FBMC/OQAM2]
\acro{tdl}[TDL]{tapped-delay line}
\acro{zf}[ZF]{zero-forcing} 
\acro{3gpp}[3GPP]{3rd generation partnership project}
\acro{5g-nr}[5G-NR]{fifth generation new radio}
\acro{5g}[5G]{5th generation}
\acro{F-ofdm}[F-OFDM]{subband filtered CP-OFDM}
\acro{aclr}[ACLR]{adjacent channel leakage ratio}
\acro{afb}[AFB]{analysis filter bank} 
\acro{bler}[BLER]{block error rate}
\acro{br}[BR]{bin resolution}
\acro{bs}[BS]{bin spacing}
\acro{cb-fmt}[CB-FMT]{cyclic block-filtered multitone}
\acro{cp-ofdm}[CP-OFDM]{cyclic prefix orthogonal frequency-division multiplexing}
\acro{cp}[CP]{cyclic prefix} 
\acro{cpu}[CPU]{central processing unit} 
\acro{dft-s-ofdm}[DFT-s-OFDM]{DFT-spread-OFDM}
\acro{dft}[DFT]{discrete Fourier transform} 
\acro{dl}[DL]{downlink} 
\acro{e-utra}[E-UTRA]{evolved UMTS terrestrial radio access}
\acro{embb}[eMBB]{enhanced mobile broadband}
\acro{evm}[EVM]{error vector magnitude}
\acro{f-ofdm}[f-OFDM]{filtered OFDM}
\acro{td-f-ofdm}[TD-F-OFDM]{time-domain filtered OFDM}
\acro{fbmc-coqam}[FBMC-COQAM]{filterbank multicarrier with circular offset-QAM}
\acro{fbmc/oqam}[FBMC/OQAM]{filter bank multicarrier with offset-QAM subcarrier modulation} 
\acro{fbmc}[FBMC]{filter bank multicarrier}
\acro{fb}[FB]{filter bank}
\acro{fc-f-ofdm}[FC-F-OFDM]{FC-based F-OFDM}
\acro{fc-fb}[FC-FB]{fast-convolution filter bank}
\acro{fc}[FC]{fast-convolution}
\acro{fd}[FD]{frequency-domain}
\acro{fft}[FFT]{fast Fourier transform} 
\acro{fir}[FIR]{finite impulse response}
\acro{fmt}[FMT]{filtered multitone} 
\acro{gb}[GB]{guard band}
\acro{gfdm}[GFDM]{generalized frequency-division multiplexing}
\acro{ibe}[IBE]{in-band emission}
\acro{ibi}[IBI]{in-band interference}
\acro{ibo}[IBO]{input back-off}
\acro{ici}[ICI]{inter-carrier interference}
\acro{idft}[IDFT]{inverse discrete Fourier transform}
\acro{ifft}[IFFT]{inverse fast Fourier transform}
\acro{isi}[ISI]{inter-symbol interference}
\acro{lpsv}[LPSV]{linear periodically shift variant} 
\acro{lptv}[LPTV]{linear periodically time-varying} 
\acro{lte}[LTE]{long-term evolution}
\acro{mcm}[MCM]{multicarrier modulation}
\acro{mcs}[MCS]{modulation and coding scheme}
\acro{mc}[MC]{multicarrier}
\acro{mse}[MSE]{mean-squared error} 
\acro{npr}[NPR]{near perfect reconstruction}
\acro{nr}[NR]{new radio}
\acro{ofdm}[OFDM]{orthogonal frequency-division multiplexing} \acro{ofdma}[OFDMA]{orthogonal frequency-division multiple access} 
\acro{ola}[OLA]{overlap-and-add} 
\acro{ols}[OLS]{overlap-and-save}
\acro{oobem}[OOBEM]{out-of-band emission mask}
\acro{oob}[OOB]{out-of-band}
\acro{oqam}[OQAM]{offset quadrature amplitude modulation}
\acro{pa}[PA]{power amplifier} 
\acro{prb}[PRB]{physical resource block}
\acro{pr}[PR]{perfect reconstruction}
\acro{psd}[PSD]{power spectral density} 
\acro{qam}[QAM]{quadrature amplitude modulation}
\acro{qpsk}[QPSK]{quadrature phase-shift keying}
\acro{ran}[RAN]{radio access network}
\acro{rbg}[RBG]{resource block group}
\acro{rb}[RB]{resource block}
\acro{rc}[RC]{raised cosine} 
\acro{rms}[RMS]{root mean squared \acroextra{[error]}}
\acro{rrc}[RRC]{square root raised cosine}
\acro{rx}[RX]{receiver}
% \acro{sblr}[SBLR]{subband leakage ratio}
\acro{scr}[SCR]{spectral confinement ratio}
\acro{sc-fdma}[SC-FDMA]{single-carrier frequency-division multiple access} 
\acro{scs}[SCS]{subcarrier spacing}
\acro{sc}[SC]{single-carrier}
\acro{sdr}[SDR]{software defined radio}
\acro{sfb}[SFB]{synthesis filter bank}
\acro{sqp}[SQP]{sequential quadratic programming}
\acro{tbw}[TBW]{transition-band width}
\acro{td}[TD]{time-domain}
\acro{tmux}[TMUX]{transmultiplexer}
\acro{to}[TO]{tone offset}
\acro{tsg}[TSG]{technical specification group}
\acro{tx}[TX]{transmitter} 
\acro{ue}[UE]{user equipment} 
\acro{uf-ofdm}[UF-OFDM]{universal filtered OFDM}
\acro{ul}[UL]{uplink} 
\acro{urllc}[URLLC]{ultra-reliable low-latency communications}
\acro{wola}[WOLA]{weighted overlap-add} 
\acro{wola}[WOLA]{windowed overlap-and-add} 
\acro{zp}[ZP]{zero prefix} 
\acro{4g}[4G]{4th Generation} 
\acro{adsl}[ADSL]{Asymmetric Digital Subscriber Line}
\acro{af}[AF]{Amplify-and-Forward} 
\acro{am/am}[AM/AM]{Amplitude Modulation/Amplitude
Modulation \acroextra{[NL PA models]}}
\acro{am/pm}[AM/PM]{Amplitude Modulation/Phase Modulation
\acroextra{[NL PA models]}} \acro{amr}[AMR]{Adaptive Multi-Rate}
\acro{ap}[AP]{Access Point} \acro{app}[APP]{A Posteriori
Probability} \acro{awgn}[AWGN]{Additive White Gaussian Noise}
\acro{bcjr}[BCJR]{Bahl-Cocke-Jelinek-Raviv \acroextra{algorithm}}
\acro{ber}[BER]{Bit Error Rate} \acro{bicm}[BICM]{Bit-Interleaved
Coded Modulation} \acro{blast}[BLAST]{Bell Labs Layered Space Time
\acroextra{[code]}} 
\acro{b-pmr}[B-PMR]{Broadband PMR} \acro{bpsk}[BPSK]{Binary
Phase-Shift Keying} 
\acro{cazac}[CAZAC]{Constant Amplitude Zero Auto-Correlation}
\acro{ccc}[CCC]{Common Control Channel}
\acro{ccdf}[CCDF]{Complementary Cumulative Distribution Function}
\acro{cdf}[CDF]{Cumulative Distribution Function}
\acro{cdma}[CDMA]{Code-Division Multiple Access}
\acro{cfo}[CFO]{carrier frequency offset} 
\acro{cfr}[CFR]{Channel Frequency Response} \acro{ch}[CH]{Cluster Head}
\acro{cir}[CIR]{Channel Impulse Response} \acro{cma}[CMA]{Constant
Modulus Algorithm} \acro{cna}[CNA]{Constant Norm Algorithm}
\acro{cqi}[CQI]{Channel Quality Indicator} \acro{cr}[CR]{Cognitive
Radio} \acro{crlb}[CRLB]{Cram\'er-Rao Lower Bound}
\acro{crn}[CRN]{Cognitive Radio Network}
\acro{crs}[CRS]{Cell-specific Reference Signal}
\acro{csi}[CSI]{Channel State Information}
\acro{csir}[CSIR]{Channel State Information at the Receiver}
\acro{csit}[CSIT]{Channel State Information at the Transmitter}
\acro{d2d}[D2D]{Device-to-Device} \acro{dc}[DC]{Direct Current}
\acro{df}[DF]{Decode-and-Forward} \acro{dfe}[DFE]{Decision
Feedback Equalizer} 
\acro{dmo}[DMO]{Direct Mode Operation}
\acro{dmrs}[DMRS]{demodulation reference signals}
\acro{dsa}[DSA]{Dynamic Spectrum Access} \acro{dzt}[DZT]{Discrete
Zak Transform} \acro{ed}[ED]{Energy Detector}
\acro{egf}[EGF]{Extended Gaussian Function}
\acro{em}[EM]{Expectation Maximization} \acro{emse}[EMSE]{Excess
Mean Square Error}
%\acro{emphatic}[EMPhAtiC]{FP7-ICT project}
\acro{epa}[EPA]{Extended Pedestrian-A \acroextra{[channel model]}}
\acro{etsi}[ETSI]{European Telecommunications Standards Institute}
\acro{eva}[EVA]{Extended Vehicular-A \acroextra{[channel model]}}
\acro{fb-sc}[FB-SC]{FilterBank Single-Carrier} 
\acro{fdma}[FDMA]{Frequency-Division Multiple Access}
\acro{fec}[FEC]{Forward Error Correction} 
\acro{flo}[FLO]{Frequency-Limited Orthogonal}
\acro{fpga}[FPGA]{Field Programmable Gate Array} 
\acro{fs-fbmc}[FS-FBMC]{Frequency Sampled FBMC-OQAM} 
\acro{ft}[FT]{Fourier Transform}
\acro{glrt}[GLRT]{Generalized Likelihood Ratio Test}
\acro{gmsk}[GMSK]{Gaussian Minimum-Shift Keying}
\acro{hpa}[HPA]{High Power Ampliﬁer}
\acro{iid}[i.i.d.]{independent and identically distributed}
\acro{i/q}[I/Q]{In-phase/Quadrature \acroextra{[complex data
signal components]}} \acro{iam}[IAM]{Interference Approximation
Method}
\acro{iota}[IOTA]{Isotropic Orthogonal Transform Algorithm}
\acro{itu}[ITU]{International Telecommunication Union}
\acro{itu-r}[ITU-R]{International Telecommunication Union
Radiocommunication \acroextra{sector}}
\acro{kkt}[KKT]{Karush-K\"uhn-Tucker} \acro{kpi}[KPI]{Key
Performance Indicator} \acro{le}[LE]{Linear Equalizer}
\acro{llr}[LLR]{Log-Likelihood Ratio} \acro{lmmse}[LMMSE]{Linear
Minimum Mean Squared Error} \acro{lms}[LMS]{Least Mean Squares}
\acro{ls}[LS]{Least Squares} 
\acro{lte-a}[LTE-A]{Long-Term Evolution-Advanced}
\acro{lut}[LUT]{Look Up Table} \acro{ma}[MA]{Multiple Access Relay
Channel} \acro{mac}[MAC]{Medium Access Control}
\acro{mai}[MAI]{Multiple Access Interference}
\acro{map}[MAP]{Maximum A Posteriori} 
\acro{mer}[MER]{Message Error Rate} \acro{mf}[MF]{Matched Filter}
\acro{mgf}[MGF]{Moment Generating Function}
\acro{mimo}[MIMO]{Multiple-Input Multiple-Output}
\acro{miso}[MISO]{Multiple-Input Single-Output}
\acro{ml}[ML]{Maximum Likelihood} \acro{mlse}[MLSE]{Maximum
Likelihood Sequence Estimation} \acro{mma}[MMA]{Multi-Modulus
Algorithm} \acro{mmse}[MMSE]{Minimum Mean-Squared Error}
\acro{mos}[MOS]{Mean Opinion Score} \acro{mrc}[MRC]{Maximum Ratio
Combining} \acro{ms}[MS]{Mobile Station}
\acro{msk}[MSK]{Minimum-Shift
Keying} 
\acro{mtc}[MTC]{machine type communications}
\acro{mu}[MU]{Multi-User} 
\acro{mud}[MUD]{MultiUser
Detection} \acro{mui}[MUI]{MultiUser Interference}
\acro{music}[MUSIC]{MUltiple SIgnal Classification}
\acro{nbi}[NBI]{NarrowBand Interference}
\acro{nc}[NC]{Non-Contiguous}
\acro{nc-ofdm}[NC-OFDM]{Non-Contiguous OFDM}
\acro{nl}[NL]{NonLinear} \acro{nmse}[NMSE]{Normalized Mean-Squared
Error} 
\acro{obo}[OBO]{Output Back-Off} 
\acro{ofdp}[OFDP]{Orthogonal
Finite Duration Pulse}\acro{omp}[OMP]{Orthogonal Matching Pursuit}
\acro{oqpsk}[OQPSK]{Offset Quadrature Phase-Shift Keying}
\acro{osa}[OSA]{Opportunistic Spectrum Access} \acro{pam}[PAM]{Pulse Amplitude Modulation}
\acro{papr}[PAPR]{Peak-to-Average Power Ratio}
\acro{pci}[PCI]{Perfect Channel Information}
\acro{per}[PER]{Packet Error Rate} \acro{pf}[PF]{Proportional
Fair} \acro{phy}[PHY]{Physical layer}  \acro{plc}[PLC]{Power Line
Communications}
\acro{pmr}[PMR]{Professional (or Private) Mobile
Radio} \acro{ppdr}[PPDR]{Public Protection and Disaster Relief}
% \acrodefplular{rb}[RB]{resource blocks}
\acro{prose}[ProSe]{Proximity Services} 
\acro{psk}[PSK]{Phase-Shift Keying}
 \acro{pswf}[PSWF]{Prolate Spheroidal Wave Function}
\acro{pts}[PTS]{Partial Transmit Sequence}
\acro{ptt}[PTT]{Push-To-Talk} \acro{pu}[PU]{Primary User}
\acro{pucch}[PUCCH]{Physical Uplink Control Channel}
\acro{pusch}[PUSCH]{Physical Uplink Shared Channel}
\acro{qoe}[QoE]{Quality of Experience} \acro{qos}[QoS]{Quality of
Service}
\acro{ram}[RAM]{Random Access Memory} \acro{rat}[RAT]{Radio Access
Technology} 
\acro{rf}[RF]{Radio Frequency} \acro{rls}[RLS]{Recursive
Least Squares}  \acro{roc}[ROC]{Receiver Operating
Characteristics}
\acro{rrm}[RRM]{Radio Resource Management}
\acro{rssi}[RSSI]{Received Signal Strength Indicator}
\acro{sc-fde}[SC-FDE]{Single-Carrier Frequency-Domain
Equalization} 
\acro{sdm}[SDM]{Space-Division
Multiplexing} \acro{sdma}[SDMA]{Space-Division  Multiple Access}
% \acro{sdr}[SDR]{Signal-to-Distortion power Ratio}
\acro{sel}[SEL]{Soft Envelope Limiter} \acro{ser}[SER]{Symbol
Error Rate} 
\acro{sfbc}[SFBC]{Space Frequency Block Code}
\acro{sic}[SIC]{Successive Interference Cancellation}
\acro{simo}[SIMO]{Single-Input Multiple-Output}
\acro{sinr}[SINR]{Signal-to-Interference-plus-Noise Ratio}
\acro{sir}[SIR]{Signal-to-Interference Ratio}
\acro{siso}[SISO]{Single-Input Single-Output}
\acro{softio}[SI-SO]{Soft-Input Soft-Output}
\acro{slm}[SLM]{Selected Mapping}
\acro{slnr}[SLNR]{Signal-to-Leakage-plus-Noise  Ratio}
\acro{sm}[SM]{Spatial Multiplexing}
\acro{sndr}[SNDR]{Signal-to-Noise-plus-Distortion Ratio}
\acro{snr}[SNR]{signal-to-noise ratio} \acro{ss}[SS]{Spectrum
Sensing} \acro{sspa}[SSPA]{Solid-State Power Amplifiers}
\acro{stbc}[STBC]{Space Time Block Code}
\acro{stbicm}[STBICM]{Space-Time Bit-Interleaved Coded Modulation}
\acro{stc}[STC]{Space-Time Coding} \acro{sthp}[STHP]{Spatial
Tomlinson Harashima Precoder} \acro{su}[SU]{Secondary Users}
\acro{svd}[SVD]{Singular Value Decomposition}
\acro{tdd}[TDD]{Time-Division Duplex}
\acro{tdma}[TDMA]{Time-Division Multiple Access}
\acro{teds}[TEDS]{TETRA Enhanced Data Service}
\acro{tetra}[TETRA]{Terrestrial Trunked Radio}
\acro{tfl}[TFL]{Time Frequency Localization} \acro{tgf}[TGF]{Tight
Gabor Frame} \acro{tlo}[TLO]{Time-Limited Orthogonal}
\acro{tmo}[TMO]{Trunked Mode Operation} 
\acro{tr}[TR]{Tone Reservation} 
\acro{ula}[ULA]{Uniform Linear
Array} \acro{v-blast}[V-BLAST]{Vertical Bell Laboratories Layered
Space-Time \acroextra{[code]}} \acro{veh-a}[Veh-A]{Vehicular-A
\acroextra{[channel model]}} \acro{veh-b}[Veh-B]{Vehicular-B
\acroextra{[channel model]}} \acro{wimax}[WiMAX]{Worldwide
Interoperability for Microwave Access} \acro{wlan}[WLAN]{Wireless
Local Area Network} \acro{wlf}[WLF]{Widely Linear Filter}
\acro{zt}[ZT]{Zak Transform}
\end{acronym}

%%% Local Variables: 
%%% fill-column: 80
%%% mode: latex
%%% TeX-master: "TSPv1_5"
%%% End:  
 
 % file with acronym definitions
  
\end{document}